\documentclass[fleqn,usenatbib]{mnras}

\usepackage[normalem]{ulem}

\usepackage[T1]{fontenc}

\DeclareRobustCommand{\VAN}[3]{#2}
\let\VANthebibliography\thebibliography
\def\thebibliography{\DeclareRobustCommand{\VAN}[3]{##3}\VANthebibliography}

\usepackage{xcolor}

\usepackage{graphicx}	% Including figure files
\usepackage{amsmath}	% Advanced maths commands
\usepackage{amssymb}	% Extra maths symbols
\usepackage{soul}
\usepackage{newtxtext,newtxmath}
\title[Stellar mass -- size relation at low mass]{Extending the evolution of the stellar mass -- size relation at $z \leq 2$ to low stellar mass galaxies from HFF and CANDELS}

\author[K. V. Nedkova et al.]
{
Kalina V. Nedkova$^{1}$\footnote{E-mail: kalina.nedkova@tufts.edu},
Boris H\"au\ss ler$^{2}$,
Danilo Marchesini$^{1}$,
Paola Dimauro$^{3}$,
Gabriel Brammer$^{4,5}$,
\newauthor
Paul Eigenthaler$^{6}$,
Adina D. Feinstein$^{7\thanks{NSF Fellow}}$,
Henry C. Ferguson$^{8}$,
Marc  Huertas-Company$^{9,10}$,\newauthor
Evelyn J. Johnston$^{11}$,
Erin Kado-Fong$^{12}$, 
Jeyhan S. Kartaltepe$^{13}$,
Ivo Labbé$^{14}$,
Daniel Lange-Vagle$^{1}$,\newauthor
Nicholas S. Martis$^{15}$,  
Elizabeth J. McGrath$^{16}$,
Adam Muzzin$^{17}$,
Pascal Oesch$^{4, 5, 18}$,
Yasna Ordenes-Briceño$^{6}$,\newauthor
Thomas Puzia$^{6}$,
Heath V. Shipley$^{19}$,
Brooke D. Simmons$^{20}$, 
Rosalind E. Skelton$^{21}$, 
Mauro Stefanon$^{22}$, \newauthor
Arjen van der Wel$^{23}$,
and 
Katherine E. Whitaker $^{5,24}$
\\
\\
$^{1}$Department of Physics \& Astronomy, Tufts University, 574 Boston Avenue, MA 02155, USA\\
$^{2}$European Southern Observatory, Alonso de Cordova 3107, Casilla 19001, Santiago, Chile\\
$^{3}$Óbservatório Nacional / MCTIC, Rua General José Cristino 77, Rio de Janeiro, RJ, 20921-400, Brazil\\
$^{4}$ Niels Bohr Institute, University of Copenhagen, Jagtvej 128, København N, DK-2200, Denmark \\
$^{5}$ Cosmic Dawn Center (DAWN), Copenhagen, Denmark\\
$^{6}$Institute of Astrophysics, Pontificia Universidad Católica de Chile, Av. Vicuña Mackenna 4860, 7820436 Macul, Santiago, Chile \\
$^{7}$Department of Astronomy and Astrophysics, University of Chicago, 5640 S. Ellis Ave, Chicago, IL 60637, USA \\
$^{8}$Space Telescope Science Institute, 3700 San Martin Drive, Baltimore, MD 21218 \\
$^{9}$LERMA, Observatoire de Paris, PSL Research University, CNRS, Sorbonne Universités, UPMC Univ. Paris 06,F-75014 Paris, France\\
$^{10}$Instituto de Astrofísica de Canarias, E-38200 La Laguna, Tenerife, Spain \\
$^{11}$N\'ucleo de Astronom\'ia de la Facultad de Ingenier\'ia y Ciencias, Universidad Diego Portales, Av. Ej\'ercito Libertador 441, Santiago, Chile\\
$^{12}$Department of Astrophysical Sciences, Princeton University, Peyton Hall, Princeton, NJ 08544, USA \\
$^{13}$School of Physics and Astronomy, Rochester Institute of Technology, 84 Lomb Memorial Drive, Rochester NY 14623, USA\\
$^{14}$Centre for Astrophysics and Supercomputing, Swinburne University of Technology, Hawthorn, Victoria 3122, Australia \\
$^{15}$NSM: NRC Herzberg, 5071 West Saanich Rd, Victoria, BC V9E 2E7, Canada\\
$^{16}$Department of Physics and Astronomy, Colby College, Waterville, ME 04961, USA\\
$^{17}$Department of Physics and Astronomy, York University, 4700 Keele St., Toronto, Ontario, M3J 1P3, Canada \\
$^{18}$Department of Astronomy, Universit\'e de Gen\`eve, Chemin Pegasi 51, 1290 Versoix, Switzerland\\
$^{19}$Department of Physics, McGill University, 3600 Rue University, Montréal, QC H3P 1T3, Canada\\
$^{20}$Department of Physics, Lancaster University, Bailrigg, Lancaster, LA1 4YB, UK\\
$^{21}$South African Astronomical Observatory, PO Box 9, Observatory, Cape Town, 7935, South Africa\\
$^{22}$Leiden Observatory, Leiden University, NL-2300 RA Leiden, The Netherlands\\
$^{23}$Sterrenkundig Observatorium, Universiteit Gent, Krijgslaan 281 S9, B-9000 Gent, Belgium \\
$^{24}$Department of Astronomy, University of Massachusetts Amherst, 710 N Pleasant Street, Amherst, MA 01003, USA
}

\date{Accepted 2021 June 7. Received 2021 June 3; in original form 2021 January 7}

\pubyear{2021}

\begin{document}
\label{firstpage}
\pagerange{\pageref{firstpage}--\pageref{lastpage}}
\maketitle

% Abstract of the paper
\begin{abstract}
We reliably extend the stellar mass – size relation over $0.2\leq z \leq2$ to low stellar mass galaxies by combining the depth of Hubble Frontier Fields (HFF) with the large volume covered by CANDELS. Galaxies are simultaneously modelled in multiple bands using the tools developed by the MegaMorph project, allowing robust size (i.e., half-light radius) estimates even for small, faint, and high redshift galaxies. We show that above 10$^7$M$_\odot$, star-forming galaxies are well represented by a single power law on the mass–size plane over our entire redshift range. Conversely, the stellar mass – size relation is steep for quiescent galaxies with stellar masses $\geq 10^{10.3}$M$_\odot$ and flattens at lower masses, regardless of whether quiescence is selected based on star-formation activity, rest-frame colours, or structural characteristics. This flattening occurs at sizes of $\sim1$kpc at $z\leq1$. As a result, a double power law is preferred for the stellar mass – size relation of quiescent galaxies, at least above 10$^7$M$_\odot$. We find no strong redshift dependence in the slope of the relation of star-forming galaxies as well as of high mass quiescent galaxies. We also show that star-forming galaxies with stellar masses $\geq$10$^{9.5}$M$_\odot$ and quiescent galaxies with stellar masses $\geq10^{10.3}$M$_\odot$ have undergone significant size growth since $z\sim2$, as expected; however, low mass galaxies have not. Finally, we supplement our data with predominantly quiescent dwarf galaxies from the core of the Fornax cluster, showing that the stellar mass—size relation is continuous below 10$^7$M$_\odot$, but a more complicated functional form is necessary to describe the relation.
\end{abstract}


%%%%%%%%%%%%%%%%%%%%%%%%%%%%%%%%%%%%%%%%%%%%%%%%%%
%%%%%%%%%%%%%%%%% BODY OF PAPER %%%%%%%%%%%%%%%%%%
\section{Introduction}

Understanding the relative importance of the different formation and evolutionary mechanisms that are responsible for shaping quiescent and star-forming galaxies at different epochs of the Universe continues to be one of the most fundamental goals of extragalactic astronomy. 
Signatures of the physical mechanisms that drive the formation pathways through which a galaxy evolves are imprinted on its structure, making galaxy structure and size key observational quantities.
While it has now been well established that both quiescent (early-type) and star-forming (late-type) galaxies were more compact at higher redshift \citep[e.g.,][]{Daddi2005ApJ,Trujillo2007, Dokkum2008}, it is still under debate if the size growth that galaxies have undergone since the early Universe is primarily driven by major mergers \citep[e.g.,][]{Naab2007ApJ,McIntosh2014MNRAS} minor mergers \citep[e.g.,][]{Buitrago2008ApJ, Naab2009ApJ, Oser2012ApJ, Newman2012ApJ}, feedback \citep[e.g.,][]{Fan2008ApJ, Damjanov2009ApJ}, secular processes, or other mechanisms.

Although galaxies evolve via complex, nonlinear processes, their structural parameters exhibit a number of tight scaling relations that aid us in understanding their evolution. The stellar mass -- size relation is particularly interesting because both stellar mass and effective radius have been shown to correlate with the quenching process at least for galaxies with stellar masses above $10^{9}$ M$_{\odot}$ \citep[e.g.,][]{Omand2014MNRAS, Chen2020ApJ}.  While the exact nature of the mechanisms that quench galaxies remains unknown, it seems there are two main quenching channels: one that quenches high mass galaxies and another that quenches low mass galaxies. \cite{Peng2010ApJ} termed these `mass quenching' and `environment quenching' where galaxies that undergo mass quenching are quenched by internal processes such as stellar feedback and AGN feedback. Environment quenching primarily impacts less massive satellite galaxies, which are more likely to be quenched by environmental processes such as ram pressure stripping or harassment from neighbouring objects.

Numerous models have been proposed to explain the quenching process and its relation to stellar mass and galaxy size. \cite{vdWel2009ApJ} suggest that star-forming galaxies evolve on the stellar mass--size plane until they reach a redshift-dependent velocity dispersion threshold. Above this velocity dispersion threshold, galaxies can no longer efficiently form stars, and so, they quench. After quenching, these galaxies undergo subsequent dry, minor mergers which result in the observed trend that quenched galaxies experience a steeper growth on the stellar mass--size plane. \cite{vanDokkum2015ApJ} also describe a simple picture explaining these trends, where galaxies evolve on the mass--size plane along `parallel tracks' according to r $\propto$ M$^{0.3}$, where the growth is predominantly driven by star formation until they reach a central density, after which they quench. They then experience steeper evolutionary tracks according to r $\propto$ M$^{2}$, where the growth is again primarily driven dry mergers. Recently, \cite{Chen2020ApJ} proposed another model in which the radius of star-forming galaxies indirectly dictates when they will quench. They argue that at fixed stellar mass, star-forming galaxies with larger sizes have smaller black holes due to their lower central densities. Therefore, larger galaxies must evolve to higher stellar masses in order to quench, ultimately resulting in a scenario where smaller star-forming galaxies quench at higher redshift. This toy model successfully explains several observed characteristics including the parallel tracks that \cite{vanDokkum2015ApJ} describe. %Analogous studies of the $\Sigma_{1}$ -- M$_*$ relation, where $\Sigma_{1}$ is stellar-mass surface density within a projected radius of 1 kpc \citep[e.g.,][]{Tacchella2015Sci,Barro2017ApJ} have shown that quenched galaxies populate a narrow locus in the $\Sigma_{1}$ -- M$_*$  plane -- referred to as the  $\Sigma_{1}$ ridgeline.

Major efforts have gone into constraining the exact behaviour of the  quiescent and star-forming sequences on the mass--size plane.  We now know that star-forming and quiescent galaxies follow different tracks \citep[e.g.,][]{vdWel2014, vanDokkum2015ApJ, Dimauro2019} such that star-forming galaxies tend to be larger in size than quiescent galaxies across a large range of redshifts and stellar masses. \cite{vdWel2014} have shown that the sizes of star-forming galaxies are proportional to the virial radius of their host dark matter halo, likely a result of conservation of angular momentum \citep{Somerville2018}. Quiescent galaxies, on the other hand, follow a steeper slope on the mass--size plane and are more compact than star-forming galaxies of similar masses \citep[e.g.,][]{Cimatti2008A&A,vdWel2014, Dimauro2019}. In recent years, \cite{Mosleh2017ApJ} and \cite{Suess2019ApJ} have argued that using half-mass, as opposed to half-light, radii results in stellar mass -- size relations which are significantly shallower and suggest that galaxies, especially star-forming ones, have not grown significantly in size since the early Universe. \cite{Suess2019ApJ} therefore claim that the evolution that is observed in the stellar mass -- size relation, when half-light radii are used, is primarily due to colour gradients. While these results pose a challenge to our current picture of galaxy evolution, multi-wavelength software, such as that developed as part of the MegaMorph project \citep{MegaMorph, Vika2013MNRAS}, which we use in this work, are well-suited to address these questions in the future.

While these observational results have provided a valuable test of evolutionary models, a key drawback of most previous studies has been their relatively high mass limits of galaxies above $\sim$ 10$^{9.5}$ M$_{\odot}$. Galaxies in this high stellar mass regime are strongly gravitationally bound systems, which protects them from environmental influences \citep[e.g.,][]{Moore1996} and therefore most likely evolve via the `mass quenching' channel. In order to understand all galaxies, both high mass and low mass, it is important to constrain the stellar mass -- size relation for low mass galaxies across a wide redshift range.

Significant improvements have also been made in terms of extending the stellar mass -- size relation to high redshift \citep[e.g.,][]{Allen2017ApJ, Hill2017ApJ} as well as to low stellar masses \citep[e.g.,][]{Shen2003,Baldry2012MNRAS,Lange2015MNRAS,Morishita2017}. Unfortunately, high redshift studies are often limited to high mass objects while studies that probe the evolution of low mass galaxies have been restricted to local galaxies \citep[see][for $z \sim 0.7$]{Morishita2017}. Thankfully, hydrodynamical simulations have allowed us to overcome the challenge of studying low mass systems at high redshift and have proven to be an invaluable test of our current understanding of galaxy evolution. Through hydrodynamical simulations, key scaling relations can be studied in a way that is impossible with observations alone. \cite{Furlong2017MNRAS} and  \cite{Genel_illustris_2018} have shown that the \textsc{EAGLE} and IllustrisTNG simulations, respectively, are remarkably successful in reproducing the observed stellar mass -- size relation. Both works found results that were consistent with \cite{vdWel2014} and were able to evolve their galaxies in time to provide key insights on how star-forming and quiescent galaxies assemble over time. But, some tensions still remain between simulations and observed galaxy sizes, especially for small galaxies \citep[e.g.,][]{Pillepich2018MNRAS}.

In order to solve these tensions, it is necessary to explore how compact, low mass galaxies evolve. In this work, we present such an analysis and extend the stellar mass -- size relation to low stellar masses $and$ to higher redshifts than previous observational studies \citep[e.g.,][]{Trujillo2004ApJ, Oesch2010, Mosleh2012, Barro2014ApJ, vdWel2014, Holwerda2015, Shibuya2015, Hill2017ApJ, Dimauro2019} by making use of the depth of the Hubble Frontier Fields (HFF) \citep{Lotz2017} and the large area probed by the CANDELS images \citep{Koekemoer2011}, as well as the multi-wavelength capabilities of the software developed as part of the MegaMorph project \citep{MegaMorph, Vika2013MNRAS}.

The paper is structured as follows. In Section \ref{sec:MegaMorph}, we explain the technical setup for the galaxy profile fitting and modelling. In Section \ref{sec:data}, we describe the galaxy sample used for this study. In Section \ref{sec:M_Re}, we show the stellar mass -- size relation for both quiescent and star-forming galaxies, and we describe the flattening of stellar mass -- size relation for quiescent galaxies in Section \ref{sec:discussion}. Finally, we summarise our findings in Section \ref{sec:conclusion}. Throughout this paper, we use a Hubble constant of H$_{0} = 70$ km s$^{-1}$ Mpc$^{-1}$ and cosmological density parameters $\Omega_{\mathrm{m}}$ = 0.3 and $\Omega_\Lambda$ = 0.7. We assume a \cite{Chabrier2003} initial mass function (IMF) for all estimates of stellar mass and all magnitudes are quoted in the AB system.

\section{\textsc{MegaMorph}} \label{sec:MegaMorph}

The tools developed as part of the MegaMorph ($\textbf{Me}$asuring $\textbf{Ga}$laxy $\textbf{Morph}$ology) project \citep[][]{MegaMorph, Vika2013MNRAS} allow robust size measurements as well as a characterisation of morphological properties to be reliably established, even for high redshift galaxies, because they use and model data at all available wavelengths simultaneously, effectively increasing the signal-to-noise ratio ($S/N$) of the data. Because of this capability, we choose to model the light profiles of all galaxies in our sample with these tools. We address key aspects of the software below, but refer the reader to \cite{MegaMorph} for details about the codes and their reliability and accuracy. %\bh{Generally speaking, you're making [the multiwavelength fitting] point 3 times, once here, once in the next sentence, and once in the next paragraph. (This is generally something I thought a few times in the paper. Try to make every point only once, unless it's important to re-iterate on it} 
%Therefore, MegaMorph presents the best set of software to model and fit the structural parameters of the galaxies in our sample 
%\bh{Given that I co-author your paper, this sounds arrogant, even if I think it's true. Maybe reword that.}.

MegaMorph builds on $\textsc{Galapagos}$ \citep{Barden2012} and $\textsc{Galfit}$ \citep{Peng2010Galfit}, but both codes have been modified to allow for multi-wavelength fitting. 
This modification allows galaxy light profiles to be more reliably fit than with other commonly used software, such as $\textsc{GIM2D}$ \citep{Simard1998,Simard2002}, $\textsc{BUDDA}$ \citep{2004BUDDA}, PyMorph \citep{Pymorph2010MNRAS}, $\textsc{Galfit}$ \citep{Peng2002AJ, Peng2010Galfit}, and $\textsc{Imfit}$ \citep{Imfit2015ApJ}. % because the MegaMorph tools are not only able to model the wavelength dependence that many structural parameters exhibit, but are also less sensitive to low $S/N$ imaging. 
$\textsc{GalfitM}$, which is based on $\textsc{Galfit}$ \citep{Peng2010Galfit}, is a two-dimensional fitting code designed to extract structural properties from galaxy images. The galaxy models allow the user to fit any number of components and functional forms. In this work, we model every object with a single S{\'e}rsic component in order to measure properties of each galaxy as a whole.  $\textsc{Galapagos-2}$ automates the source detection, the two-dimensional S{\'e}rsic profile modelling using $\textsc{GalfitM}$, and catalogue creation. Both codes have been adapted from their original versions to fit multiwavelength data by replacing the galaxy model parameters with wavelength-dependent functions -- namely Chebyshev polynomials of the first kind \citep{1965ChebyshevPolyBook}. While the tools developed as part of the MegaMorph project allow structural parameters to vary systematically with wavelength, user specifiable limits and parameters must be appropriately chosen. In particular, the degrees of freedom with which each parameter is allowed to vary as a function of wavelength must be specified. This is done in order to balance the advantage of multi-band fitting with constraining the model parameters to change with wavelength in a physically meaningful way. We explicitly discuss the degrees of freedom with which we allow the magnitudes, sizes, and S{\'e}rsic indices to vary below. 

Spectral energy distributions (SEDs) have a complex wavelength dependence and therefore cannot be reproduced with low-order polynomials. In our fitting, we ensure that SEDs can be accurately recovered by giving the fitting function in $\textsc{GalfitM}$ full freedom. 
%For the CANDELS sample, there are a different number of bands available for each field, therefore the order of the polynomial that we allow varies across fields. However, we still allow full freedom for the magnitudes, following the same idea. 
While other parameters also have some wavelength dependence, full freedom is not always necessary nor, in fact, advisable, in order to make use of the advantages of multi-band fitting. For instance, the measured size of a galaxy depends on the wavelength at which the observation is made \citep[e.g.,][]{Evans1994MNRAS, LaBarbera2010MNRAS, MegaMorph, Vulcani2014MNRAS}, such that galaxies are often found to be much smaller when measured in redder bands when compared to bluer bands. This is because sizes measured in bluer bands are more sensitive to the younger stellar population within galaxies. These, generally speaking, tend to reside in galaxy disks, which are typically more extended than their respective bulge components. Sizes measured in redder bands instead reflect the extent of the older stars, which are generally found in the very central part of galaxies and/or their spheroidal components, i.e., bulges. As a result, sizes measured in blue bands tend to be larger than those measured at longer wavelengths. %\textcolor{red}{}\bh{Is it worth referring back to Suess here in context of comparing galaxies at different redshifts, as they claim the measured effects are colour gradient effects? Same has been found by a relatively recent Kriek paper, I think.} 
To allow for this change in size in our galaxy models, we decide to allow some variation with wavelength. As we have sufficiently many bands available, we allow the effective radius to vary as second order Chebyshev polynomial, enabling us to recover the size's smooth wavelength dependence. S{\'e}rsic indices also have a strong wavelength dependence for similar reasons.
%Since the disk component is more prominent at short wavelengths, S{\'e}rsic indices measured from bluer bands tend to be roughly equal to one, representing the exponential light profile of the disk component. 
If a galaxy consists of both a bulge and a disk component, as most spiral galaxies do, then the S{\'e}rsic index measured at longer wavelengths will reflect the light profile of the bulge, while the S{\'e}rsic index measured in the bluer bands would reflect the light profile of the disk \citep{Vulcani2014MNRAS}. The S{\'e}rsic index of such a galaxy would then increase with wavelength. Because of this dependence and following the same arguments with the number of images/bands available, the S{\'e}rsic indices are also allowed to vary as second order Chebyshev polynomials. Finally, for the centre positions of the profiles, as well as axis ratios and position angles, we choose to fit constant values, with no wavelength dependence, by fitting zeroth order Chebyshev polynomials.

%\bh{mention axis ratios and positing angles for completeness? TECHNICALLY, axis ratios could change (bulges in the red are rounder than disks), but it's a small-ish effect and hard to nail down). Maybe even position?}

We further discuss the advantages of fitting our data with a multi-wavelength approach in the following section, and show fitting results from the software developed as part of the MegaMorph project in \S \ref{sec:Galapagos} .
% ---------------% 
\section{Data} \label{sec:data}

We use both CANDELS and HFF data to construct magnitude-limited samples of star-forming and quiescent galaxies over a large stellar mass and redshift range. In this section, we describe the data used in this work, from imaging/aperture photometry in \S \ref{sec:data_hff} and \S \ref{sec:data_candels} for the HFF and CANDELS fields, respectively, to size and stellar mass estimates in \S \ref{size_est} and \S \ref{stellarmass}. We present several methods for distinguishing star-forming and quiescent galaxies in \S \ref{sf_vs_q} and show galaxy light profile models from the MegaMorph tools in \S \ref{sec:Galapagos}.

\subsection{HFF} \label{sec:data_hff}
The HST Frontier Fields (HFF) program \citep{Lotz2017} provides a unique data set that allows structural parameters to be obtained for bright objects as well as faint, high-redshift galaxies. The Frontier Fields consist of six cluster fields centred on strongly lensed galaxy clusters that were imaged in parallel with six blank fields. The HFF-DeepSpace photometric catalogues \citep{Shipley2018} combine images from the Advanced Camera for Science (ACS) and Wide Field Camera 3 (WFC3) with K$_s$ imaging, which was taken as part of the `K-band Imaging of the Frontier Fields' (KIFF) project \citep{KIFF}, from the Very Large Telescope (VLT) HAWK-I and Keck-I MOSFIRE instruments. These data were combined in a consistent way to provide coverage in up to 17 filters spanning wavelengths from the UV to NIR. The HFF-DeepSpace catalogues also contain post-cryogenic Spitzer imaging at 3.6$\mu$m and 4.5$\mu$m from the Infrared Array Camera (IRAC), as well as any available archival IRAC 5.8$\mu$m and 8.0$\mu$m data. In addition to the photometric catalogues, \cite{Shipley2018} provide catalogues of photometric redshifts and stellar population properties. We refer the reader to \cite{Shipley2018} for more details on the properties of the HFF-DeepSpace catalogues and how they were constructed, but we briefly address key aspects of the catalogue construction that are relevant for this work.

We use seven of the WFC3 bands, which are consistently available for all parallel and cluster fields (i.e., F435W, F606W, F814W, F105W, F125W, F140W, and F160W). We choose to exclude any ground-based and Spitzer imaging in order to keep the image quality consistent across all bands. The images that we use have pixel scales of 0.06$^{''}$/pixel. The depth of each field is listed in Tables 7 and 8 of \cite{Shipley2018} for the F814W and F160W bands, respectively. The HFF-DeepSpace catalogues are unique in that they model bright cluster galaxies (termed bCGs, although it should be noted that this terminology is different from the traditional use of BCG referring to the brightest cluster galaxy) together with intra-cluster light (ICL) and remove them from the images. This is done in order to obtain information about background or underlying objects, which is crucial for this work. We test if the removal of the ICL has any impact on the galaxy sizes that we obtain by comparing the size distributions of galaxies in the parallel fields to those in the cluster fields. We find that the size distributions are similar and consistent, indicating that the ICL subtraction does not introduce significant systematics. 

We model the light profiles of galaxies from the `bCG subtracted' images in order to gain information about objects that would otherwise be outshone by neighbouring bCGs making reliable fitting of their light profiles difficult, if not impossible. In an additional set of fits, we model the bCGs from the original images separately. This ensures that our sample consists of both the bright bCGs and any faint objects that may lie `behind' them. For the parallel fields, we find no significant improvement between using the bCGs subtracted images versus the original images in terms of the number of objects recovered; however, we follow the same approach on those fields anyway, for consistency. An additional benefit of using the bCG subtracted images is that detecting the faint objects in the vicinity of these large bCGs  can be drastically improved by removing those brightest objects first. Without removing them, it is simply impossible to find a SExtractor \citep{sextractor1996A&AS} setup that works on these small and faint objects, reducing our sample size by 10-15\% from this effect alone.

To avoid biasing measured sizes and magnitudes, we exclude any objects from the cluster fields that may be lensed. \textsc{Galapagos-2}, has two features to ensure this. The first is meant to flag `detections' that are not real. Mostly, these are hot pixels at the edge of the field, but this also allows to ensure that galaxies are detected as one object, rather than being split up, e.g., due to internal structure. These objects are removed before starting the galaxy light profile modelling. The second allows flagging objects as `important, yet not real/wanted'. A prime example of this are detected diffraction spikes of stars, which need to be modelled as secondary objects (see \citealt{Barden2012}  for primary, secondary, and tertiary nomenclature), in order to not influence the fit results of nearby objects. 
However, they do not portray real objects and as such are only treated as secondary/tertiary objects, never as primary objects. 
They are removed from the object catalogue by the end of the fitting process. 
This latter feature allows us to sensibly deal with lensed objects, especially arcs.  
%We select any arcs in the image to be modeled only as neighboring objects, never as primary object, as their fit values could not be trusted due to their lensed nature and are therefore excluded from our final sample. 
\begin{figure}
    \centering
    \includegraphics[scale=.485]{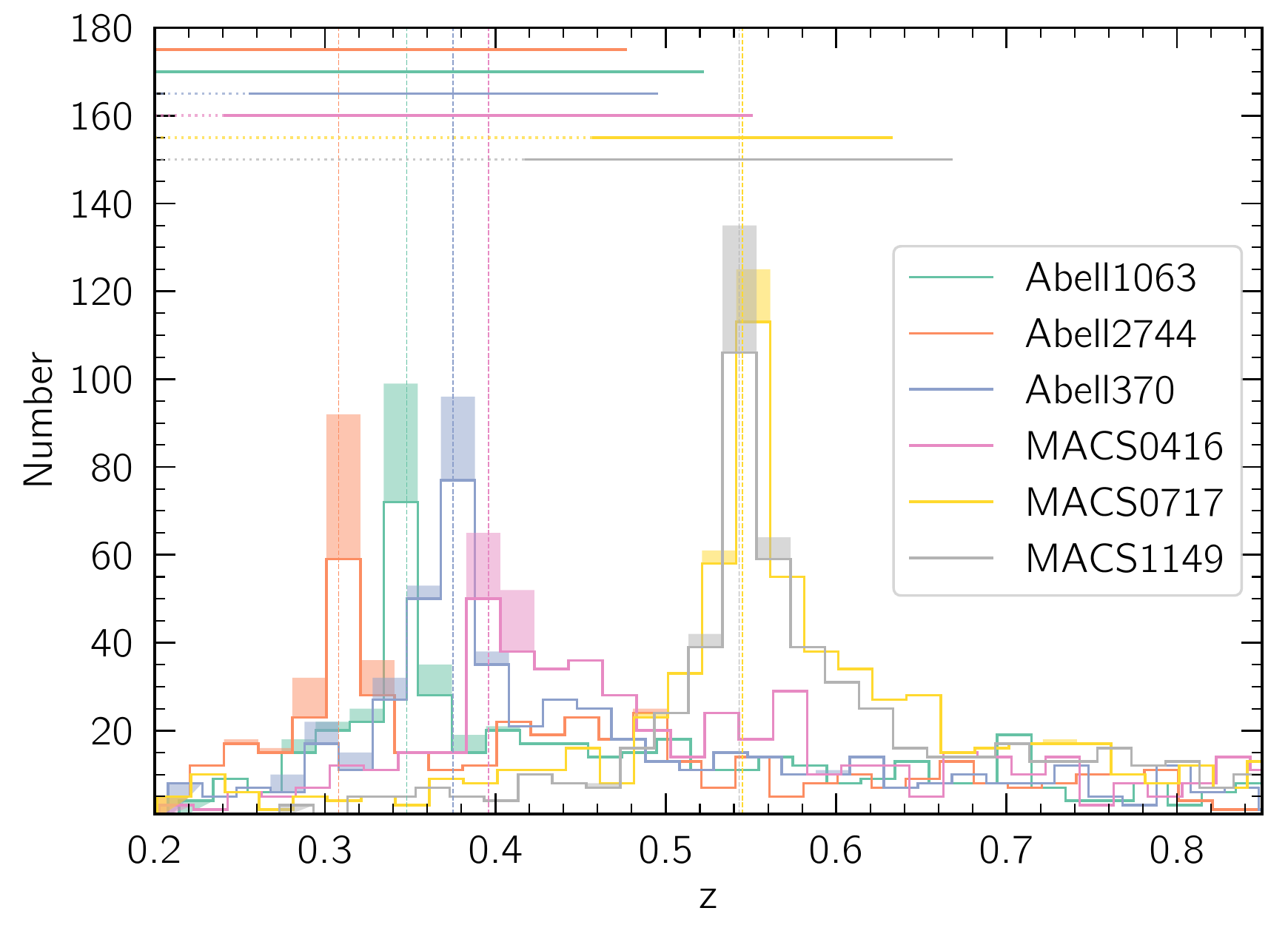}
    \caption{The redshift distribution of all six HFF cluster fields. Histograms are colour coded by cluster, with the corresponding vertical dashed lines indicating the spectroscopic redshift of each cluster. Galaxies with photometric redshifts within $3 \times \sigma_{\mathrm{NMAD}}$ of the cluster redshift (horizontal lines) are classified as cluster members. Our final sample includes all cluster member galaxies as well as galaxies with redshifts lower than the cluster redshift and is indicated by the dotted horizontal lines. We exclude objects with redshifts higher than $z_\mathrm{upper}$ listed in Table \ref{tab:HFFcluster_properties} in order to avoid biasing any measurements by including lensed galaxies. The parts of the histogram that are filled in represent the redshift distribution of the bCGs.}
    \label{fig:z_hist}
\end{figure}

In our sample analysed in subsequent sections, we further limit the redshift of the objects in the cluster fields to be either consistent with the cluster redshift or lower, in order to avoid any lensing effects. Motivated by \cite{Morishita2017}, we use the normalised median absolute deviations ($\sigma_{\mathrm{NMAD}}$), which have been derived by \cite{Shipley2018} for the HFF, to find cluster members. We define any galaxy with a photometric redshift that satisfies 

\begin{equation}
    3 \times \sigma_{\mathrm{NMAD}} \geq \frac{|z - z_{\mathrm{clu}}|}{1 + z_{\mathrm{clu}}}
    \label{eq:clusterMembers}
\end{equation}
 as having a redshift that is consistent with the cluster redshift. %\bh{is the factor of 3 common in HFF? It's quite generous.} \kn{not sure what the motivation of the factor of 3 is. Just what \cite{Morishita2017} use. Their sigma_NMAD is much smaller though}. 
 In Equation \ref{eq:clusterMembers}, $z$ is the photometric redshift of each galaxy, $z_{\mathrm{clu}}$ is the spectroscopic redshift of the cluster, and $\sigma_{\mathrm{NMAD}}$ is the normalized median absolute deviation, which is obtained by comparing estimated photometric redshifts and confirmed published spectroscopic redshifts from the literature. $\sigma_{\mathrm{NMAD}}$ is different for each cluster field and is listed in Table \ref{tab:HFFcluster_properties}.

\begin{table}
\caption{For each HFF cluster field, we list the spectroscopic redshift ($z_\mathrm{spec}$), the normalised median absolute deviation ($\sigma_{\mathrm{NMAD}}$), and the redshift limit below which we include objects ($z_\mathrm{upper}$).}
\centering

\begin{tabular}{ l c c c}
\hline
%\multicolumn{1}{l}{ \ } & \multicolumn{5}{c}{ HFF Cluster Fields} \\ 

Cluster Field & $ \ z_\mathrm{spec} \ $ & $ \ \sigma_{\mathrm{NMAD}}$ \  &  $ \ z_\mathrm{upper} \ $ \\
\hline 

Abell1063 & 0.348 & 0.043 & 0.522 \\
Abell2744 & 0.308 & 0.043 & 0.477 \\
Abell370 & 0.375 & 0.029 & 0.522 \\
MACS0416 & 0.396 & 0.037 & 0.550 \\
MACS0717 & 0.545 & 0.019 & 0.632 \\
MACS1149 & 0.543 & 0.027 & 0.668 \\
\hline
\end{tabular}
\label{tab:HFFcluster_properties}
\end{table}

Figure \ref{fig:z_hist} shows the redshift distribution of the cluster galaxies and redshift cuts we apply in an effort to exclude galaxies whose magnitudes or sizes may be affected by lensing. For each cluster, a clear peak in the number of galaxies can be seen at the spectroscopic redshift (indicated by the vertical dotted lines) of each cluster. The spectroscopic redshifts of the clusters are reported in Table \ref{tab:HFFcluster_properties}. The redshift distribution of the bCGs are shown as filled histograms. As expected, the majority of bCGs have redshifts that are consistent with the spectroscopic redshifts of the cluster field they belong to. We do not restrict the redshift for the HFF parallel fields as no significant lensing effects are to be expected.

\begin{figure*}
    \centering
    \includegraphics[width=1\textwidth]{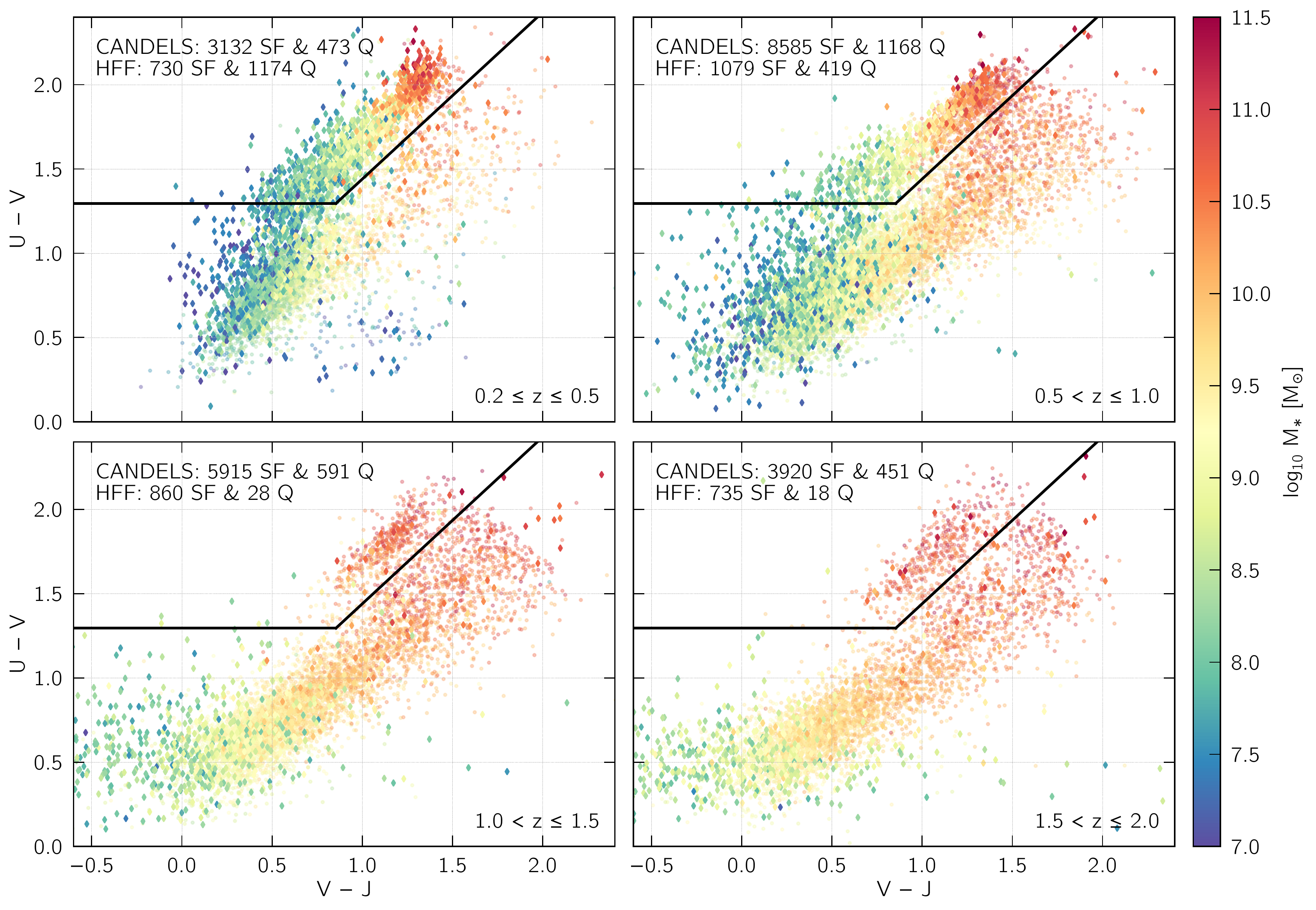}
    \caption{The UVJ diagram of the entire sample divided into four redshift bins. Objects from the HFF parallel and cluster fields are shown as diamonds and galaxies from the CANDELS fields are shown as dots. All galaxies are colour-coded by their stellar mass, which is estimated by the FAST code \citep{Kriek2009} and corrected for the difference in the F160W magnitude from \textsc{GalfitM} and the photometric catalogues (see text for further details). The redshift range of each panel is indicated at the bottom right corner and the number of star-forming and quiescent objects from HFF and CANDELS is indicated in the top left.}
    \label{fig:UVJ_mass}
\end{figure*}

\subsection{CANDELS and 3D-HST} \label{sec:data_candels}
We measure the structural parameters from the Hubble Space Telescope (HST) imaging from the Cosmic Assembly Near-infrared Deep Extragalactic Legacy Survey (CANDELS) for the galaxies in all five CANDELS fields (i.e., GOODS-S, GOODS-N, COSMOS, EGS, and UDS) \citep{Koekemoer2011}. While the CANDELS fields are not as deep as the HFF fields, CANDELS allows us to study a much larger galaxy sample due to the area covered by this survey. Each object from the 0.03$''$/pixel CANDELS images is matched by position to galaxies in the 3D-HST photometric catalogues \citep{Skelton2014}, limiting the maximum spacial separation to 0.2$''$. 
We note here that the 0.03$''$/pixel images provided by CANDELS that we use in this work are not the same as those used by \cite{vdWel2012ApJS} and \cite{Dimauro2018MNRAS}, who derive galaxy properties from the 0.06$''$/pixel CANDELS images. This choice has no effect on the fitting parameters as we find excellent agreement in the derived magnitude, size, and S{\'e}rsic index with both \cite{vdWel2012ApJS} and \cite{Dimauro2018MNRAS}.

\subsection{Sample Selection} \label{sampleSelection}

Our initial sample consists of 39685 objects from the HFF and 106663 objects from CANDELS after spatially matching objects from the images to the galaxies in the HFF-DeepSpace \citep{Shipley2018} and 3D-HST photometric catalogues \citep{Skelton2014}, and applying the redshift limit described in \S \ref{sec:data_hff} to galaxies in the HFF cluster fields. We first require that galaxies have \texttt{FLAG\_GALFIT=2}, meaning that they have been successfully modelled and fit with the fitting software,  \textsc{GalfitM} and \textsc{Galapagos-2},  which are discussed in detail in \S \ref{sec:MegaMorph}. This reduces our HFF and CANDELS samples to 22090 and 105717 objects, respectively. We note that a significantly higher fraction of HFF galaxies were unsuccessfully modelled compared to CANDELS galaxies. This is because we only model HFF objects that are within the F160W footprint since we will apply a completeness cut in the H-band anyway. 
%Any object that is not imaged in the F160W band is not modelled unless it has an expanded Kron ellipse that overlaps with that of a primary object \citep{Barden2012}, where the primary object would need to have F160W coverage. In this case, the object outside the F160W footprint is modelled as a neighbour. We have not placed the same constraint on the CANDELS objects as the run time of this sample was not a concern.
All CANDELS objects have been modelled, not just those with F160W coverage; however, we will apply a magnitude cut in the H-band for CANDELS as well, effectively requiring that all CANDELS objects in the final sample are within the F160W footprint, too.

For the separate bCG run, in which 330 galaxies were successfully modelled, we visually inspect the fits to ensure that they are reliable and found no major issues, apart from the typical effects of fitting two-component systems with one-component models.
We do not visually inspect all objects from the `bCG subtracted' images, as this would be an unfeasible task and the modelling for these `standard' galaxies is much closer to the tests carried out in \citep{MegaMorph} and other publications. We have, however, also looked at a subset of these ($\sim$ 700 galaxy models), to ensure that \textsc{Galapagos-2} works as intended. For each galaxy in the HFF,  we fit morphological parameters in seven bands and we require that each galaxy that is modelled is covered by at least three of those seven images. 
% Question from KEW -- why not use 3 bands for CANDELS too? 
For the CANDELS fields, we have a different number of bands available for each field, with a minimum of four bands for UDS. Therefore, requiring that galaxies are covered by at least three bands, as we do for HFF galaxies, would be a strict constraint. Hence, we require that every galaxy that is modelled in CANDELS, have sufficient data (i.e., $>$30\% of pixels within the primary ellipse are \textit{not} masked) in at least two images. Galaxies that do not satisfy this have unsuccessful models, and are hence removed by our first quality cut.

For galaxies which have been successfully modelled, we then require that they have a spectroscopic or a reliable photometric redshift in the range $0.2\leq z \leq2$. 
For our CANDELS sample, we also use redshifts derived from grism spectra \citep{Momcheva2016ApJS} when available. We include objects that have \texttt{use$\_$phot=1}, which removes stars (i.e., objects with \texttt{star$\_$flag=1}), sources close to bright stars, and objects with  $S/N <3$ from the photometry aperture in the F160W band. These cuts leave 11438 and 71798 objects in the HFF and CANDELS samples, respectively.

To ensure that we only select objects that are bright enough to be modelled, we include only HFF galaxies that are one magnitude brighter than the 90$\%$ detection completeness limit in the F160W band when the injected mock objects are not allowed to overlap with detected objects. These completeness limits are reported in Table 8 of \cite{Shipley2018}. For CANDELS, we exclude any object $>$24.1mag in the F160W band, following the same idea. After applying these magnitude cuts, we have 8686 HFF galaxies and 27555 CANDELS galaxies. Additionally, we apply cuts based on the model parameters, which are all motivated by \cite{MegaMorph}. These cuts are as follows: (i) While we allow S{\'e}rsic indices to vary between 0.2 and 12 in the fitting, we exclude objects with a S{\'e}rsic index below 0.205 and above 8 (in any band) because these objects have run into the  S{\'e}rsic index constraint, are often found to be point sources, or are unreliably fit based on visual inspection. (ii) The half light radius in the modelling is restricted to be between 0.3 and 400 pixels to avoid using results from objects that are unphysical. Therefore, in the sample selection, we only include objects with effective radii between 0.305 and 395 pixels to exclude any objects that have ran into either of these constraints. For the separate bCG run, we remove the 400 pixel size limit since the bCGs can be -- and are -- larger than this. 
(iii) We require that the magnitude measured from all bands is within 5 magnitudes of the magnitude measured from our source extractor runs (\texttt{MAG\_BEST} + an empirically derived offset between bands). 
Although this constraint is rather lenient, it is again intended to ensure that we are not using objects that have unreliable light profile fits. 
% I also apply a cut on the magnitude (between 0 & 40)
% as well as an axis ratio cut (0.001 to 1)
% But these are so lenient that I won't mention them.
This ensures a sample with reliable fitting parameters of 6632 HFF galaxies and 24736 CANDELS galaxies.
Finally, we limit the stellar mass of the galaxies that we use, but we provide a detailed discussion of the stellar mass cuts in \S \ref{stellarmass}.

\subsection{Effective Radius} \label{size_est}
For each galaxy, we measure the half-light radius,
or effective radius, along the major axis in each observed band using \textsc{GalfitM} and \textsc{Galapagos-2}, which are discussed in \S \ref{sec:MegaMorph}. Half-light radii have been used in the literature to study galaxy sizes at least since \cite{deVaucouleurs}. It has long been known, however, that using alternative radii can result in significantly different scaling relations \citep[see e.g.,][for recent results]{Graham2019PASA,Trujillo2020MNRAS}. In addition to using half-light radii, we have also derived the stellar mass -- size relations for star-forming and quiescent galaxies using radii that contain 20\% and 80\% of the total light by converting the measured half-light radii into these values by analytically integrating the Sérsic profile of each object. We find that the stellar mass -- size relations remain qualitatively the same in that there is still a flattening for the quiescent sample, while the star-forming relation is well represented by a single power law. Apart from the expected shift in normalisation, the slopes of the curves remain largely unchanged, with one notable exception: for quiescent galaxies, the slope of the mass  -- size relation at the low mass end is weakly dependent on the radius definition that is used. This change, however, is small and well within the scatter of the relations and indicates a change of Sérsic index with galaxy mass, examining which is not part of this work.

Using the Chebyshev polynomials returned by \textsc{GalfitM}, we are able to derive the rest-frame 5000\AA \ size, in order to have a redshift independent measure of galaxy size, allowing a clean comparison over all redshifts. We additionally derive the rest-frame 4000\AA \ and 6000\AA \ size of each galaxy; however, we find that the sizes at the three rest-frame wavelengths are consistent. Therefore, the derived stellar mass -- size relation does not depend strongly on the rest-frame wavelength that is used, at least in the 4000 -- 6000\AA \ range. Throughout the remainder of this paper, we choose to use the 5000\AA \ size  for easier comparison to the literature and because it can be obtained without extrapolating the polynomial for higher redshift galaxies. 

\subsection{Stellar Mass} \label{stellarmass}

The 3D-HST photometric catalogues \citep{Skelton2014, Momcheva2016ApJS} contain photometric and grism redshift estimates as well as stellar population parameters determined by the FAST code, which fits stellar population synthesis models to the measured SEDs of galaxies to infer several galactic properties \citep{Kriek2009}. 
The HFF-DeepSpace catalogue \citep{Shipley2018} also provides stellar masses that have been derived in a similar way. 
We use these FAST-derived stellar masses with their 1$\sigma$ errors, which are determined by using the Monte Carlo simulation option in FAST. 
We test how the stellar mass changes depending on the metallicity by running FAST for the HFF data with two options: (i) fixing the  metallicity of all galaxies to solar metallicity and (ii) allowing the metallicity to vary. 
We find no significant difference in the derived stellar mass between the two runs, therefore we choose to use the run with fixed metallicity in order to be consistent with the way the CANDELS data were modelled.

As the galaxy sizes that we measure are obtained from modelling Sérsic profiles that integrate the profile out to infinity, while the stellar masses in the HFF-DeepSpace and 3D-HST catalogues are determined from aperture photometry that miss some light at large radii, a correction must be applied to the stellar mass estimates in order for them to be consistent with the profiles used to measure galaxy sizes. 
Following \cite{vdWel2014}, we correct for the difference between the F160W flux from the photometric catalogues and the F160W magnitude as measured with \textsc{GalfitM}, in the final stellar mass that we use.
However, we find that this correction only has an effect on the largest galaxies, for which the corrected masses are on average a factor of 1.08 larger. 
There is no significant effect on the masses of the majority of objects, and the conclusions in this paper are not significantly changed by this correction.

\begin{figure*}
    \centering
    \includegraphics[width=1.0\textwidth]{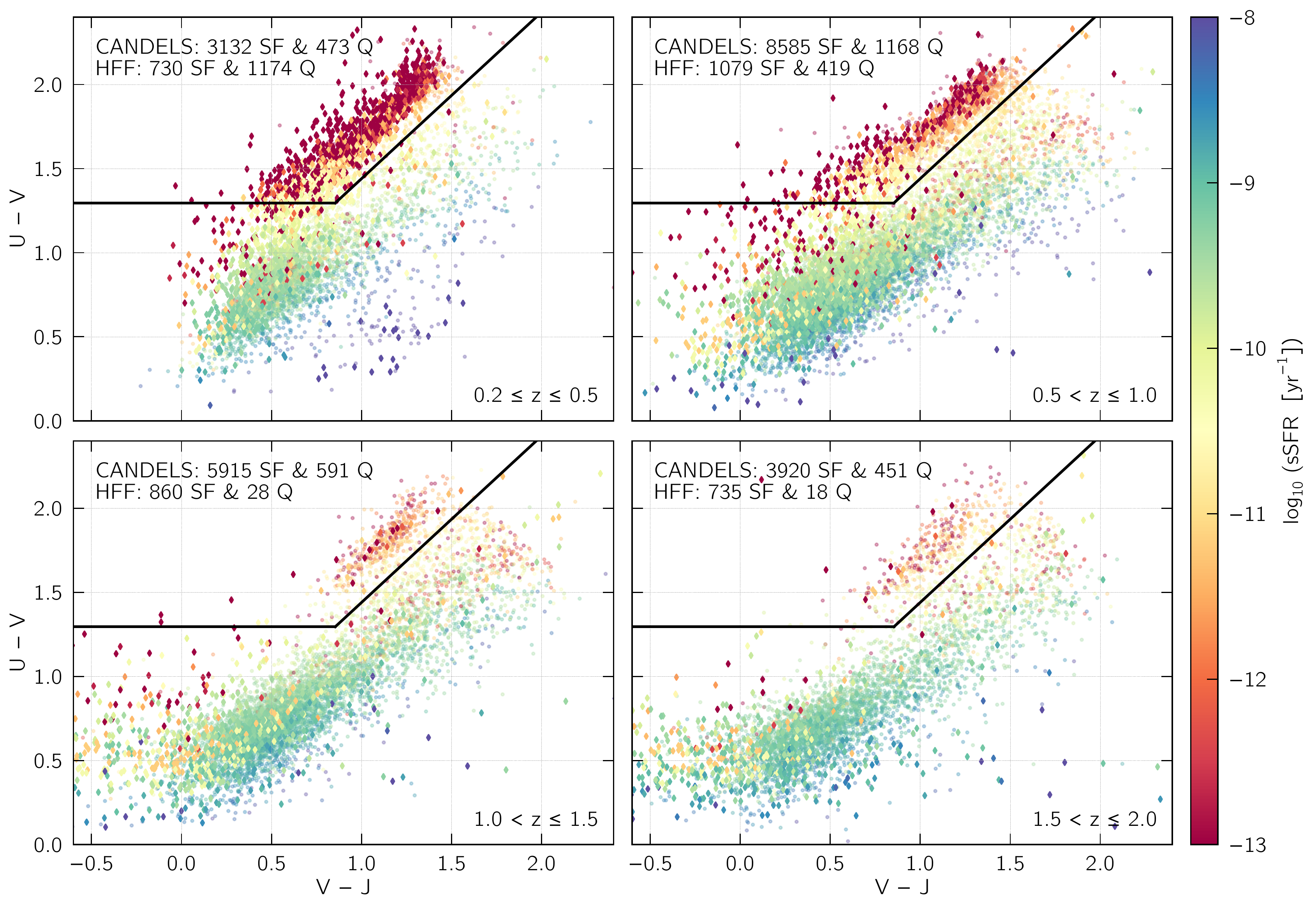}
    \caption{Same UVJ diagram shown in Figure \ref{fig:UVJ_mass}, except the galaxies are colour-coded by their specific star-formation rates (sSFRs), estimated by the FAST code \citep{Kriek2009}. The redshift bin is indicated for each panel in the bottom right corner and the number of star-forming and quiescent objects from HFF and CANDELS, when galaxies are separated according to the UVJ diagram, is indicated in the top left. In general, galaxies with low measured sSFR lie within the quiescent region, as expected; however there is a small number of galaxies with low sSFR that fall outside the quiescent wedge. Upon checking, we find that most of these galaxies have uncertain sSFR measurements, although it is possible that some are post-starburst galaxies, that have recently quenched their star-formation and therefore still have bluer colours (see text). }
    \label{fig:UVJ_ssfr}
\end{figure*}

In our final sample, we include galaxies with corrected stellar masses above 10$^{7}$ M$_{\odot}$. 
We obtain stellar mass uncertainties from the upper and lower 68$^{\rm{th}}$ percentiles that the FAST code \citep{Kriek2009} returns and we exclude objects for which the stellar mass uncertainty is $\geq$2 dex. 
This cut is a lenient one, as it is intended to only remove galaxies for which the SED modelling is highly unreliable. 
After applying these last quality cuts, the final sample consists of 5043 HFF and 24235 CANDELS galaxies. A higher fraction of HFF objects are removed by the stellar mass cuts because the HFF consists of more low mass galaxies compared to CANDELS, and therefore, the M$_* \geq $10$^{7}$ M$_{\odot}$ cut that we apply removes more HFF objects. 
In Figures \ref{fig:UVJ_mass} and \ref{fig:UVJ_ssfr}, we further provide the number of star-forming and quiescent galaxies that fall into each redshift bin, noting that the HFF consists of many quiescent galaxies at $z\leq1$, but only 28 and 18 in  $1.0 < z\leq1.5$ and $1.5 < z\leq2.0$, respectively. This is expected as our sample does not consist of any cluster field galaxies at $z\geq1$ as a result of the redshift limits that we impose in \S \ref{sec:data_hff}.

In Figure \ref{fig:UVJ_mass}, we show the UVJ diagram colour-coded by stellar mass for four redshift bins spanning 0.2$\leq z \leq$2.0. Rest-frame colours are derived using EAZY \citep{eazy_brammer}\footnote{\url{https://github.com/gbrammer/eazy-photoz}}. For galaxies in both the HFF, shown as diamonds, and in the CANDELS fields, shown as dots, the most massive galaxies have redder U$-$V and V$-$J colours, as expected from other studies. This suggests that the stellar mass estimates are reliable. We also note that we detect more low mass galaxies at low redshifts than at high redshifts, which is also expected. The UVJ boundaries that separate the star-forming and quiescent galaxies, shown as black lines in Figure \ref{fig:UVJ_mass}, are exactly matched to those used by \cite{vdWel2014}.

For any magnitude-limited sample, the minimum stellar mass at which galaxies can be observed depends on their stellar mass-to-light ratio (M/L) and redshift. The M/L depends on the stellar populations and is therefore reflected in galaxy colours, meaning that quiescent galaxies will have higher M/L because of their older stellar populations and redder colours (e.g., \citealt{Marchesini2009ApJ}). Therefore, the mass-completeness limit of quiescent galaxies will be higher than that of star-forming galaxies. We do not consider a mass complete sample in this work since we do not attempt to study the number density of the galaxies in our sample and because we want to extend the stellar mass--size relation to low mass galaxies. We place a conservative magnitude limit so that the size measurements are reliable even for galaxies with stellar masses that fall below our mass completeness limit. Nevertheless, there will, naturally, be some biases, particularly against low-surface-brightness (LSB) galaxies, which could possibly comprise a large fraction of the overall low-mass galaxy population \citep[e.g.,][]{Wright2020arXiv}. To test this effect, we select 120 LSB galaxies from the Abell1063 parallel field. Determining which galaxies are labelled as LSB systems is strongly dependent on the limits of the surveys that are used \citep[e.g.,][]{Disney1976Natur}. For this test, we define LSB galaxies as those with effective radii larger than 20 pixels and apparent magnitudes fainter than 25.5mag in the F125W band, as these lie well below the surface-brightness of the majority of objects in the Abell1063 parallel field. For each of the 120 LSB galaxies, we then generate 10 mock galaxies with the same magnitude and size as the original, but with randomly selected position angles and axis ratios. The mock galaxies are randomly assigned a Sérsic index of $n=1$ or $n=4$. We then inject these mock LSB galaxies into the F125W image and we recover 1025 ($\sim$ 85\%) of the injected objects with SExtractor. The majority of the LSB galaxies that are not detected happen to be overlapping with bright foreground objects. This shows that any bias against these `large' LSB galaxies, which could systematically shift a stellar mass -- size relation, should be negligible.

Another class of objects that our sample could be potentially biased against are compact galaxies. Although compact galaxies are easier to detect than large objects of similar brightness because their flux is more concentrated and therefore peaks well above the background, it can be difficult to distinguish compact galaxies from point sources. In the HFF-DeepSapce \citep{Shipley2018} and 3D-HST \citep{Skelton2014} catalogues, compact objects are classified as stars based on the tight correlation in size and magnitude that point sources follow. Both studies show that point sources can be cleanly separated from extended sources for mag$_\mathrm{F160W} \leq 25$. Objects fainter than this are assigned a different flag (i.e., \texttt{star\_flag} = 2) and are hence not removed by our quality cuts discussed in \S \ref{sec:data}. Because of this conservative classification of stars, it is unlikely that compact galaxies are classified as point sources, and removed. We additionally test how often the modelling failed (i.e., \texttt{FLAG\_GALFIT=1}) and how often compact objects ran into fitting constraints compared to their more extended counterparts. We find that the modelling of compact objects is no more likely to fail than it is for more extended objects, but compact objects are $\sim$3 times more likely to run into fitting constraints. This is expected as the fitting constraints are specifically chosen to remove point sources and any galaxies for which the modelling is unphysical. These results suggest that compact galaxies are not removed from the sample, but faint compact point sources likely are by the cuts on the model parameters, giving us confidence that the results of this study will not be strongly impacted biases against LSB or compact objects.

\subsection{Selecting star-forming and quiescent galaxies} \label{sf_vs_q}

\begin{figure*}
    \centering
    \includegraphics[width=1.0\textwidth]{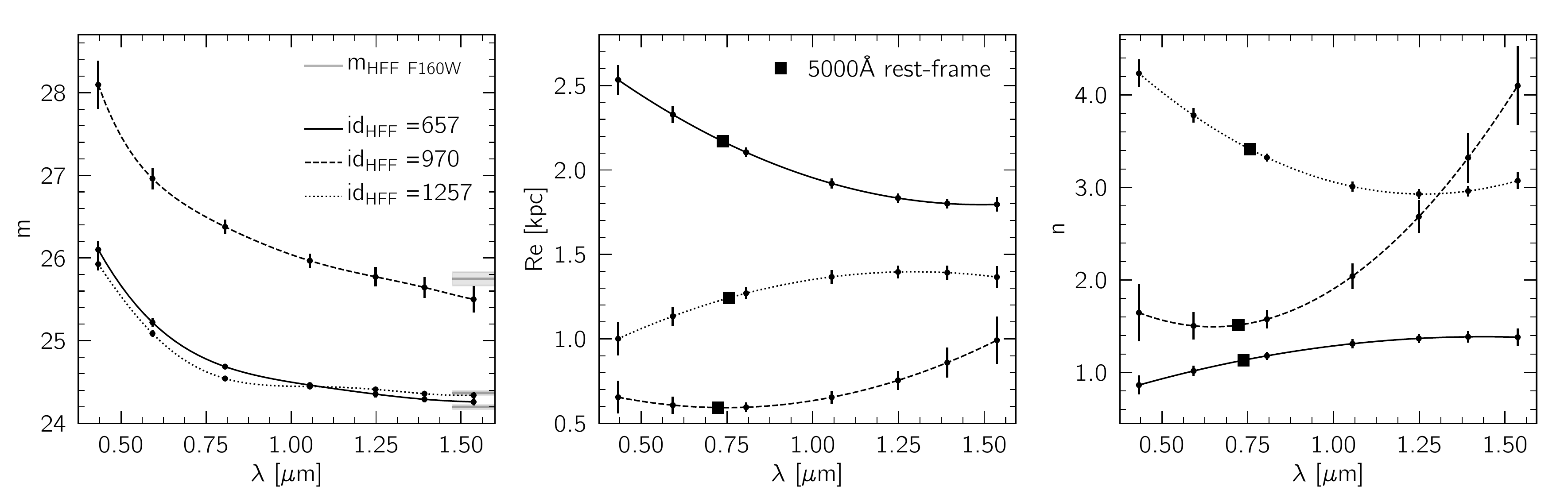}
    \caption{Example fitting results of the apparent magnitude (m), effective radius (Re), and S{\'e}rsic index (n) as a function of wavelength for three galaxies in the Abell1063 parallel field, all at roughly the same redshift. These three galaxies, are chosen to display the variety of functional forms that can be fit for different galaxy parameters. For each galaxy, the fit is shown in a different line style as indicated in the legend and the black squares indicate the 5000\AA \ rest-frame values that can be read from the Chebyshev polynomials. The left panel shows the apparent magnitude of each object, as derived by \textsc{GalfitM}, with the F160W magnitude from the HFF-DeepSpace catalogue \citep{Shipley2018} indicated as a horizontal grey line around 1.6$\mu$m, where the uncertainty is shown as a shaded grey region. Although these H-band magnitudes are measured with different methods, they are consistent, indicating that the modelling is reliable, even for faint objects, such as galaxy id$_{\mathrm{HFF}}=970$. In the modelling, the magnitudes are allowed to vary freely as a function of wavelength (hence, rest-frame magnitudes are not derived directly from the Chebyshev polynomial), while for the size and S{\'e}rsic index of each galaxy, which are shown in the middle and right panel, the Chebyshev polynomial fit is allowed to vary as a second order polynomial. All of the errors shown are the ones derived by $\textsc{GalfitM}$ and are therefore underestimated by a factor of 2 to 2.5 (see \S \ref{sec:M_Re}). The magnitude errors are increased by a factor of 5, to illustrate that the faintest galaxy has the largest uncertainties.} 
    \label{fig:Gala_params}
\end{figure*}

Quiescence can be estimated based on a variety of galaxy properties including colour, star formation activity, galaxy structure, and morphology. However, great care must be taken when distinguishing quiescent galaxies from star-forming ones since quiescence does not mean that there is \textit{no} residual star formation. Quiescent galaxies can also appear blue in colour if they have only recently stopped forming stars and have disk-like structures \citep{Graham_Guzman2003}.
All of these caveats mean that it is likely that different galaxies are identified as quiescent depending on which criteria are used. It is therefore crucial that star-forming and quiescent galaxies are carefully separated using a robust method. 

Often, galaxies are identified as quiescent based on their position on the UVJ diagram \citep[e.g.,][]{Labbe2005,Wuyts2007,Williams2009ApJ}. This method is powerful because the U$-$V colour allows galaxies with red colours, which indicates old stellar populations, to be selected while the V$-$J colour is used to differentiate galaxies that have old stellar populations from dusty star-forming galaxies \citep{Whitaker2012ApJ}. Unfortunately, selecting quiescent galaxies from the UVJ diagram often misses galaxies that have recently ceased their star-formation \citep[e.g.,][]{Marsan2015ApJ}. 
For instance, post-starburst galaxies will still be relatively blue in colour despite having very little ongoing star-formation because -- while they are no longer forming stars -- there are still short-lived stars within those galaxies. An added complication of using the UVJ diagram to select star-forming and quiescent galaxies is the lack of sufficiently dusty templates in EAZY \citep{eazy_brammer}, resulting in the cut-off at the top right corner in each panel of Figure \ref{fig:UVJ_mass}. Although properties in the HFF-DeepSpace catalogues are measured with the most up-to-date EAZY templates, the properties of the CANDELS galaxies are not corrected for this effect. Therefore, we do not apply a vertical cut in V$-$J, following, e.g., \citet{vdWel2014} and \citet{Martis2016}.

A popular alternative is selecting galaxies directly based on their star-formation activity. While this method gets around some of the challenges of selecting quiescent galaxies based on the UVJ diagram, it heavily relies on reliable SED fits, which have been shown to suffer when modelling unique galaxies. We show the UVJ diagram for our sample colour coded by the specific star-formation rate in Figure \ref{fig:UVJ_ssfr}. There is a small fraction ($\sim$ 2\%) of objects that have low specific star-formation rates 
but do not lie within the quiescent region on the UVJ diagram (i.e., galaxies that are colour coded red but lie outside the quiescent wedge). While some of these objects could be recently quenched galaxies that have not yet moved into the quiescent region, we find that the majority of these objects have uncertain sSFRs. The median uncertainty on $\log{(\rm sSFR)}$ of all the data shown in Figure \ref{fig:UVJ_ssfr} is $\sim$0.4 dex, while the median uncertainty of the galaxies that have low sSFR but fall outside the quiescent wedge is $\sim$1.2 dex.

\begin{figure*}
    \centering
    \includegraphics[width=1.0\textwidth]{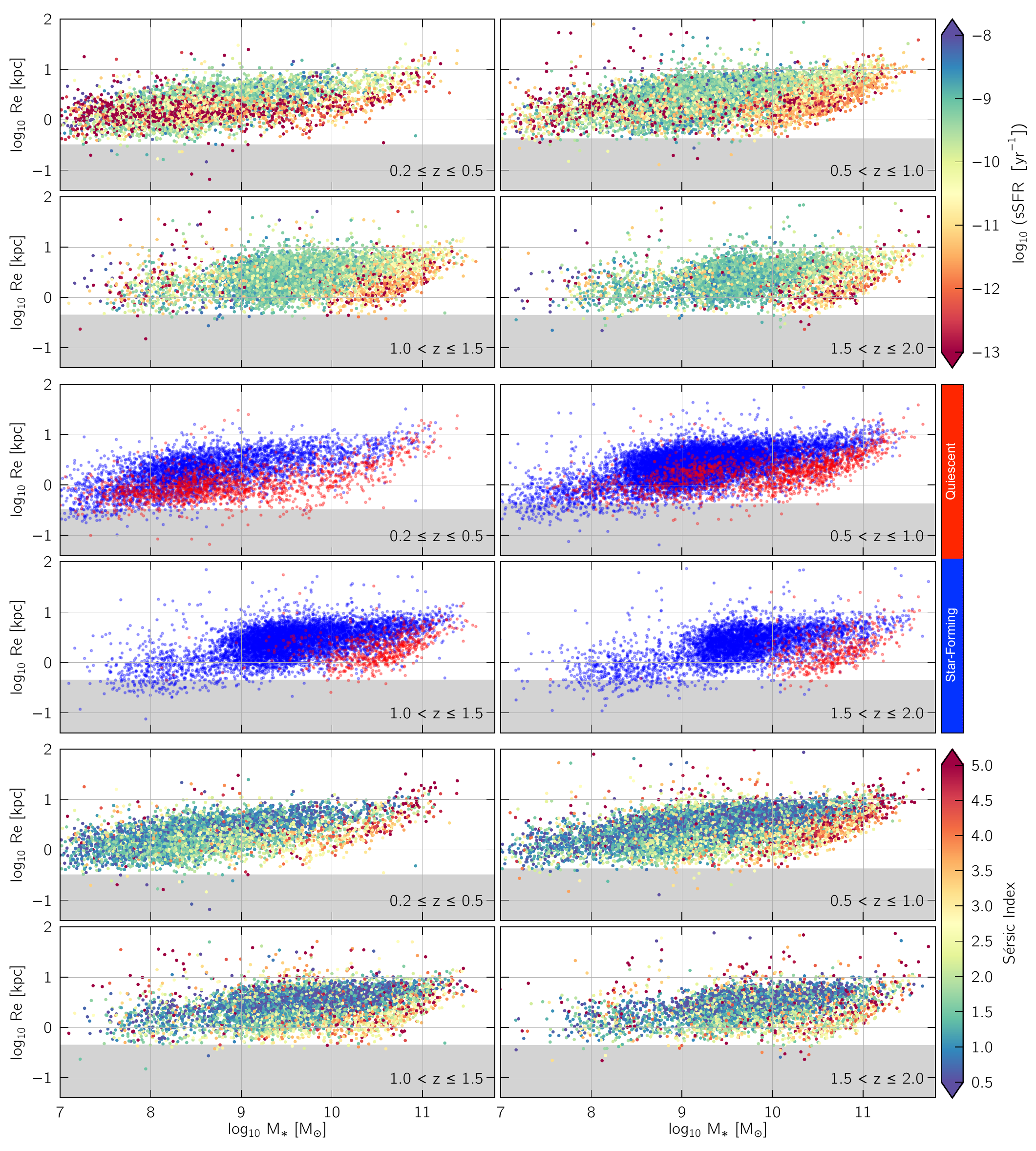}
    \caption{The stellar mass -- size relations for different redshift bins for quiescent and star-forming galaxies. The sizes reported are rest-frame 5000\AA \ sizes as derived in \S \ref{size_est}. The top four panels are colour-coded by specific star-formation rate (sSFR). The middle four panels are colour-coded based on each galaxy's position on the UVJ diagram, with the quiescent galaxies shown on top of the star-forming galaxies to highlight the flattening seen at $z \leq1$. Finally, the bottom four panels are colour-coded by S{\'e}rsic index. The colour-coding shows that there is good agreement between the three selection criteria that we test for separating star-forming and quiescent galaxies.  The grey areas in each panel indicate sizes at R$_\mathrm{e}<\mathrm{FWHM}_{\mathrm{F160W}}/2$ at the maximum redshift in each panel, to indicate potentially difficult sizes to measure (see text).} \label{fig:MASS-SIZE}
\end{figure*}

Quiescent galaxies and star-forming galaxies are also believed to have intrinsically different structure, making it possible to distinguish the two based on structural properties \citep[e.g.,][]{Shen2003, Ravindranath2004ApJ604L9R}. Quiescent galaxies are more generally thought to be spheroids with relatively concentrated light profiles, while star-forming galaxies are more disk-like. We therefore also show the stellar mass -- size relation as a function of S{\'e}rsic index, where classical quiescent galaxies are described by a \cite{deVaucouleurs} profile (i.e., $n=4$)  and star-forming spiral galaxies are well described by an exponential light profile (i.e., $n=1$). %It is rare that a real galaxy is modelled well with one of these classical models, therefore we consider anything above $n=2.5$ to a bulge-dominated galaxy and anything below $n =2.5$ to be a disk-dominated system, as is commonly done in the literature \citep[e.g.,][]{Bruce2014MNRAS, Lange2015MNRAS}.
Of the three methods for selecting quiescence, this is the method most prone to misclassification because reliably deriving S{\'e}rsic indices, especially for non-local galaxies, is difficult, and is only a rough approximation for disk and bulge dominated objects. Another group of missed cases are quenched disk galaxies \citep[e.g.,][]{McGrath2008ApJ, Salim2012ApJ, Carollo2016ApJ} since galaxies can quench while retaining their structure. We therefore argue that this method is the least reliable, but in order to understand how the selection of quiescence impacts the results of the stellar mass--size relation, we separate quiescent from star-forming galaxies based on all three criteria and compare them in Figure \ref{fig:MASS-SIZE}. %In the remaining part of the paper, we select quiescent and star-forming galaxies based on their rest-frame UVJ colours.

% --------------------- MegaMorph --------------------- 

\subsection{\textsc{MegaMorph} Galaxy Models} \label{sec:Galapagos}

\begin{figure*}
    \centering
    \includegraphics[width=1.0\textwidth]{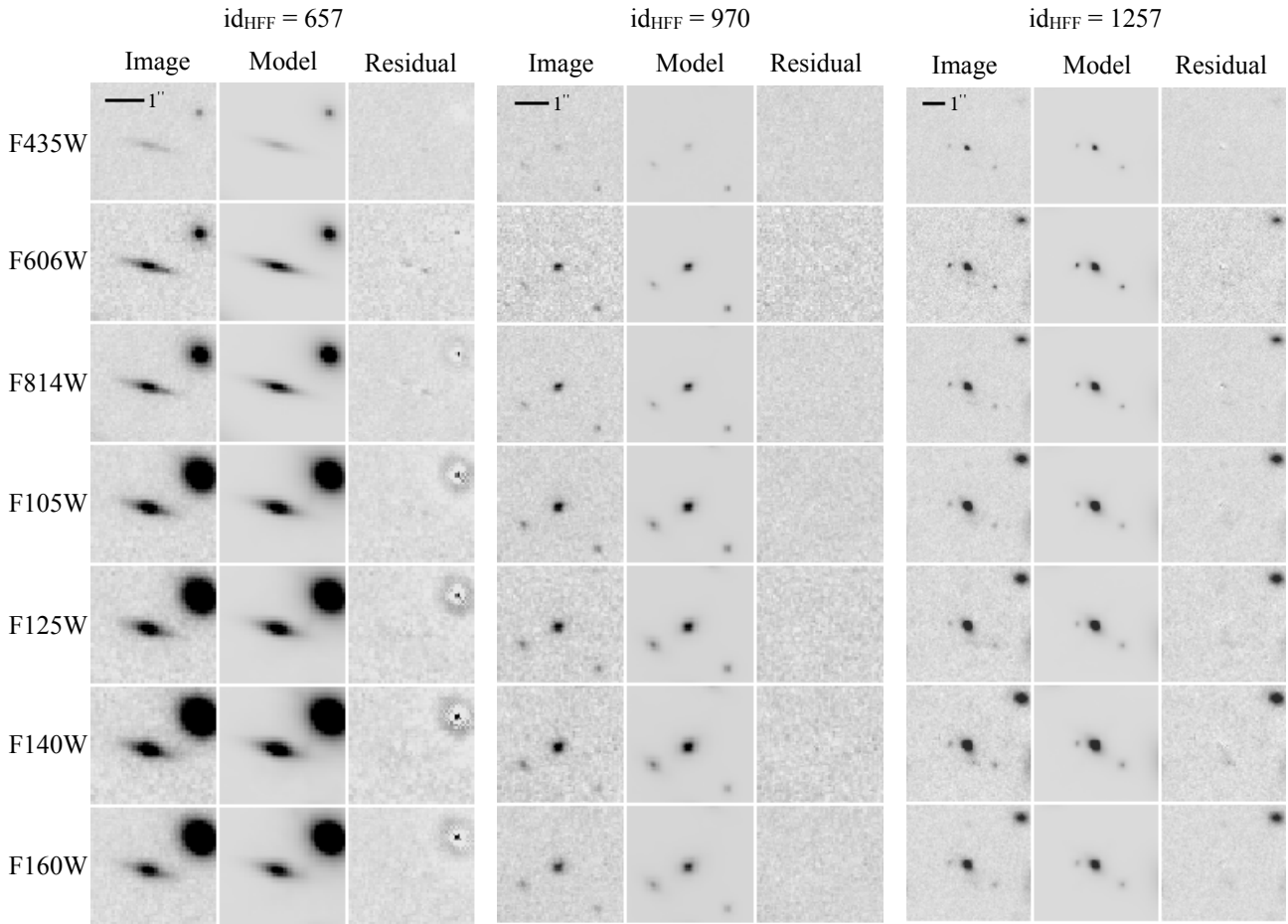}
    \caption{Fitting results for the three galaxies from the Abell1063 parallel field for which we show the parameters in Figure \ref{fig:Gala_params}. The image, model, and residual are shown for each of the seven bands that we use for the HFF. All images are oriented such that up is North and left is East. %\bh{I am seriously impressed by how empty the residuals are, though, especially given that we fit single-sersic models to clearly 2-component systems, at least in the first case}.
    For all examples shown here, the remaining flux in the residuals at each pixel around the primary object is less than 10$\%$ of the flux in the corresponding pixel of the image. This amount of residual flux is to be expected and is generally found when fitting smooth profiles to real galaxies.}
    \label{fig:galfitm}
\end{figure*}

As discussed in \S \ref{sec:MegaMorph}, from the wavelength dependence of the parameters that we fit, we can identify many properties of each galaxy that might otherwise be missed with single band fitting. We illustrate the importance of modelling morphological parameters as a function of wavelength in Figure \ref{fig:Gala_params}, where we show three example galaxies. These galaxies' parameters show very different wavelength dependencies and have been specifically selected in order to showcase the wide variety of functional forms that we allow our fitting routine to recover. We also show the images, models, and residuals for these same galaxies in Figure \ref{fig:galfitm}. Although we did not consider the stellar masses nor the redshifts of the example galaxies, they are all low mass galaxies at $z\sim0.5$. Galaxies id$_\mathrm{HFF} = 657$, id$_\mathrm{HFF} = 970$, and id$_\mathrm{HFF} = 1257$ have corrected stellar masses of $10^{8.27}$ M$_\odot$,  $10^{7.83}$M$_\odot$, and $10^{8.21}$ M$_\odot$, respectively. Hence, the quality of the models and residuals shown in Figure \ref{fig:galfitm} are typical for low mass galaxies.

In the left panel of Figure \ref{fig:Gala_params}, we test how apparent H-band magnitudes derived with the MegaMorph tools compare to those from the HFF-DeepSpace catalogue \citep{Shipley2018}. The  \textsc{GalfitM} apparent magnitudes are expected to be different from those in the HFF-DeepSpace catalogue, since they are derived in different ways. The magnitudes in the HFF-DeepSpace catalogue are measured with the AUTO aperture photometry, in which the extent of a galaxy is defined and all of the flux within that area is summed. A small AUTO-to-total correction is then applied. As previously discussed in \S \ref{stellarmass},  $\textsc{GalfitM}$, on the other hand, first models galaxy profiles and then these profiles are integrated to infinity in order to obtain total-S{\'e}rsic magnitudes in each band. Despite these differences, it can be seen that the H-band magnitudes are consistent for the three example galaxies in Figure \ref{fig:Gala_params}, indicating that the modelling results are robust.

Given the reliability of the modelling, we now go on to investigate the properties of the three example galaxies that we can infer from the wavelength dependence of their parameters. Galaxy id$_{\mathrm{HFF}}=657$, shown with solid lines, is very similar to galaxy id$_{\mathrm{HFF}}=1257$ in terms of brightness. However, these galaxies' sizes and Sérsic indices indicate that they are in fact very different. Perhaps the most striking property of galaxy  id$_{\mathrm{HFF}}=657$ is its large effective radius. This galaxy has a size that is larger at short wavelengths than at longer wavelengths, indicating that this is most likely a multi-component, star-forming object, where the short wavelengths reflect the size of the disk component and the longer wavelengths reflect the size of the bulge. On the other hand, this galaxy's S{\'e}rsic index is roughly equal to one at all wavelengths, which is indicative of a blue, disk-dominated system. The presence of a bulge is not well motivated from the  S{\'e}rsic index data alone. As it is more difficult to measure reliable S{\'e}rsic indices than reliable sizes, we argue that this object is likely a two-component system. From Figure \ref{fig:galfitm}, this galaxy appears more extended at shorter wavelengths, which is consistent with Figure \ref{fig:Gala_params}, where we show that the effective radius along the major axis is decreasing with wavelength.
In fact, the visual impression supports that this is an edge-on disk system with a significant, large, and round, bulge component.

Galaxy id$_{\mathrm{HFF}}=1257$, shown with a dotted line, has parameters that are consistent with a quiescent, elliptical galaxy. This object has a large S{\'e}rsic index at all wavelengths, suggesting that this is an elliptical galaxy, which classically have deVaucouleurs profiles with n$=4$ \citep[e.g.][]{Vulcani2014MNRAS, Kennedy2015MNRAS}.

Finally, galaxy id$_{\mathrm{HFF}}=970$, shown as a dashed line in Figure \ref{fig:Gala_params} and in the central column of Figure \ref{fig:galfitm}, is the faintest and smallest galaxy of the three examples, and therefore has the largest measurement uncertainties. 
%These characteristics can also be seen in Figure \ref{fig:galfitm}. 
This object is slightly smaller at shorter wavelengths than at longer wavelengths, which is what would be expected for a bulge-dominated galaxy. Like galaxy id$_{\mathrm{HFF}}=657$, this is most likely a two-component object based on the strong wavelength dependence of the S{\'e}rsic index shown in the right panel of Figure \ref{fig:Gala_params}. The S{\'e}rsic index at the bluest wavelength is $n \approx 1.6$, while at longer wavelengths, the S{\'e}rsic index is large, again a result of the disk component being most prominent at shorter wavelength while the bulge component being more prominent at longer wavelengths.

%All of these details can give us crucial information about the intrinsic properties of these galaxies, which could be lost with single band fitting. These details could be obtained by using single-band fits at different wavelengths and interpolating the values; however, this would only be feasible for the brightest galaxies. 
%\bh{Proposed:
All of these details can give us crucial information about the intrinsic properties of galaxies, which could be lost with single band fitting. In principle, in noise-less data, these details could be obtained by using single-band fits at different wavelengths and interpolating the values using the same polynomials; however, as data are not noiseless, this would only be feasible for the very brightest galaxies and not for the majority of objects we are after.
Additionally, multi-wavelength fitting gives us the advantage of being able to obtain the 5000\AA \ rest-frame parameters from the Chebyshev polynomials directly, allowing for a fairer comparison across redshifts. 
We indicate the rest-frame 5000\AA \ sizes and Sérsic indices for the three example galaxies shown in Figure \ref{fig:Gala_params} as black squares. 
We note that we do not derive rest-frame magnitudes in this way since we allow full freedom in order to recover the SEDs, which can lead to Runge's phenomenon, in which the polynomial is unconstrained in between the fixed points where data is available, making estimated magnitude values unfeasible.

 As can be seen from all three galaxy models in Figure \ref{fig:galfitm}, neighbouring objects are modelled along with the primary object. For object id$_{\mathrm{HFF}}=657$, there is a bright, neighbouring spheroidal galaxy in the upper right corner. The residuals of the neighbouring object are characteristic of a multi-component galaxy that has been modelled with a single S{\'e}rsic profile, exactly as we have done. Although the spheroidal neighbouring galaxy itself is not particularly well modelled in the centre, we are able to reliably recover the primary galaxy's parameters, as can be seen from the residuals of the primary object. Another interesting feature of this figure is the bright object ``next'' to galaxy id$_{\mathrm{HFF}}=1257$, again in the upper right corner. This object is not modelled and can be clearly seen in the residual of every band. This object is too far away from object id$_{\mathrm{HFF}}=1257$ to contribute to its flux, so it is masked out for the modelling. The residuals seen for these three galaxies are examples of normal residuals as these objects are not chosen to have excellent models nor particularly clean residuals.

Uncertainties in the magnitude, size, and S{\'e}rsic index measurements are produced directly by $\textsc{GalfitM}$. In Figure \ref{fig:Gala_params}, we show the $\textsc{GalfitM}$-derived uncertainties for each parameter in each band. Unfortunately, these uncertainties have been shown to be significantly underestimated \citep{Haeussler2007}. Since the 3D-HST \citep{Skelton2014} and HFF-DeepSpace \citep{Shipley2018} catalogues already provide reliable magnitudes and uncertainties for the objects in the CANDELS and HFF fields, respectively, we do not make an effort to obtain reliable uncertainties for the measured magnitudes but use those reported values instead. Additionally, because we do not use the uncertainties on the S{\'e}rsic index measurements for any of our results, we focus on deriving reliable uncertainties only for galaxy sizes. \cite{Haeussler2007} simulated images with similar properties to the HST $\textsc{Gems}$ \citep{Rix2004ApJS} dataset, which contained over 40,000 simulated galaxies of various magnitudes, sizes, position angles and axis ratios, and the light profiles of these galaxies were fit with $\textsc{Galfit}$.
%\bh{Also, as a further comment: We identified this at the time to Galfits internal assumptions about poisson noise being unrealistic, ignoring correlated noise, etc. In all our fits (certainly in CANDELS, and I think in HFF), we feed in sigma images, right? That should largely take those assumptions out of the equation. Maybe that's in fact the reason we find a factrr of 2-3 rather than the 10 I found in the past. Never thought of that.} 
\cite{Haeussler2007} found that $\textsc{Galfit}$ substantially underestimates the true uncertainties of the fit and suggest that the uncertainties estimated by the $\textsc{Galfit}$ are not reflecting the Poisson noise of the images.
They find that the $\textsc{Galfit}$ uncertainties are underestimated by a factor of $\sim10$ or more. Since \textsc{GalfitM} uses the same assumptions to derive uncertainties as \textsc{Galfit}, the errors estimated by both codes should be consistent. However, it is important to point out that the simulated galaxies in \cite{Haeussler2007} were all fit and modelled in one band. 
\cite{MegaMorph} performed a similar analysis for multi-wavelength fitting in 9 bands for simulated galaxies with properties similar to $\textsc{GAMA}$ \citep[Galaxy And Mass Assembly;][]{Driver2011MNRAS} images. While they do not present an analysis of the error bars in that work, subsequent checks on this issue revealed that
%.... [the error bars are not as badly underestimated] Using this multi-wavelength approach, they find that 
the uncertainty for the multi-wavelength fitting is only underestimated by a factor of $\sim 2 - 2.5$. Therefore, we increase all size errors estimated by $\textsc{GalfitM}$ by a factor of 3 in order to be conservative. To obtain the uncertainties on the 5000\AA \ rest-frame size, we linearly interpolate the uncertainties on the two bands that enclose the rest-frame wavelength. Given our redshift range, the rest-frame 5000\AA \  always falls within our wavelength range (i.e., $<$ 1.6$\mu$m);
therefore, we do not extrapolate any rest-frame parameters or uncertainties.

% ----------------------------% 
% THE Re-M* RELATION
% ----------------------------% 

\section{Stellar Mass -- Size Relation} \label{sec:M_Re}
The stellar mass -- size relation is shown in Figure \ref{fig:MASS-SIZE}, where the sample is divided into four redshift bins in the same way as in Figures \ref{fig:UVJ_mass} and \ref{fig:UVJ_ssfr}. Sizes smaller than the FWHM$_\mathrm{F160W}$/2 at the maximum redshift of each bin are indicated in grey. Size measurements smaller than this limit are not as robust as larger sizes; however, further investigation of the simulations carried out in \cite{MegaMorph} reveals that galaxy sizes smaller than FWHM/2 can be reliably measured if the point spread function (PSF) of the image is well known, which for HST data is the case. In their simulations, the FWHM of the PSF was $\sim$3.5 pixels, and galaxies were simulated down to effective radii of $\sim$1pixel, well below the FWHM/2 limit. No systematic effects on the measured image sizes were seen, implying that the measured sizes below this limit are not artificially shifted. Furthermore, in our sample, there are only 94 galaxies that have effective radii smaller than FWHM$_\mathrm{F160W}$/2. Although we are able to measure the small sizes of these galaxies, they do not comprise a large portion of the total sample and, therefore, would not heavily influence the stellar mass — size relation that we derive. We also note that the FWHM$_\mathrm{F160W}$/2 limit is well below the galaxy size distribution of our objects. To significantly change the result of this work, a strong (factor of a few in size measurements) and systematic (all small objects would have to be fit larger) measurement error would have to be observed. No such effect was found in any galaxy regime, specifically not for small objects.

In Figure \ref{fig:MASS-SIZE}, we differentiate between star-forming and quiescent galaxies based on three criteria: the sSFR, UVJ diagram, and S{\'e}rsic index as discussed in \S \ref{sf_vs_q}. In Figure \ref{fig:MASS-SIZE}, the top four panels are colour coded according to the sSFR as derived from FAST. The middle four panels are colour coded according to each galaxy's position on the UVJ diagram. Finally, the bottom four panels are colour coded by the S{\'e}rsic index following the idea that objects with high S{\'e}rsic index are predominantly passive ellipticals. As discussed in \S \ref{sf_vs_q}, none of these selection techniques are perfect, as they rely intrinsically on good photometric measurements, which becomes tricky especially for faint objects. %Selecting quiescent galaxies based on sSFR additionally requires very good SED models and is prone to misclassification of unique galaxies \citep[e.g.,][]{daCunha2008MNRAS}. Selecting quiescent galaxies based on their position on the UVJ diagram, on the other hand, is likely to miss galaxies that have recently ceased their star-formation, such as post-starburst galaxies as discussed in \S \ref{sf_vs_q}. Deriving S{\'e}rsic indices, especially for faint, high redshift galaxies is very difficult, making the separation of star-forming and quiescent galaxies according to the S{\'e}rsic index the least reliable separation method. However, as this divide is often used in the literature, we show it here for comparison. 
Despite all of these difficulties, Figure \ref{fig:MASS-SIZE} shows that there is generally good agreement between the three selection criteria. This is not necessarily expected, since the sSFRs and rest-frame colours are derived from SED models, while the Sérsic indices are obtained from modelling light profiles with \textsc{GalfitM}. As these quantities are independently derived, the level of consistency between these selection methods suggests that the light profile modelling \textit{and} the SED modelling are reliable. Naturally, there are some differences between the selection methods. For instance, the flattening of the quiescent sample is less pronounced with the Sérsic cut, as there appear to be only a few low mass galaxies with high Sérsic indices; however, the flattening is visible if galaxies with Sérsic indices $>2.5$ are considered to be quiescent, as is generally done in the literature \citep[e.g.,][]{Bruce2014MNRAS, Lange2015MNRAS}  such that flattening can be seen for the galaxies which are colour-coded as yellow points in the figure.

From Figure \ref{fig:MASS-SIZE}, the star-forming and quiescent sequences can be seen to occupy the same regions of the stellar mass -- size plane for all three separation criteria even though individual galaxies may be classified differently by the different methods. 
It is also worth noting that there is a cloud of quiescent galaxies at the very low mass end (i.e. below 10$^{8}$ M$_{\odot}$) that can be seen especially in the $0.2 \leq z \leq 0.5$ and $0.5 < z \leq 1.0$ redshift bins when the quiescent sample is selected based on the sSFR (top panels).  This cloud of galaxies cannot be seen when we select galaxies based on the UVJ diagram or S{\'e}rsic index. Further investigation of these objects reveals that their SEDs are visually consistent with being young, blue, low-mass galaxies. However, due to the rather featureless SEDs of young, low-mass galaxies and the relatively large measurement errors in these faint objects, the derived sSFRs suffer from large uncertainties. In fact, many of these objects would be classified as star-forming if the lower 1$\sigma$ sSFR were used instead of their best-fit sSFR.
Combined with the uncertainties on the Sérsic indices previously discussed, we therefore choose to separate star-forming and quiescent galaxies based on their positions on the UVJ diagram for the subsequent sections of this paper because this appears to be the most robust method for our sample.

Above $\sim$10$^{10}$ M$_{\odot}$, which is the stellar mass regime that most other studies in the literature have focused on, the quiescent sequence on the stellar mass -- size plane can be easily distinguished from the star-forming sequence at all redshifts. In other words, the quiescent galaxies follow a distinct trend on the mass-size plane that is different from the one that the blue, star-forming galaxies follow. This distinction can be made regardless of whether star-forming and quiescent galaxies are separated based on sSFR, position on the UVJ diagram, or S{\'e}rsic index. In this high mass regime, the quiescent galaxy sequence shows a steeper slope than the star-forming population across all redshifts, consistent with previous works \citep[e.g.,][]{vdWel2014,Dimauro2019, Mowla2019ApJ}. Consequently, at a given stellar mass, high mass quiescent galaxies tend to be smaller in size than their star-forming counterparts, again consistent with the literature. These trends however, do not hold true for less massive objects.

%For intermediate to low stellar mass galaxies, the trends observed for high mass galaxies do not hold.  
Below $\sim$10$^{10}$ M$_{\odot}$, the quiescent and star-forming galaxy sequences are not quite as clearly distinguishable from each other. This could partly be an effect due to the fact that it is more difficult to separate less massive galaxies into star-forming and quiescent because it is generally harder to measure the properties of less massive objects. In spite of this, we recover similar trends for the star-forming and quiescent sequence across all selection methods, suggesting that this behaviour is real. The stellar mass -- size relation of low mass quiescent galaxies appears flat and then steepens for high mass quiescent galaxies, while the star-forming galaxies continue to grow in size as they grow in mass from 10$^7$ to 10$^{11.5}$ M$_\odot$. We will quantitatively discuss these relations in the following sections.
The flattening of the stellar mass -- size relation for quiescent galaxies has been shown in previous works \citep[e.g.,][]{Cappellari2013MNRAS, Berg2014, Norris2014MNRAS,Lange2015MNRAS, Whitaker2017ApJ}; however, most of these are low redshift studies or are limited to relatively high stellar mass galaxies.
In this work, we extend the stellar mass -- size relation to include low stellar mass galaxies while also quantitatively exploring its flattening as a function of redshift.
Unfortunately, at redshifts $z>1$, our number of quiescent galaxies with stellar masses below 10$^9$ M$_\odot$ is limited. This is, in fact, why we use the multi-band tools discussed in \S \ref{sec:Galapagos}, as they allow us to push our analysis to fainter, less massive objects than was possible in previous studies. Nevertheless, some evidence of a flattening remains at these higher redshifts. 

\subsection{Quiescent Mass -- Size Relation }
We first analyse the quiescent galaxy sample shown in Figure \ref{fig:MASS-SIZE}, selected using the UVJ diagram. In \S \ref{Q_HighMass}, we discuss the high mass end (i.e., $\geq 10^{10.3}$ M$_\odot$) only, for easier comparison with the literature. In \S \ref{Q_AllMass}, we discuss the behaviour of the quiescent sequence over the entire stellar mass range analysed in this work (i.e., $\geq 10^{7}$ M$_\odot$). Finally, in \S \ref{sec:Q_evolution}, we show the redshift evolution of the best-fit trends and discuss what this implies for how quiescent galaxies build up their mass and grow over cosmic time. 

\subsubsection{The High Mass End } \label{Q_HighMass}
We begin by fitting the high mass end of the stellar mass -- size relation for quiescent galaxies as this mass regime has been extensively studied in the literature \citep[e.g.,][]{Maltby2010MNRAS,vdWel2014,Kuchner2017A&A,Dimauro2019}. These previous works have shown that the stellar mass -- size relation is well represented by a single power law at the high mass end for both quiescent and star-forming galaxies. We therefore fit a power-law function of the form:
\begin{equation}
    R_e = A \Big( \frac{M_*}{5 \times 10^{10} \ \mathrm{M}_\odot} \Big) ^ {B}
    \label{eq:singlepl}
\end{equation}
 following \cite{vdWel2014}, \cite{Mowla2019ApJ}, and \cite{Dimauro2019} in order to make a direct comparison. In Equation \ref{eq:singlepl}, $R_e$ is the 5000\AA \ rest-frame half-light radius along the major axis in kpc and $M_{*}$ is the stellar mass in M$_{\odot}$ and corrected as described in \S  \ref{stellarmass}; $A$ and $B$ are best-fit parameters that describe the trend. As this relation is linear in the log-log space in which we fit the data, we refer to $B$ as the slope and $\log_{10}(A)$ as the intercept at $5 \times 10^{10}$ M$_\odot$. 

\begin{figure}
    \centering
    \includegraphics[width=0.46\textwidth]{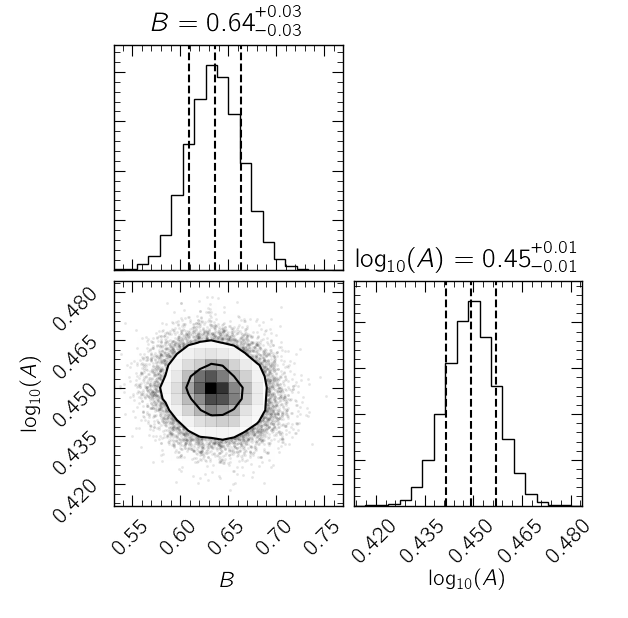}
    \caption{Corner plot for the estimated parameters for high mass quiescent galaxies in the redshift bin 0.5 $<$ $z$ $\leq$ 1.0 according to Equation \ref{eq:singlepl}. We choose to show the corner plot for this redshift bin because we have the largest sample size in this range and therefore, smaller uncertainties; however, the parameters exhibit similar behaviour in the other redshift bins as well. For the 1D posterior distributions, the vertical dashed lines mark the 16$^\mathrm{th}$, 50$^\mathrm{th}$, and 84$^\mathrm{th}$ percentiles, where the 16$^\mathrm{th}$ and 84$^\mathrm{th}$ percentiles of each distribution are indicated as the lower and upper uncertainties of each parameter estimation, respectively. These values are reported for the quiescent fits in Table \ref{tab:from_eq2}. For the 2D posterior distribution, the contours mark the 1$\sigma$ and 2$\sigma$ levels.}
    \label{fig:Q_highM_cornerplot}
\end{figure}
 
 The best-fit parameters, $A$ and $B$, of our model are determined using a Bayesian inference with a Markov Chain Monte Carlo (MCMC) approach \citep{emcee2013PASP}. We assume uniform priors for both parameters and each galaxy's contribution to the fit is weighted by its uncertainty in mass and size, which are described in \S \ref{stellarmass} and \S \ref{sec:Galapagos}, respectively, by taking the covariance of the two uncertainties. The parameters are fit in the $\log_{10}$ M$_*$ -- $\log_{10}$ Re space, as in the other works that we compare our results to. Motivated by the parameters obtained in previous studies, we allow  $\log_{10} (A)$ and $B$ to vary over [-0.5, 2] and [0, 2], respectively. We set 50 random walkers, and perform 10,000 MCMC iterations, which allow the parameters to converge on the best-fit value for all four fits, one for each redshift bin. The resulting corner plot \citep{corner} for the fitting routine in the 0.5 $<$ $z$ $\leq$ 1.0 range is shown in Figure \ref{fig:Q_highM_cornerplot}.

\begin{figure*}
    \centering
    \includegraphics[width=1.0\textwidth]{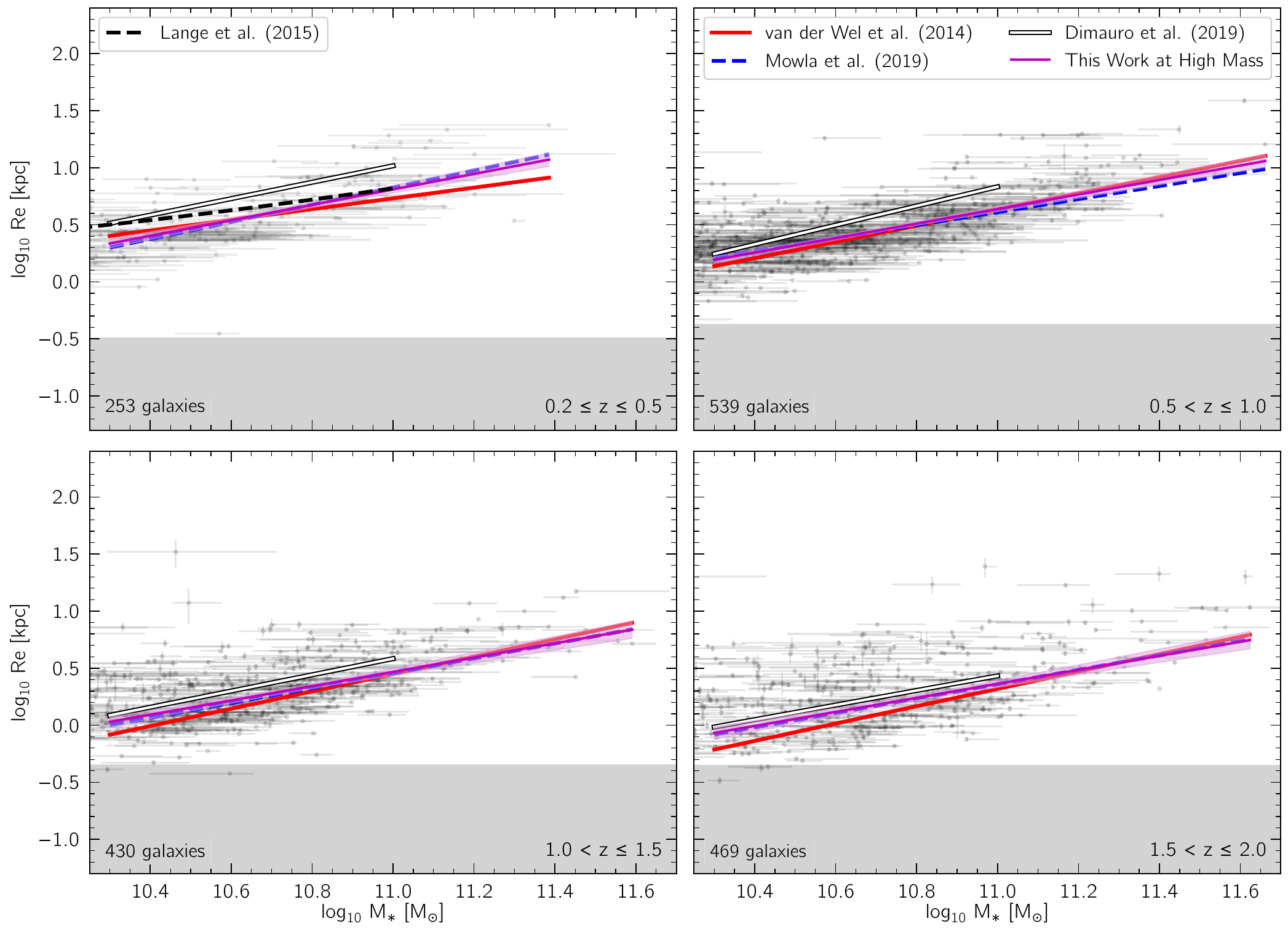}
    \caption{The stellar mass -- size relations for high mass quiescent galaxies. The best-fit power law is shown in magenta and the 1$\sigma$ level that includes 68\% of all models is indicated by the shaded region. In order to make a fair comparison to \protect\cite{Dimauro2019} (white line), \protect\cite{vdWel2014} (red line) and \protect\cite{Mowla2019ApJ} (blue dashed line), we use the same lower stellar mass limit of $10^{10.3}$ M$_\odot$. We additionally compare to \protect\cite{Lange2015MNRAS} (black dashed line in the lowest redshift bin only), who fit a different functional form to their sample, which we discuss in detail in \S \ref{Q_AllMass}, and study local galaxies with $0.01<z<0.1$. As in Figure \ref{fig:MASS-SIZE}, the grey areas indicate sizes at $R_{\rm e} < {\rm FWHM_{\rm F160W}}$/2 at the maximum redshift of each panel. }
    \label{fig:Q_highM_M_Re}
\end{figure*}

 In Figure \ref{fig:Q_highM_M_Re}, we show the stellar mass -- size relations across four redshift bins for the high mass quiescent galaxy population, where the number of objects that are fit in each redshift bin are indicated in the bottom left corner. Our best-fit models are shown in magenta where the shaded region signifies the 1$\sigma$ level within which 68\% of all models fall. 
 Across the entire redshift range that we explore, we find generally good agreement with the stellar mass -- size relations from \cite{vdWel2014}, \cite{Mowla2019ApJ}, and \cite{Dimauro2019}, which are also shown in Figure \ref{fig:Q_highM_M_Re}. Finally, we also compare our results in the lowest redshift bin to \cite{Lange2015MNRAS}, who fit the stellar mass -- size relation with a different functional form, which we discuss further in \S \ref{Q_AllMass}.

\subsubsection{Entire Mass Range } \label{Q_AllMass}

For the full stellar mass range that we are exploring, the stellar mass -- size relation for quiescent galaxies flattens, as can be seen from Figure \ref{fig:MASS-SIZE}. 
In order to capture this flattening at the low mass end, we fit the quiescent galaxy mass -- size relation with a double power-law function motivated by \cite{Shen2003} 
and \cite{Lange2015MNRAS}:
\begin{equation}
\begin{split}
R_e &= \gamma \big( M_* \big)^{\alpha} \Big( 1 + \frac{M_*}{10^{\delta}} \Big)^{\beta - \alpha}  
% \log_{10}R_e &= \log_{10}\gamma + \alpha \log_{10} M_* + (\beta - \alpha)  \log_{10}\Big( 1 + 10^{\log_{10} M_* -\delta} \Big)
\end{split}
\label{eq:doublepl}
\end{equation} 
where $R_e$ is the 5000\AA \ rest-frame half-light radius in kpc and $M_{*}$ is the corrected stellar mass in M$_{\odot}$, as in Equation \ref{eq:singlepl}; $\alpha$ and $\beta$ describe the slope at the low and high mass end, respectively; $\gamma$ is the normalisation, or in other words, the effective radius at a stellar mass of 10$^0$ M$_\odot$. Finally, 10$^\delta$ is the stellar mass at which the second derivative of the function is at a maximum. Since it is difficult to assign an intuitive, physical meaning to this parameter, we simply refer to it as $\delta$. This parameter can be considered to be the distinction between high and low mass galaxies \citep[e.g.,][]{Shen2003, Lange2015MNRAS}; however, we find that this value does not align well with the visual transition from one slope to the other. For example, in the highest redshift bin, the most probable $10^{\delta}$ value is $\sim 10^{11.27}$ M$_\odot$, which would imply calling even very massive galaxies `low mass', ultimately making such a cut a poor choice. For quiescent galaxies, we choose to make the distinction between high mass and low mass galaxies at 10$^{10.3}$ M$_{\odot}$ 
%\bh{where such terms are being used} 
since this is the stellar mass range we use in \S \ref{Q_HighMass} following \cite{vdWel2014}, \cite{Mowla2019ApJ}, and \cite{Dimauro2019}.

 \begin{figure}
    \centering
    \includegraphics[width=0.48\textwidth]{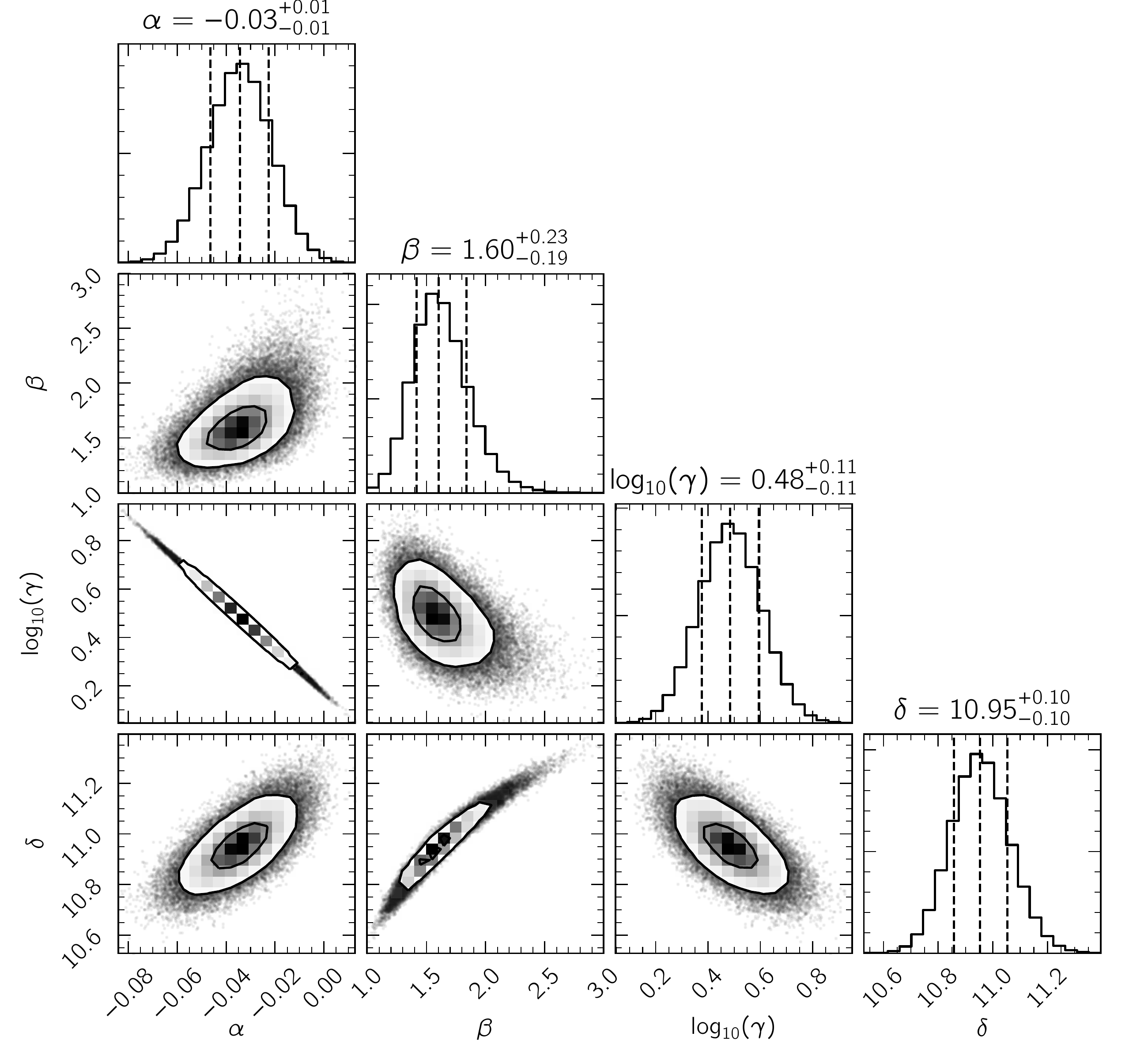}
    \caption{Corner plot for the four parameter MCMC fitting routine for the 0.5 $<$ $z$ $\leq$ 1.0 redshift bin. The vertical dashed lines in the 1D posterior distributions indicate the 16$^\mathrm{th}$, 50$^\mathrm{th}$, and 84$^\mathrm{th}$ percentiles, where the 16$^\mathrm{th}$ and 84$^\mathrm{th}$ percentiles of each distribution are indicated as the lower and upper uncertainties of each parameter estimation, respectively. These results are reported for the entire redshift range used in this work in Table \ref{tab:from_eq2}. For the 2D posterior distributions, the contours mark the 1$\sigma$ and 2$\sigma$ levels.}
    \label{fig:Q_allM_cornerplot}
\end{figure}

\begin{figure*}
    \centering
    \includegraphics[width=1.0\textwidth]{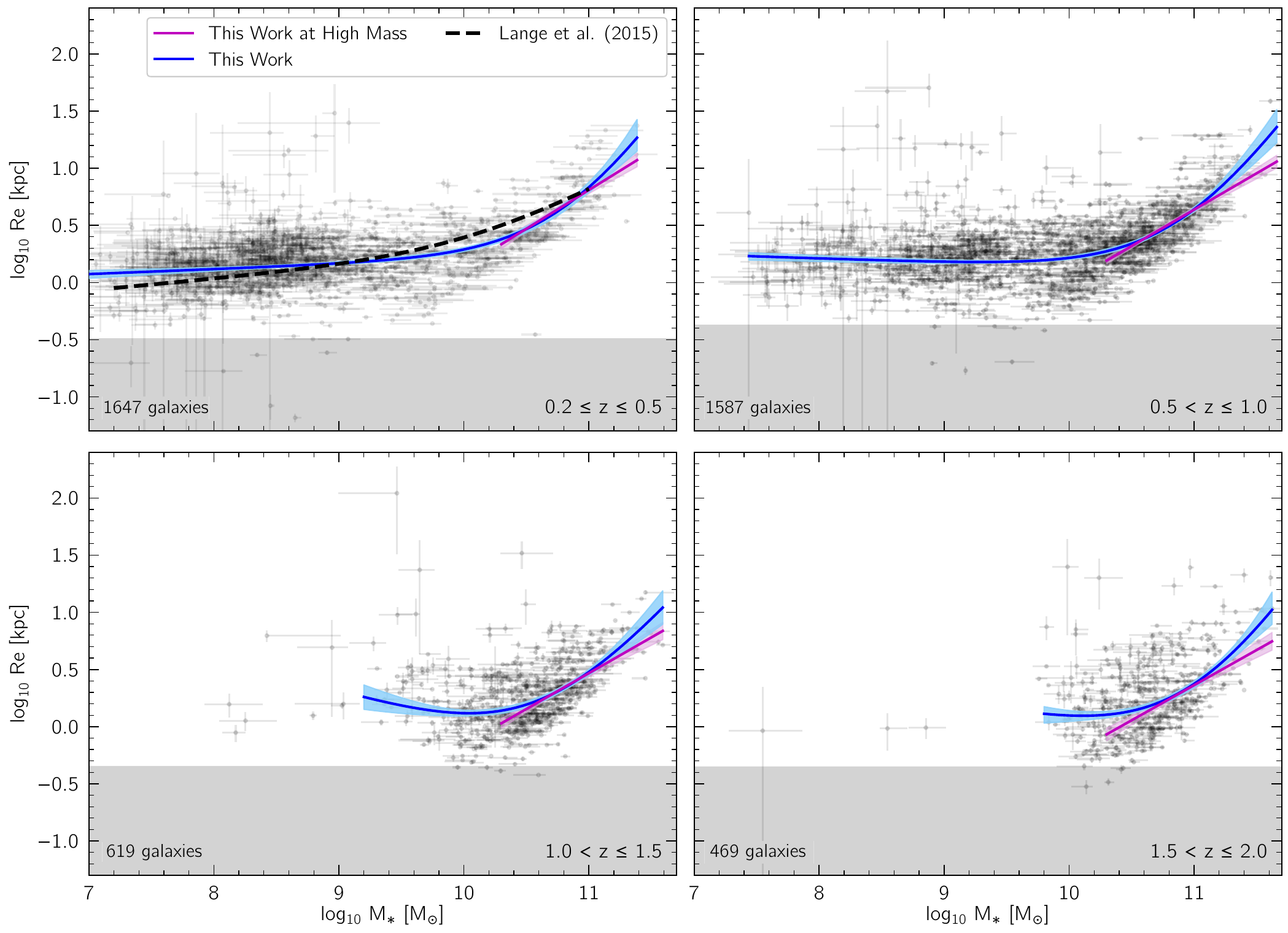}
    \caption{The stellar mass -- size relation of quiescent galaxies, where the single power-law models for the high mass end are indicated in magenta and the double power-law fits are shown in blue. The number of galaxies fit with the double power-law fit are indicated in the bottom left corner of each redshift bin. As in previous figures, the grey areas indicate sizes at R$_{\rm e} < {\rm FWHM_{\rm F160W}}$/2 at the maximum redshift of each panel.}
    \label{fig:Q_M_Re}
\end{figure*}

The double power-law function, shown in Equation \ref{eq:doublepl}, is fit to the quiescent galaxy samples using an MCMC approach in a way that is similar to how the single power law is fit to the high mass quiescent galaxies. Namely, we assume uniform priors for all parameters and the galaxies are weighted by their uncertainties in mass and size. All parameters are fit in log space, and because of this we choose to fit $\log_{10}(\gamma)$ as opposed to $\gamma$. Although the same posterior probability distributions can be recovered for the parameters by allowing them to vary over a variety of ranges as long as they are large enough, we choose to let the parameters vary over the following ranges:
\begin{itemize}
  \item The low mass slope: $\alpha \in  [-2, 2]$
  \item The high mass slope: $\beta \in  [0, 10]$
  \item The normalisation, i.e., $\log_{10}$ Re at 1 M$_\odot$: $\log_{10}(\gamma) \in  [-10, 10]$
  \item log$_{10}$ of the stellar mass at which the second derivative of the function is at a maximum: $\delta \in  [9, 13]$
\end{itemize}
The parameter spaces of the low mass and high mass slopes, as well as the normalisation, are derived empirically. The parameter space of 10$^\delta$ is constrained from 10$^9$ to 10$^{13}$, since this is the stellar mass range within which we expect the low mass slope to transition to the high mass slope. 

For the four parameter fitting, we set 100 random walkers, and perform 30,000 MCMC iterations. The resulting corner plot \citep{corner} is shown in Figure \ref{fig:Q_allM_cornerplot}. 
For the double power law fit (Eq. \ref{eq:doublepl}), we see a strong anti-correlation between $\alpha$ -- the low mass slope -- and $\log_{10} (\gamma)$ -- the normalisation. This indicates that if the low mass slope is a large negative value, then the resulting model would have a large size at $10^{0}$ M$_{\odot}$. This behaviour can be seen in Table \ref{tab:from_eq2}, where the low mass slope is most negative for the $1.0<z\leq1.5$, so in turn, we see the largest normalisation in this redshift range as well. If the low mass slope is zero, then the normalisation would be the size ($\log_{10}$Re) at $10^{\delta}$. Additionally, $\beta$ (the high mass slope) and $\delta$ ($\log_{10}$ of the stellar mass at which the second derivative is at a maximum) are correlated. This behaviour explains why the four parameter model always has a steeper high mass slope than the two parameter model, as can be seen in Figure \ref{fig:Q_M_Re}. %\bh{not sure I can follow that argument without explanation}. 
If we set 10$^\delta$ M$_\odot$ as the lower stellar mass limit, instead of using 10$^{10.3}$ M$_\odot$, the resulting two parameter model has a slope that is consistent with the high mass slope that we recover with the four parameter model. 
%\bh{Ok, got it now, but the last sentence is a little complicated, needed to read a few times.}

We show the stellar mass -- size relation for quiescent galaxies over the full stellar mass range explored in this work in Figure \ref{fig:Q_M_Re}. The best-fit model is shown in blue with the 1$\sigma$ level, which includes 68\% of all possible models, in light blue. 
We additionally show the high mass model with its corresponding 1$\sigma$ level from Figure \ref{fig:Q_highM_M_Re} in magenta. Generally, we find that the `high mass' and `all mass' models are in good agreement since they overlap and display similar behaviour above $10^{10.3}$ M$_\odot$. This indicates that the two independently derived models are consistent, lending confidence to our result. The best-fit model from \cite{Lange2015MNRAS} is also shown in the first panel of Figure \ref{fig:Q_M_Re} as a dashed black line.  Although this model has somewhat different characteristics from the model that we obtain, it is important to note that \cite{Lange2015MNRAS} use GAMA data \citep{Driver2011MNRAS}, so that their stellar mass -- size relations are for $z\sim0$ galaxies. Furthermore, the \cite{Lange2015MNRAS} stellar mass -- size relation shown here is derived in the g-band, where star-forming and quiescent galaxies are separated with a $u-r$ colour cut. Hence, it is not surprising that there is some difference between their results and ours. Nonetheless, the overall shape of the two curves is generally consistent.

\begin{figure}
    \centering
    \includegraphics[width=0.48\textwidth]{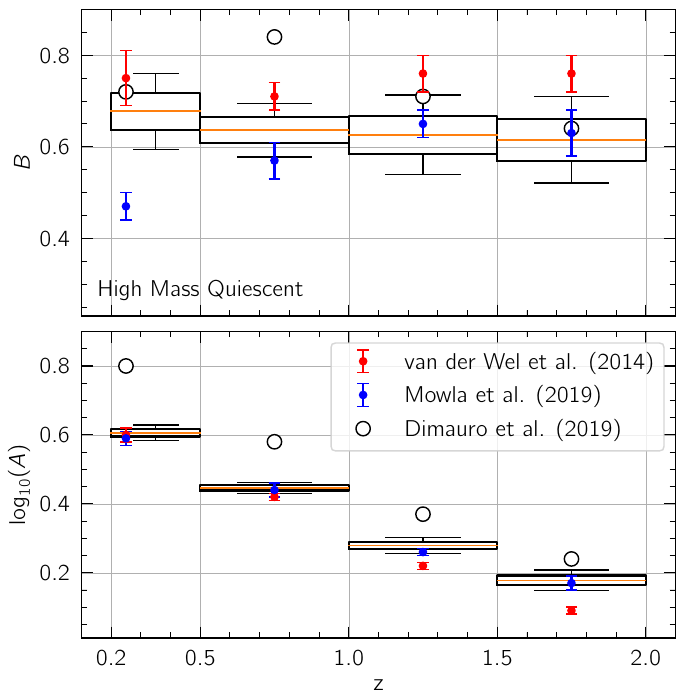}
    \caption{The evolution of the parameters  $B$ and $\log_{10}(A)$ with redshift from the single power-law fit shown in Equation \ref{eq:singlepl} for high mass quiescent galaxies (i.e., M$_{*}$ $\geq 10^{10.3}$ M$_{\odot}$). For the posterior probability distribution in each redshift bin, we indicate the median with an orange line, the 1$\sigma$ confidence level as a box, and the 2$\sigma$ confidence level an error bar. The values shown here are reported in Table \ref{tab:from_eq1}. We additionally compare these results to other high mass studies, as indicated in the legend. We find that the slope of the best-fit relation for high mass quiescent galaxies remains constant with redshift, consistent with previous works. There is a clear decrease in the intercept -- $\log_{10}(A)$ -- with redshift. $\log_{10}(A)$ describes the size at a fixed stellar mass of $5\times10^{10}$ M$_\odot$ and its evolution with redshift indicates that quiescent galaxies at higher redshift were more compact, as expected.}
    \label{fig:Q_evolution_highMass}
\end{figure}

\begin{figure*}
    \centering
    \includegraphics[width=1.0\textwidth]{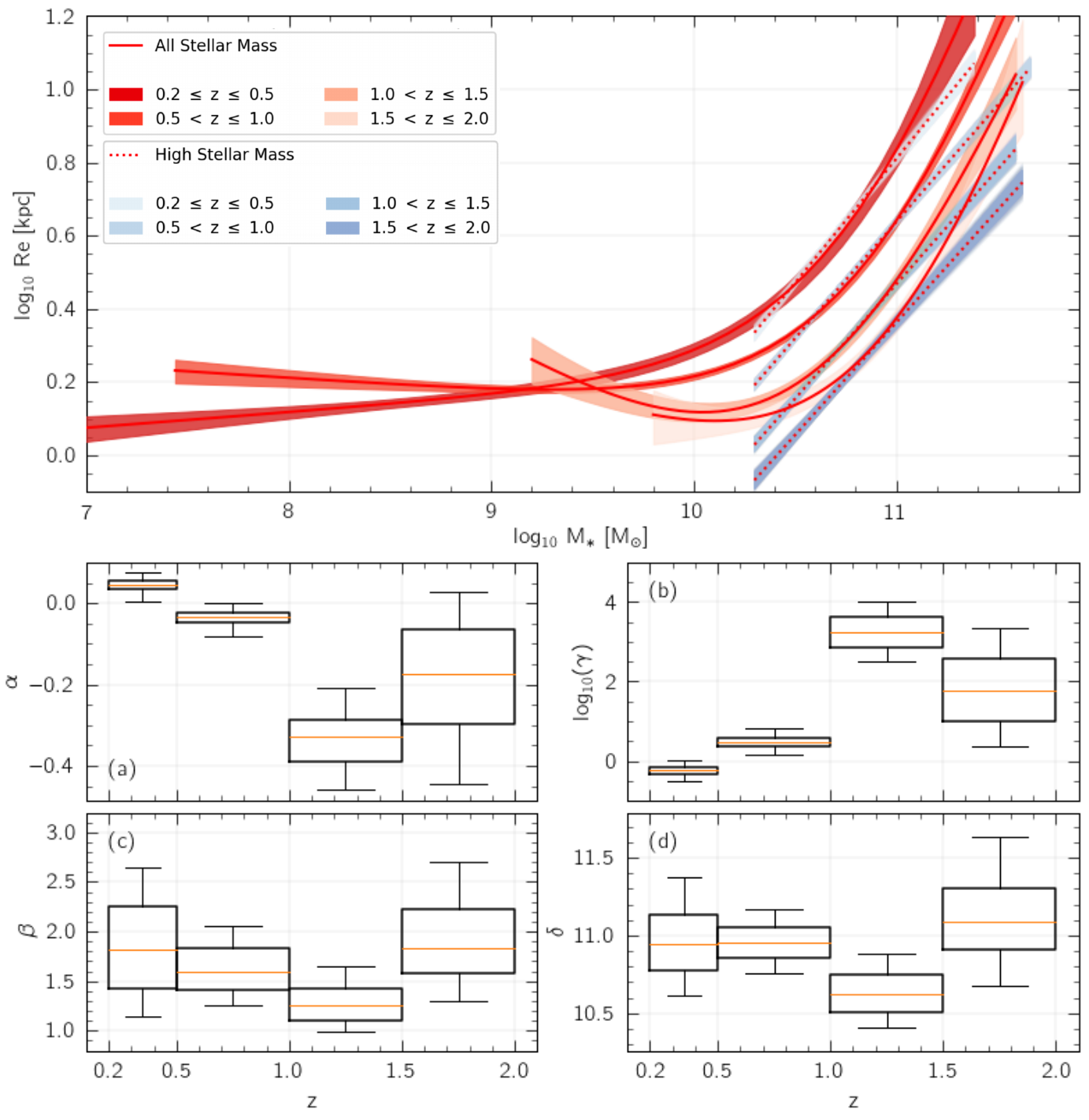}
    \caption{The evolution of the stellar mass – size relation for the quiescent sample. In the top panel, we show the quiescent best-fit trends from each of our four redshift ranges, where the concentration of the colour for the 1$\sigma$ level indicates the redshift bin, as shown in the legend. The best-fit model for each redshift bin is shown as a solid red line while the high mass best-fit models are shown as dashed lines. The redshift evolution of $\alpha$, $\beta$, $\log_{10}\gamma$, and $\delta$ are shown in panels (a), (b), (c), and (d), respectively.  The median, 1$\sigma$ and 2$\sigma$ levels for each of these parameters are shown as an orange line, box, and error bar, respectively. As in Figure \ref{fig:Q_allM_cornerplot}, $\alpha$ and $\log_{10} \gamma$ show anti-correlation while $\beta$ and $\delta$ are correlated. Unsurprisingly, the uncertainties for the parameters that describe the low mass end (i.e., $\alpha$ and $\log_{10}\gamma$) are large in the highest two redshift bins, in which there are few low mass quiescent galaxies. }
    \label{fig:Q_evolution}
\end{figure*}

From the quiescent galaxy data, it appears that a model with a positive slope at the low mass end could be better representative of the data. Given the small number of low mass quiescent galaxies at $z>1$, the MCMC models are primarily driven by intermediate mass galaxies. A negative low mass slope would imply that quiescent galaxies with stellar masses below $\sim 10^{10}$ M$_\odot$ are more extended than their higher mass counterparts. 
This result is difficult to explain physically, so we test how the model behaves if it is restricted to have a positive low mass slope. We find that the general shape of the curve is similar and the behaviour at the high mass end remains largely unchanged when the low mass slope is forced to be positive. Given that we can reproduce the general behaviour of the model with and without constraining the low mass slope, we opt to show the model in which the parameters are unconstrained, but refer explicitly to the mass range in which this model is valid and has been tested in our data. Interestingly, \cite{Genel_illustris_2018} also predict an increasingly negative low mass slope with redshift when reproducing the stellar mass -- size relation with Illustris.

\begingroup
\setlength{\tabcolsep}{6pt} % Default value: 6pt
\renewcommand{\arraystretch}{1.3} % Default value: 1

\begin{table*}
\caption{The estimated parameters for quiescent galaxies according to the single power law shown in Eq. \ref{eq:singlepl} for the high mass galaxies and the double power law shown in Eq. \ref{eq:doublepl} for quiescent galaxies over $10^7 - 10^{11.5}$ M$_\odot$. The high mass parameters correspond to the magenta lines in Figure \ref{fig:Q_M_Re}, and the all mass parameters correspond to the blue curves. As the `All Mass' parameters have posterior probabilities that are not symmetric around the median, we report the 16$^\mathrm{th}$ and 84$^\mathrm{th}$ percentiles as lower and upper limits. The corner plots for the $0.5<z\leq1.0$ redshift range can be seen in Figures \ref{fig:Q_highM_cornerplot} and \ref{fig:Q_allM_cornerplot}, for the high mass and all mass parameters, respectively. We additionally provide the root-mean-square error (RMSE) for each model. In the right-most column, we report the stellar mass below which we do not place confidence in our model due to a lack of data. For the first two redshift bins, this value is the stellar mass of the lowest-mass galaxy in that bin. At higher redshift, this is visually set to a value below which we do not have a significant number of data points. }
\centering
\begin{tabular}{c c c c c c c c c c}
\hline
\multicolumn{1}{l}{ \ } & 
\multicolumn{3}{c}{High Mass} &
\multicolumn{5}{c}{All Mass} &
\multicolumn{1}{c}{Lower Limit on}\\ 

$z$ & $\log_{10}$($A$) & $B$& RMSE&  $\alpha$ & $\beta$ & $\log_{10}(\gamma)$  & $\delta$ & RMSE & $\log_{10}$ M$_*$ [M$_\odot$] \\
\hline 

0.2$\leq{z}\leq$0.5 & 
$0.61{\pm0.01}$ & $0.68{\pm0.04}$& 0.23& 
$0.04^{+0.01}_{-0.01}$&$1.82^{+0.44}_{-0.39}$ & $-0.24^{+0.09}_{-0.09}$
& $10.94^{+0.19}_{-0.17}$  & 0.23
& 7.00 \\

0.5$<$ $z$ $\leq$1.0 &
$0.45{\pm0.01}$&$0.64{\pm0.03}$& 0.24&
$-0.03^{+0.01}_{-0.01}$&$1.60^{+0.23}_{-0.19}$ & $0.48^{+0.11}_{-0.11}$
& $10.95^{+0.10}_{-0.10}$ & 0.23 & 7.44\\

1.0$<$ $z$ $\leq$1.5 & 
$0.28{\pm0.01}$ & $0.63{\pm0.04}$ & 0.22 & 
$-0.33^{+0.06}_{-0.06}$ & $1.25^{+0.17}_{-0.14}$
& $3.24^{+0.39}_{-0.38}$ & $10.63^{+0.12}_{-0.12}$ 
& 0.25 & 9.20 \\

1.5$<z\leq$2.0 &
$0.18{\pm0.01}$ & $0.61{\pm0.05}$ & 0.28 & 
$-0.17^{+0.11}_{-0.12}$ & $1.84^{+0.40}_{-0.26}$
& $1.76^{+0.81}_{-0.74}$ & $11.09^{+0.21}_{-0.18}$ 
& 0.29 & 9.80 \\

\hline
\end{tabular}
\label{tab:from_eq2}
\end{table*}
\endgroup

\subsubsection{Quiescent Galaxy Evolution} \label{sec:Q_evolution}

After galaxies quench, they continue to undergo subsequent evolution by merging with other galaxies and/or minor episodes of star-formation activity. Whereas major mergers, involving galaxies of similar mass, will lead to comparable growth in both size and mass, minor mergers can leave the overall galaxy profile shape largely unchanged, while leading to substantial size evolution. These results have been shown by a number of studies \citep[e.g.,][]{Buitrago2008ApJ,Bezanson2009ApJ} and are also well-supported by simulations \citep[e.g.,][]{Naab2009ApJ, Oser2012ApJ}. We find that the average size evolution which galaxies experience once they quench depends strongly on their stellar mass, such that above stellar masses of $10^{10.3}$ M$_\odot$, quenched galaxies experience a steep evolution on the mass -- size plane with slopes of $\sim 0.7$ at $z \leq2$. \cite{vanDokkum2015ApJ} argued that the evolution of quiescent galaxies with stellar masses above 10$^{10}$ M$_{\odot}$ is primarily driven by dissipationless, dry mergers. \cite{Newman2012ApJ} showed that this mechanism requires a high rate of occurrence of minor mergers to account for the observed size growth of quiescent galaxies.
While multiple dry mergers can significantly increase the size of a galaxy, they do not significantly impact the total stellar mass \citep[e.g.,][]{Bezanson2009ApJ,Hopkins2010MNRAS, vanDokkum2010ApJ}. This causes the quiescent sequence to follow a steep slope on the mass--size plane and 
our results are hence consistent with this picture at the high mass end. 
The evolution of the best-fit parameters with redshift in this high mass regime is shown in Figure \ref{fig:Q_evolution_highMass}. 
The slope, shown in the top panel, does not show a strong evolution, while the intercept, $\log_{10}(A)$, shows significant evolution with redshift, representing the upwards shift of the relation over time. These results are consistent with many other studies and imply that high mass quiescent galaxies were more compact in the early Universe than they are today.

 Quenched low mass galaxies, however, exhibit an almost flat relation on the stellar mass -- size plane. 
 Because of this, the quiescent galaxy sequence has to be fit with a double power law function, which is shown in Equation \ref{eq:doublepl}. 
 We show the redshift evolution of the double power law fit and parameters in Figure \ref{fig:Q_evolution}. 
 In the top panel, all models are shown, where the shading of the $1\sigma$ region indicates the redshift according to the legends. A clear evolution can be seen for the stellar mass -- size relation of quiescent galaxies. 
 In panel (a), we show the evolution of the slope at the low mass end, $\alpha$, and in panel (b), we show the evolution of the normalisation, $\log_{10}\gamma$, which is anticorrelated to $\alpha$. In panel (c), we show the evolution of the slope at the high mass end, $\beta$. The slope at the high mass end remains relatively constant with increasing redshift, which is consistent with the behaviour of the slope when only high mass quiescent galaxies are fit with a single power law (top panel of Figure \ref{fig:Q_evolution_highMass}). Finally, in panel (d), we show the evolution of $\delta$. Since this parameter is correlated to the slope at the high mass end ($\beta$), we do not see a strong redshift evolution for $\delta$.

\subsection{Star-forming Mass -- Size Relation }
The star-forming galaxy relation on the mass--size plane is analysed in a similar fashion to the quiescent relation, such that in \S \ref{SF_HighMass}, we discuss the high mass end (i.e., $\geq 10^{9.5}$ M$_\odot$) only and in \S \ref{SF_AllMass}, we discuss the trends exhibited over the entire mass range available to us. In \S \ref{sec:SF_evolution}, we show and discuss the redshift evolution of star-forming galaxies on the stellar mass--size plane.

\subsubsection{The High Mass End} \label{SF_HighMass}

\begin{figure*}
    \centering
    \includegraphics[width=1.0\textwidth]{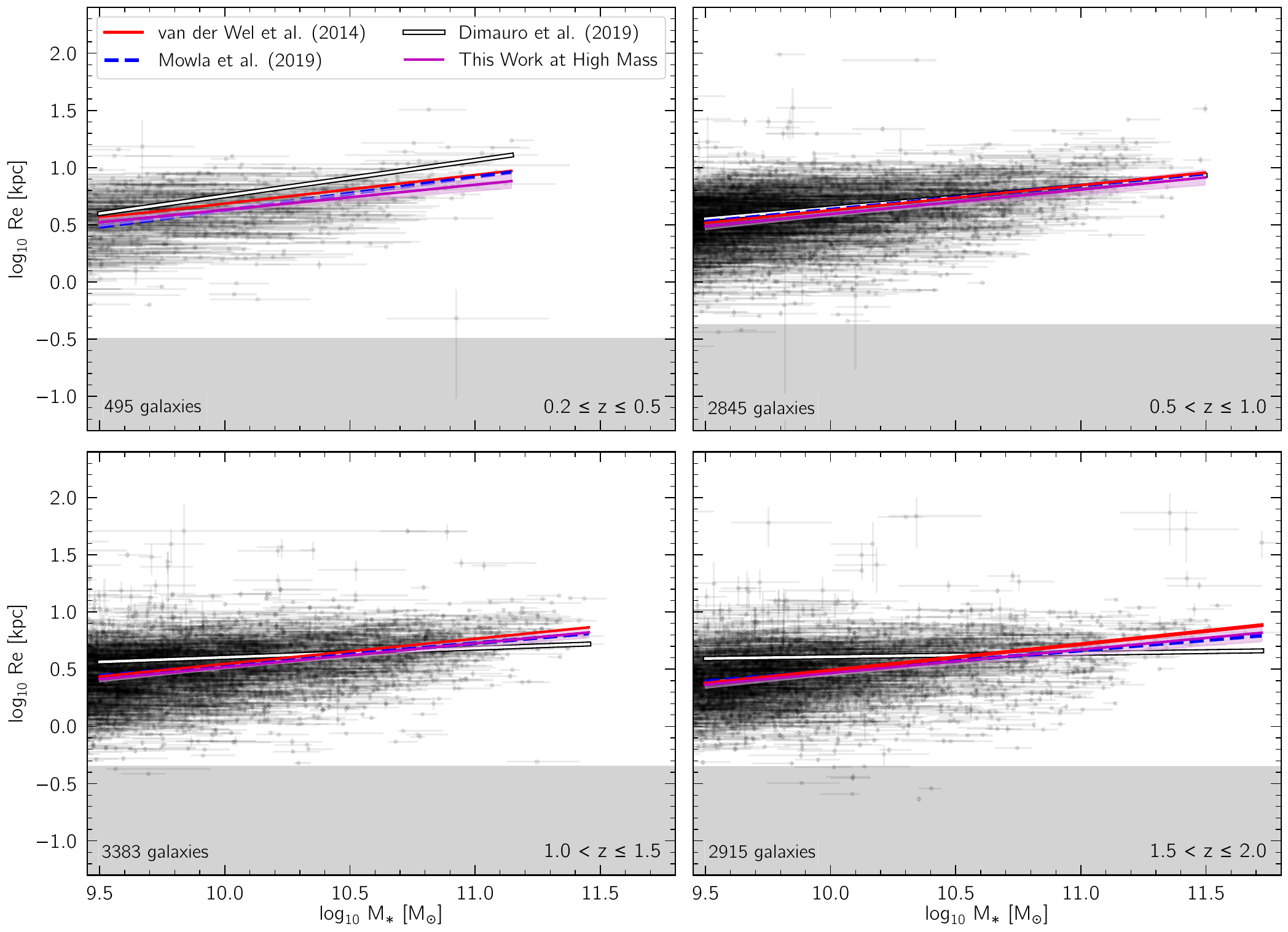}
    \caption{ Same as Figure \ref{fig:Q_highM_M_Re}, but for the star-forming galaxy sample. The best-fit line to all star-forming galaxies with stellar masses above $10^{9.5}$ M$_\odot$ is shown in magenta. Relations from \protect\cite{vdWel2014}, \protect\cite{Mowla2019ApJ}, and \protect\cite{Dimauro2019} are shown as indicated for comparison. As in previous figures, the grey areas indicate sizes at R$_{\rm e}<{\rm FWHM_{\rm F160W}}$/2 at the maximum redshift of each panel. See the main text for discussion and comparison to other works.}
    \label{fig:SF_highM_M_Re}
\end{figure*}

 In an effort to derive consistent and fair comparisons to previous studies, we first fit high mass star-forming galaxies only. We model all star-forming galaxies with stellar masses above $10^{9.5}$ M$_{\odot}$ to cover the same mass range as \cite{vdWel2014} and \cite{Dimauro2019}. The stellar mass – size relation of these high mass star-forming galaxies is shown in Figure \ref{fig:SF_highM_M_Re}. The best-fit power-law, using Equation \ref{eq:singlepl}, is shown as a magenta line with a shaded magenta region indicating the 1$\sigma$ level that includes 68\% of all MCMC models. These models are obtained in the same way as the high mass quiescent best-fit parameters.

 At the high mass end, our results are consistent with \cite{vdWel2014}, shown as a red line in Figure \ref{fig:SF_highM_M_Re}. We also find very good agreement with \cite{Mowla2019ApJ}, who extended the CANDELS data presented by \cite{vdWel2014} to include very massive galaxies with stellar masses above $2 \times 10^{11}$ M$_{\odot}$ from the COSMOS-DASH program \citep{dash_momcheva}. %\bh{Does that not mean that the blue lines should be longer in the plot??}. \kn{their lines are shown to 11.5Msun}
 This is perhaps not particularly surprising as we already find consistent results with \cite{vdWel2014} at the high mass end. However, our results for high mass star-forming galaxies differ significantly from \cite{Dimauro2019}, especially at high redshift. At $z \geq 1$, \cite{Dimauro2019} find a very shallow slope of $\beta \approx 0$ whereas we find that the slope remains roughly equal to $0.2$ across the entire redshift range considered in this work (see also Fig. \ref{fig:SF_evolution}). This difference is surprising given that \cite{Dimauro2019} use the same method for deriving rest-frame sizes from CANDELS data (i.e., they use the software, \textsc{GalfitM} and \textsc{Galapagos-2}, and  obtain the rest-frame 5000\AA \ size from the Chebyshev polynomial). Likewise, they also derive their stellar masses using FAST \citep{Kriek2009} and divide their star-forming and quiescent populations using the UVJ diagram. 
 
In order to understand this difference, we compare the sizes measured in the F160W band for galaxies that are included in this study and in \cite{Dimauro2019}. % \bh{(no plot shown)}. Will include in appendix
  We choose to compare the measured F160W sizes as opposed to the rest-frame size in order to avoid any differences caused by different redshift estimates.
  We find excellent size agreement with no systematic offsets, suggesting that the MegaMorph tools are returning consistent results and are therefore not the root of this discrepancy. \cite{Dimauro2019} use redshift estimates from \cite{Dahlen2013ApJ} whereas we use redshift from the 3D-HST catalogues \citep{Skelton2014, Momcheva2016ApJS}, which have the added benefit of including redshifts derived from grism spectra.
  While there is scatter 
  %\bh{how large?} 
  between the redshifts that we use and those used by \cite{Dimauro2019}, we find that the redshifts are generally consistent, again, with no systematic differences.

  It is important to also note that we take a simpler approach to modelling the trend exhibited by the star-forming galaxies compared to \cite{vdWel2014} and \cite{Dimauro2019}. 
  These previous studies use a model that takes into account possible mis-classifications of the star-forming and quiescent galaxies using the UVJ diagram. Although this is an important issue to consider, we recover similar results as \cite{vdWel2014} using our simpler modelling method.  This indicates that fitting the sequence with an MCMC approach, with fewer free parameters, can perform just as well and might be preferred due to its simplicity. As the galaxy sample that \cite{Dimauro2019} use is publicly available \citep{Dimauro2018MNRAS}, we fit their star-forming sample with our MCMC approach. In doing this, we also obtain a slope that becomes shallower with redshift, reaching $\beta = 0.06$ at $1.5 < z < 2.0$. Given that we recover a similarly shallow slope at high redshift, we attribute the different results to a different sample selection rather than the modelling used. We discuss these differences in more detail in Appendix \ref{Paola_comp}. %\bh{Excellently split up! What needs to be said is there, the rest in the Appendix!}

\subsubsection{Entire Mass Range } \label{SF_AllMass}
 
\begin{figure*}
    \centering
    \includegraphics[width=1.0\textwidth]{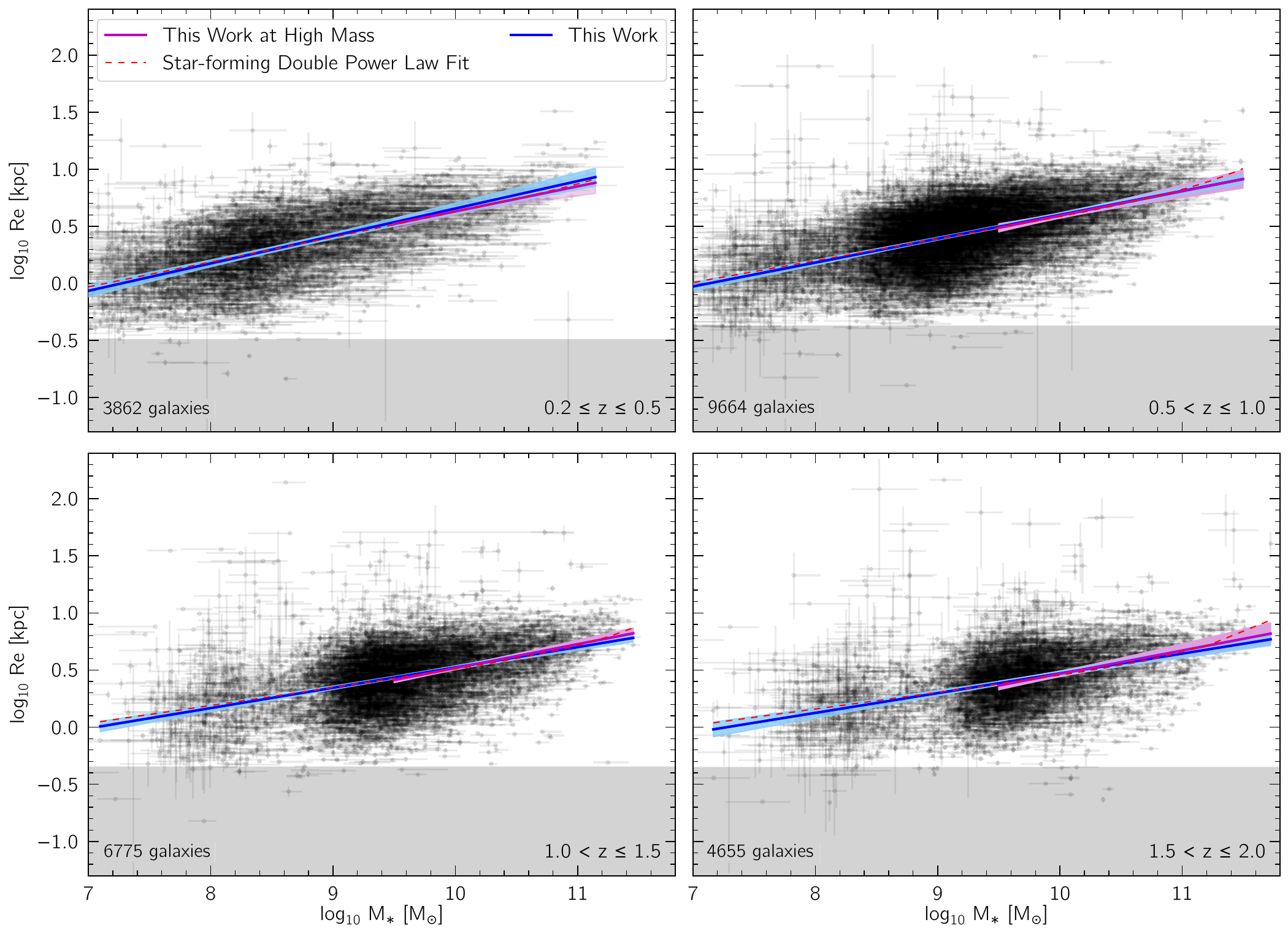}
    \caption{Same as Figure \ref{fig:Q_M_Re}, but for the star-forming galaxy sample. The magenta and blue lines with the corresponding 1$\sigma$ levels show the models for the high mass star-forming galaxies (i.e., M$_* \geq 10^{9.5}$ M$_\odot$) and the entire stellar mass range  (i.e., M$_* \geq 10^{7}$ M$_\odot$), respectively.  The red dashed line is the best-fit double power law model, which is derived in a similar way to the double power law model for the quiescent galaxies (see text for details). The star-forming double power law fit is in excellent agreement with the all-mass fit at the low mass end and with the high mass fit at the high mass end. The blue and magenta lines from the 4-parameter and linear models are also in good agreement -- they fall within 1$\sigma$ of each other at all redshifts. Therefore, the double power law model is over-fitting the data and is unnecessary, but shown here as a means to indicate that a single power law fit is a good representation of the data. As in previous figures, the grey areas indicate sizes at R$_{\rm e}< {\rm FWHM_{\rm F160W}}$/2 at the maximum redshift of each panel. }
    \label{fig:SF_M_Re}
\end{figure*}

 Star-forming galaxies appear to follow a single power law over the full stellar mass range explored in this work. We therefore fit the star-forming galaxy population in all four redshift bins according to Equation \ref{eq:singlepl}, in addition to fitting it with a double power-law function, shown in Equation \ref{eq:doublepl}, as we did for the quiescent sample. The single power law best-fit models are shown in Figure \ref{fig:SF_M_Re} as a blue lines, with the 1$\sigma$ limits shown in light blue. We additionally show the star-forming high mass fits and the corresponding 1$\sigma$ limits from Figure \ref{fig:SF_highM_M_Re} in magenta. As can be seen from Figure \ref{fig:SF_M_Re}, the blue and magenta fits either overlap entirely or are consistent within 1$\sigma$, indicating that the high mass fit is consistent with the fit over the entire stellar mass range studied in this work, across $0.2 \leq z \leq 2.0$. This also shows that the fit over our entire mass range well represents the high mass end, giving us confidence in such simple result over such a large mass range.
 
 We fit the star-forming sample in the same way as the quiescent sample, with one notable exception -- we do not constrain the parameter $\delta$. By not constraining it, $\delta$ can occur at stellar masses much greater than the stellar mass range that is available to us, ultimately, allowing the MCMC model to essentially fit a single power law relation, even though a double power law functional form is assumed. The resulting double power law fits are shown as red dashed lines in Figure \ref{fig:SF_M_Re}. In all redshift bins, the double power law is consistent with the single power law fit, highlighting that a double law fit to the star-forming population is unnecessary. 
 
\subsubsection{Star-Forming Galaxy Evolution} \label{sec:SF_evolution}

\begin{figure*}
    \centering
    \includegraphics[width=1.0\textwidth]{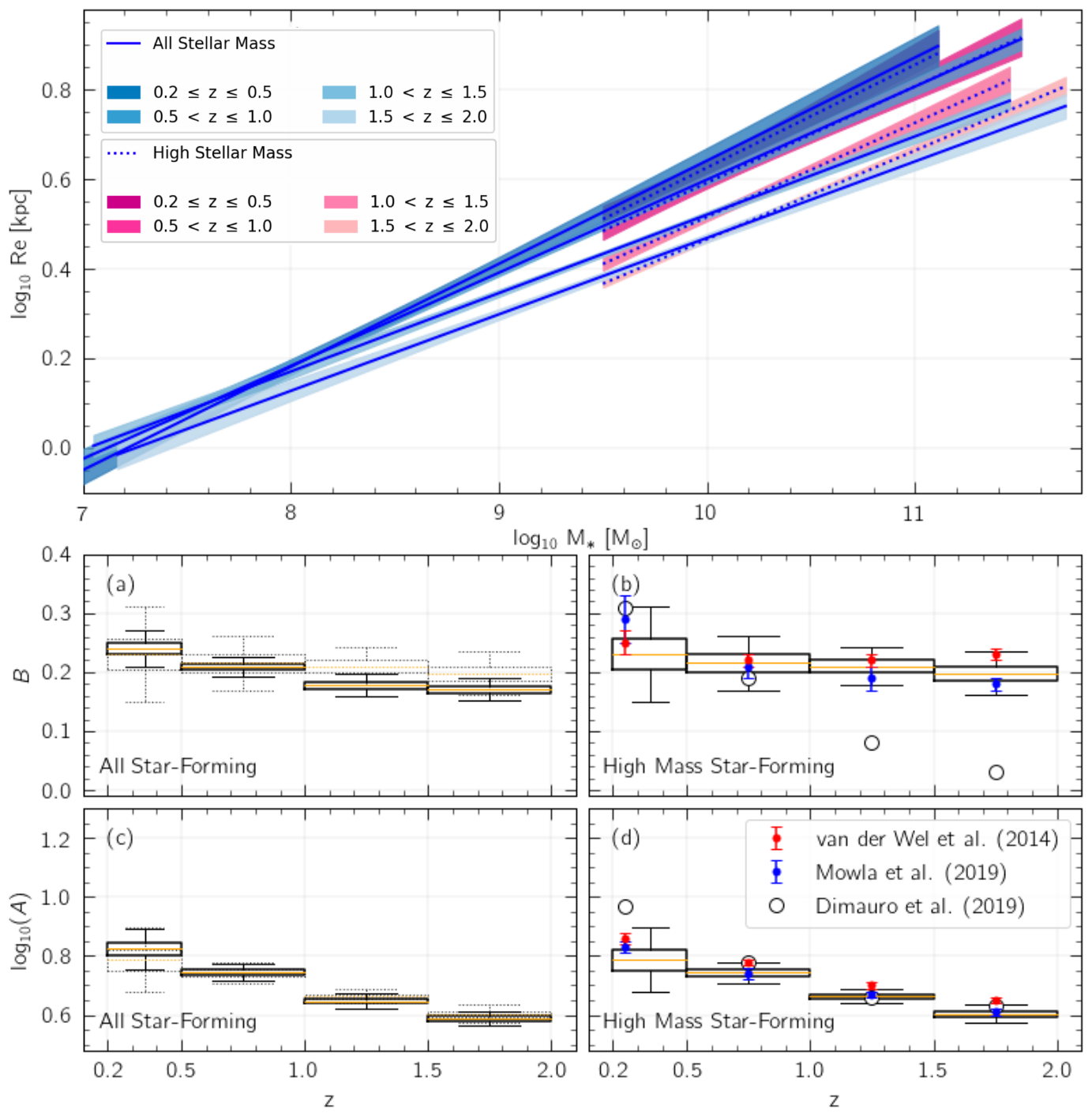}
    \caption{The evolution of the stellar mass – size relation and the best-fit parameters for the star-forming sample. The top panel shows the best-fit model for each redshift bin as a solid blue line while the high mass best-fit models are shown as dashed lines. The 1$\sigma$ levels for each model are shown as blue and magenta regions, where the concentration of the colour indicates the redshift bin. The bottom four panels display the redshift dependence of the parameters $B$, which is the slope of the fit and $\log_{10}(A)$, which is the intercept. Panel (a) shows the evolution of $B$ when the entire mass range is fit, and panel (b) shows the evolution when only the high mass galaxies are modelled. Likewise, panels (c) and (d) show the evolution of $\log_{10}(A)$ over the whole mass range and for the high mass range, respectively. The high mass range contains galaxies with stellar masses $10^{9.5}$ M$_\odot \leq $ M$_{*} \leq 10^{11.5}$ M$_\odot$ in order to provide a direct comparison to \protect\cite{vdWel2014} and \protect\cite{Dimauro2019}. At each redshift, we show the median, the 1$\sigma$ confidence level, and the 2$\sigma$ confidence levels as an orange line, box, and error bars, respectively. Our results are consistent with \protect\citet{vdWel2014} (in red) for both the slope and normalisation, while \protect\cite{Dimauro2019} (in white) find a much steeper evolution in the slope. Our results are also in very good agreement with those from \protect\citet{Mowla2019ApJ} (in blue).}
    \label{fig:SF_evolution}
\end{figure*}

While they are star-forming, galaxies are constantly building up their stellar mass via new star formation. It is now well understood that galaxies grow in size over cosmic time; however, certain physical processes, such as gas inflows to the centre of the galaxy can also cause galaxies to become more compact \citep{Zolotov2015MNRAS, Tacchella2017} . This interplay between structure and star formation can be studied and better understood by constraining the stellar mass -- size relation. 

The average galaxy growth with cosmic time is visible in the redshift evolution of the stellar mass -- size relation above $\sim 10^{9}$ M$_\odot$ shown in the top panel of Figure \ref{fig:SF_evolution}, where the zero point is decreasing with increasing redshift (see panel d in Fig. \ref{fig:SF_evolution}). Interestingly, this trend appears not to hold true for low mass galaxies, as shown in the top panel of Figure \ref{fig:SF_evolution}. In fact, galaxies with $\sim 10^{7} - 10^{8}$ M$_\odot$ appear to remain in the same region of the stellar mass -- size plane from 0.2 $\leq z \leq 2$, suggesting that low mass galaxies are not growing much in size with cosmic time, unlike their more massive counterparts. 
\begingroup
\setlength{\tabcolsep}{6pt} % Default value: 6pt
\renewcommand{\arraystretch}{1} % Default value: 1

\begin{table*}
\caption{The estimated parameters for star-forming galaxies according to the single power law shown in Eq. \ref{eq:singlepl} for each redshift bin. We fit the high mass end only (i.e., M$_{*}$ $\geq$ 10$^{9.5}$ M$_{\odot}$ for the star-forming galaxies) in order to provide direct comparisons to previous studies. The high mass parameters correspond to the magenta lines in Figure \ref{fig:SF_M_Re}. The last three columns show the best-fit model parameters and the RMSE for star-forming galaxies over our entire mass range. We find good agreement between our high mass best-fit parameters and the best-fit parameters derived for our entire mass range, indicating that the models are reliable. In Figure \ref{fig:SF_evolution}, we show the evolution of these parameters with redshift.}
\centering
\begin{tabular}{c c c c c c c}
\hline
\multicolumn{1}{l}{ \ } & 
\multicolumn{3}{c}{High Mass} &
\multicolumn{3}{c}{All Mass} \\ 

% \multicolumn{1}{l}{ \ } & 
% \multicolumn{1}{c}{\underline{(M$_{*}$ $\geq$ 10$^{9.5}$ M$_{\odot}$) }}
% \multicolumn{4}{l}{\underline{(M$_{*}$ $\geq$ 10$^{7}$ M$_{\odot}$) }}\\ 

$z$ & $\log_{10}$($A$) & $B$ & RMSE & $\log_{10}$($A$) & $B$  & RMSE \\
\hline 

0.2$\leq{z}\leq$0.5 & 
$0.78{\pm0.03}$ & $0.22{\pm0.03}$&$0.21$&
$0.82{\pm0.02}$&$0.24{\pm0.01}$ &$0.25$\\

0.5$<$ $z$ $\leq$1.0 &
$0.74{\pm0.02}$&$0.21{\pm0.02}$&$0.23$&
$0.75{\pm0.01}$&$0.21{\pm0.01}$&$0.25$\\

1.0$<$ $z$ $\leq$1.5 & 
$0.66{\pm0.01}$ & $0.21{\pm0.01}$ &$0.23$& 
$0.65{\pm0.01}$ & $0.18{\pm0.01}$&$0.25$\\

1.5$<z\leq$2.0 &
$0.61{\pm0.01}$ & $0.20{\pm0.02}$&$0.24$ & 
$0.59{\pm0.01}$ & $0.17{\pm0.01}$&$0.26$\\

\hline
\end{tabular}
\label{tab:from_eq1}
\end{table*}
\endgroup

Our best-fit models as well as the evolution of our best-fit parameters and their uncertainties estimated by our MCMC analysis are shown in panels a-to-d in Figure \ref{fig:SF_evolution}. The top panel shows the best-fit curve for each redshift bin from Figure \ref{fig:SF_M_Re} shown in different shades of blue as indicated by the legend. Similarly, we show the high mass models (where we limit the star-forming sample to M$_*$ $\geq$ 10$^{9.5}$M$_\odot$) as dotted lines with magenta $1\sigma$ regions. The high mass fit and the fit over all mass are in excellent agreement, where the dotted line falls within $1\sigma$ region of the solid blue line. In panels (a) and (c), we show the slope and intercept as a function of redshift, respectively, when the entire mass range is used. In panels (b) and (d), we show the redshift evolution of the high mass best-fit parameters using the same lower mass limit as \cite{vdWel2014}, \cite{Dimauro2019}, and \cite{Mowla2019ApJ}.  We note that the measurements for our lowest redshift bin span $0.2 \leq z \leq 0.5$, while measurements from previous studies include redshifts down to $z \sim 0$. This difference in the redshift ranges is indicated by the offset mean redshift of the studies we compare to from the mean of our lowest redshift bin at $z = 0.35$ shown in panels (b) and (d). %\kn{not sure if this is clear -- just want to point out that for the 1st zbin, the blue, red \& white dots are centered differently from mine.} \bh{Just say it. Our redshift coverage is 0.2-0.5, the other works used 0 to 0.5, which leads to the offset in average redshift used in Figure ...}
We also show the result from panel (b) and (d) as more transparent box plots in panels (a) and (c), respectively, in an effort to compare the high mass and all mass fits. The slope, $\beta$ becomes slightly steeper when the entire mass range is used, while the intercept, $\log_{10} \alpha$, is consistent for both fits.

Unsurprisingly, the normalisation, or intercept (i.e., the $\log_{10}$ ($A$) parameter) shows a strong evolution with redshift indicating that at fixed mass, galaxies have undergone significant size growth since $z \sim 2$. This evolution is present when we include our entire mass range as well as when we model only the high mass star-forming galaxies, as shown in panels (c) and (d). 
The slope, $B$, also shows some indication of evolution with redshift, although the evolution is not as strong and in fact appears to be absent when only the high mass star-forming sample is considered. 

Interestingly, from panel (b) in Figure \ref{fig:SF_evolution}, we see that \cite{vdWel2014} and \cite{Mowla2019ApJ} derive very different best-fit parameters for the slope of model from \cite{Dimauro2019}. While \cite{Dimauro2019} find a rather steep evolution in the slope, \cite{vdWel2014} and \cite{Mowla2019ApJ} find none. The results from our work are more consistent with \cite{vdWel2014} and \cite{Mowla2019ApJ}, as can be seen from both similarity in the power laws in Figure \ref{fig:SF_M_Re} and from the proximity of the best-fit parameters in Figure \ref{fig:SF_evolution}. 
As mentioned before, we discuss the difference between our own results and the results by \cite{Dimauro2019} in Appendix \ref{Paola_comp} in detail.

% -------------- DISCUSSION -------------- 
\section{Discussion} \label{sec:discussion}
We present in this paper detailed measurements of the
stellar mass--size relation of star-forming and quiescent
galaxies over 0.2 $\leq z \leq 2$. In this section, we provide a discussion based on the results from the previous section, covering topics from possible selection effects in \S \ref{sec:selection_effects} to providing possible interpretations of the results in  \S \ref{sec:interpretations} and comparing these findings to dwarf galaxy studies in 
\S \ref{sec:dwarfs}.

\subsection{Selection Effects} \label{sec:selection_effects}
Before we discuss the implications of our results and their connection to low redshift studies, we first consider possible selection effects. In this work, we have combined galaxies from the cluster and parallel fields imaged as part of the HFF program with galaxies from the CANDELS fields to measure the stellar mass -- size relation and its redshift evolution. As briefly noted in \S \ref{sec:data_hff} and \S \ref{sec:data_candels}, the HFF and CANDELS data have different depths. The HFF data is deeper and therefore allows us to probe the low mass end of the stellar mass – size relation. As a result, the HFF and CANDELS data have different stellar mass distributions, with the HFF comprising the majority of low mass objects, and CANDELS providing the majority of intermediate- and high-mass objects since it samples a larger area. Despite this, we find that the size distributions are consistent with each other and that both the HFF and CANDELS samples are consistent with the best-fit trends derived in \S \ref{sec:M_Re}. Hence, although we have combined data with different depths, we do not expect this difference to affect the stellar mass – size relations that we measure.

We have also used a combination of photometric and spectroscopic redshifts. By design, spectroscopic redshift estimates are much more reliable than photometric redshift estimates, although the typical photometric uncertainties for both HFF and CANDELS are small. \cite{Shipley2018} report a scatter of $\sim$0.02$-0.05$(1+z) for the HFF-DeepSpace catalogues depending on the specific HFF pointing and  \cite{Bezanson2016ApJ...822...30B} report an average scatter $\sim$0.0197$\pm$0.0003(1+z) for the 3D-HST catalogues. In our final sample, we have spectroscopic redshift estimates for 3441 galaxies from CANDELS and 606 galaxies from the HFF. As a result of the redshift cuts that we impose on the HFF cluster fields, discussed in \S \ref{sec:data_hff}, our HFF sample is devoid of galaxies with spectroscopic redshift estimates at $z \geq1$.
We find that using only objects with spectroscopic redshifts does not quantitatively change the results of this study, but using the larger photometric sample yields better statistical significance, specifically at high redshift, where spectroscopic redshift estimates are very rare.

Although we exclude any galaxies that are possibly lensed (see \S \ref{sec:data_hff}), \cite{Yang2021MNRAS} recently presented the stellar mass -- size relation of strongly lensed, high mass galaxies in the HFF cluster fields. After applying a lens reconstruction technique, they find that the strongly lensed galaxies lie on relations that are consistent with those derived by \cite{vdWel2014}. As we also find results that agree well with \cite{vdWel2014}, our results are also consistent with \cite{Yang2021MNRAS} even though we have removed the galaxies that they study from our sample.

Finally, we note that while we have included galaxies from both dense and less-dense environments, only $\sim$ 10\% of the entire sample presented here comes from the HFF cluster fields. While galaxy evolution depends on the environment such that dense, cluster environments are believed to accelerate galaxy evolution \citep[e.g.,][]{Shankar2013MNRAS}, the majority of galaxies presented in this work are not evolving due to cluster-specific processes, because only a small fraction of the total sample comes from dense environments. In fact, the effect of environment on the observed stellar mass -- size relation is still debated. Several works have shown that galaxy sizes and masses exhibit weak, if any, environmental dependence \citep[e.g.,][]{Weinmann2009MNRAS,Maltby2010MNRAS,Marc2013ApJ, Shankar2014MNRAS, Sweet2017MNRAS}. On the other hand, \cite{Kuchner2017A&A} report larger sizes for early-type galaxies and smaller sizes for massive late-type galaxies in clusters in comparison to the field at $z\sim 0.44$, consistent with higher redshift studies \citep[e.g.][]{Papovich2012ApJ, Delaye2014MNRAS}. Given the low percentage of cluster galaxies that we use in this study, and that the effect of the environment is subtle, we do not expected the different environments to significantly influence the stellar mass -- size relations that we recover in this work.

\subsection{Interpretation of the star-forming and quiescent stellar mass--size relations} \label{sec:interpretations}
We show the stellar mass -- size relation for all galaxies in our sample in Figure \ref{fig:MASS-SIZE} as well as for quiescent galaxies only in Figure \ref{fig:Q_M_Re} and star-forming galaxies in Figure \ref{fig:SF_M_Re}. Our results confirm previous findings that the star-forming and quiescent galaxy populations lie on distinct regions on the stellar mass--size plane \citep[e.g.,][]{vdWel2014,vanDokkum2015ApJ, Morishita2017, Dimauro2019}. While the mass -- size relation of star-forming galaxies is well represented by a single power-law  across 10$^{7}$ M$_{\odot}$ to 10$^{11.5}$ M$_{\odot}$, quiescent galaxies show a clear flattening at the low mass end. By taking the median size of galaxies with stellar masses below $10^{9.5}$ M$_\odot$, we clearly observe that the flattening occurs at around 1 kpc out to $z\sim1$, with some indication that this flattening exists out to $z=2$ although the number of low mass quenched galaxies is limited above $z\sim1$.

In the local Universe, it has been known that a flattening occurs in the stellar mass -- size relation of quiescent galaxies for some time \citep[e.g.,][]{Shen2003, Misgeld2011MNRAS, Lange2015MNRAS, Eigenthaler2018ApJ}. At higher redshifts, \cite{Huang2017ApJ} have shown that if the half-light radius scales linearly with the dark-matter radius, as is expected from galaxy formation models, then the stellar mass -- size relation of quiescent galaxies should be nonlinear, with a bend at a few $\times 10^{10}$ M$_\odot$ out to at least $z\sim 3$. In contrast, for star-forming galaxies, they obtain a linear relation on the size--mass plane. Our results for both star-forming and quiescent galaxies agree very well with the findings and predictions of \cite{Huang2017ApJ}. It is hence likely that the flattening in the stellar mass -- size relation of quiescent galaxies that we observe is a result of the baryonic radius scaling linearly with the dark-matter radius.

\begin{figure*}
    \centering
    \includegraphics[width=1.0\textwidth]{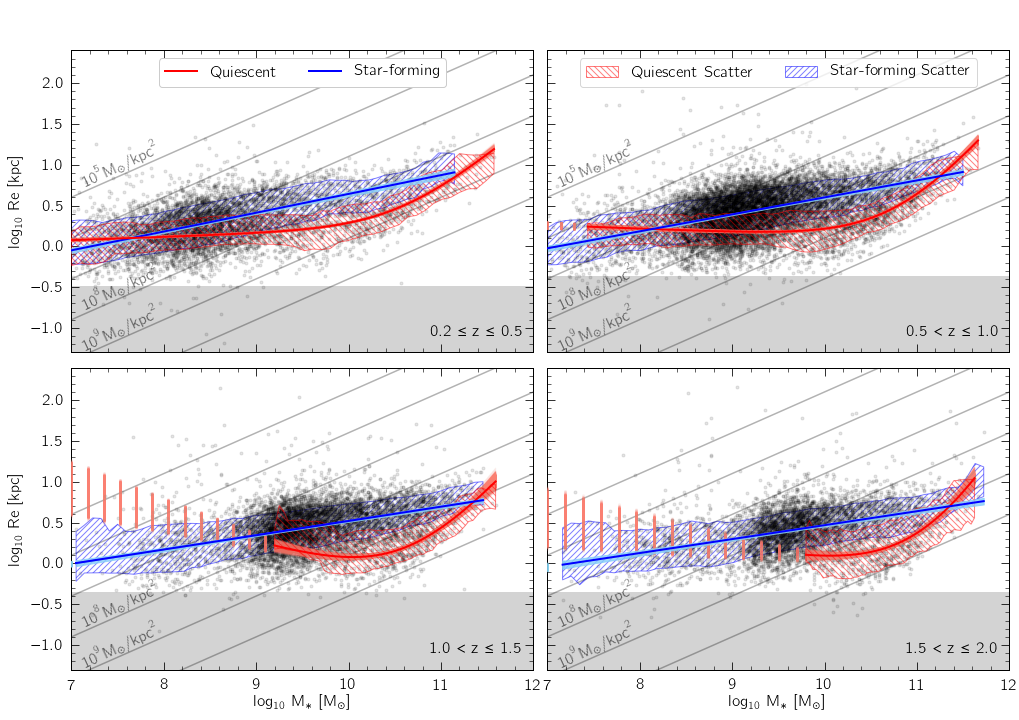}
    \caption{Comparison of the star-forming and quiescent best-fit curves from Figures \ref{fig:SF_M_Re} and \ref{fig:Q_M_Re}, respectively. We show the entire sample (star-forming and quiescent) that falls into each redshift bin as black points. All models are shown as solid lines with a shaded region representing the 1$\sigma$ confidence level over the stellar mass range over which the models are robust. The red and blue dashed regions indicate the scatter of the quiescent and star-forming galaxies, respectively. The stellar mass range over which we calculate the scatter is the same as the range over which we show our models. The grey indicate constant effective mass surface density ($\Sigma$). All models have also been extrapolated to 10$^7$M$_\odot$, where the extrapolation is shown as vertical lines indicating the 1$\sigma$ spread of the model at low stellar masses.  %\bh{Much of the following is again text for the main text, not caption.} Although the star-forming and quiescent galaxy models differ for stellar masses below $\sim$ $10^{8.5}$ M$_\odot$, in all redshift bins, the scatter is larger than the difference of the models. This indicates that the star-forming and quiescent sequences occupy the same region of the stellar mass -- size plane at low stellar mass. 
    As in previous figures, the grey areas indicate sizes at R$_{\rm e}< {\rm FWHM_{\rm F160W}}$/2 at the maximum redshift of each panel.}
    \label{fig:comp}
\end{figure*}

In Figure \ref{fig:comp}, we directly compare the models of the quiescent galaxy relation to the models of the star-forming stellar mass -- size relation. In all four redshift bins, the star-forming models are shown in blue, while the models for the quiescent galaxies are in red. The corresponding blue and red shaded regions indicate the 1$\sigma$ level, which include 68\% of all MCMC models. Even with the 1$\sigma$ levels, we find that the quiescent and star-forming galaxies populate distinct stellar mass--size relations. For each redshift bin, we also show the scatter around each model as a dashed region. 
These dashed regions show the running upper and lower 68$^{\mathrm{th}}$ percentiles. Up to this point, we have not addressed the scatter around our best-fit models for either the star-forming nor the quiescent samples. From Figures \ref{fig:Q_M_Re} and \ref{fig:SF_M_Re}, it can be seen that there is indeed a significant amount of scatter in galaxy sizes at a given stellar mass \citep{Ruhland2009ApJ695,Somerville2018}. The observed scatter in galaxy size could be due to radial stellar migration. For instance, \cite{El-Badry2016ApJ} have shown that processes such as stellar feedback can cause galaxy size to fluctuate significantly in just a few hundred million years. In Figure \ref{fig:comp}, we quantify the amount of scatter around the best-fit trends and compare this scatter to the difference in the star-forming and quiescent best-fit models. We additionally show lines of constant effective mass surface density ($\Sigma = $ M$_* / (2\pi$ Re$^2$)) following \cite[e.g.,][]{Misgeld2011MNRAS,Eigenthaler2018ApJ}. These lines indicate the median mass density within the half-light radius, such that mass assembly along these lines implies density-invariant growth. Any galaxies which lie on a stellar mass--size relation which is parallel to these lines are building up mass both inside and outside their half-light radius.

For $z > 1$, we do not place much confidence in the quiescent best-fit models at low stellar masses, as we do not have many low mass quiescent galaxies at these redshifts to constrain the model. The best-fit model is extended to $10^{9.2}$M$_\odot$ for the $1.0 < z \leq 1.5$ bin and to $10^{9.8}$M$_\odot$ for  $1.5 < z \leq 2$, as shown in Table \ref{tab:from_eq2}. We have included the extrapolated model below these stellar masses down to $10^{7}$M$_\odot$, where the width of the red extrapolated region shows the 1$\sigma$ confidence level. We have done this to highlight that the high redshift models are  ill-constrained at low stellar masses.

For the first two redshift bins, shown at the of top Figure \ref{fig:comp}, below $\sim 10^{9}$ M$_\odot$, the star-forming and quiescent models intersect, such that the quiescent galaxy model has a shallower slope than the model for star-forming galaxies. Although the models are not the same at this low mass end, the star-forming and quiescent galaxies occupy the same region of the stellar mass -- size plane. The scatter around the models is larger 
than the difference between the models, indicating that star-forming and quiescent galaxies could be evolving through the same processes in this stellar mass regime. In the $\sim 10^{9} - 10^{10.5}$ M$_\odot$ stellar mass range, the star-forming and quiescent galaxies occupy very different regions of the mass -- size plane, suggesting that galaxies in this mass regime are building up their mass via different mechanisms depending on whether they are actively forming stars or not. Finally, massive galaxies, with M$_* \geq 10^{10.5}$ M$_\odot$, also follow different stellar mass -- size relations depending on whether they are quiescent or star-forming. Massive quiescent galaxies follow a steep stellar mass -- size relation, indicating that on average, quiescent galaxies are rapidly growing in size, most likely due to minor mergers, while massive star-forming galaxies appear to lie on the same linear relation as their less massive counterparts. 
This suggests that star-forming galaxies across $\sim 10^{7} - 10^{11.5}$ M$_\odot$ are primarily building up their mass and size by forming new stars, while quiescent galaxies evolve through different mechanisms depending on their stellar mass. 

Indeed, comparing the stellar mass -- size relations to the lines of constant $\Sigma$ reveals a few interesting features that support this argument. 
For $0.2\leq z \leq 2$, the stellar mass -- size relation of star-forming galaxies is shallower than the lines of constant $\Sigma$. This implies that star-formation allows galaxies to assemble mass predominantly inside their effective radius. Quiescent galaxies with low to intermediate stellar masses also follow stellar mass -- size relations that are shallower than the lines of constant $\Sigma$, indicating that they too are primarily assembling their mass inside the half-light radius. Massive quiescent galaxies instead lie on relations that are steeper, suggesting that they are assembling mass in their outskirts. Indeed, it is known that these massive systems primary grow via minor mergers \citep[e.g.,][]{Buitrago2008ApJ, Naab2009ApJ, Oser2012ApJ}, which cause galaxies to assemble mass in the outer regions.

To gain more insight into the underlying processes that drive these different phases of growth, in a future paper we will present the decomposition into bulge+disk components for the galaxies in the sample presented in this work to understand how different galaxy components build up their mass (Nedkova et al., in prep).

\subsection{Link to Dwarf Galaxies} \label{sec:dwarfs}

\begin{figure}
    \centering
    \includegraphics[width=0.47\textwidth]{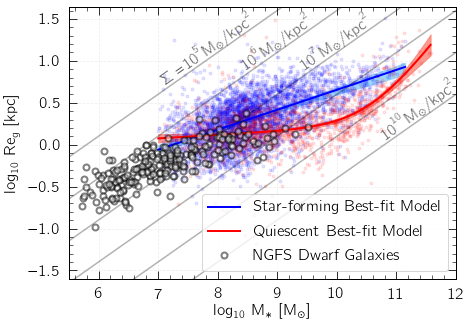}
    \caption{The stellar mass -- size relation of the star-forming (shown in blue) and quiescent (shown in red) galaxies from the lowest redshift bin, i.e., $0.2 \leq z \leq 0.5$, where the half light radius has been measured in the rest-frame g-band instead of at 5000\AA. The stellar mass -- size relation of dwarf galaxies from the  Fornax galaxy cluster presented in \protect\cite{Eigenthaler2018ApJ} are shown as empty black circles. Lines of constant effective mass surface density ($\Sigma$) are drawn in grey and labelled.}
    \label{fig:dwarfs}
\end{figure}

It is interesting to consider why the flattening of the quiescent sequence appears to occur at an effective radius of $\sim$1 kpc and whether this flattening continues to exist when the stellar mass -- size relation is extended to stellar masses below $10^{7}$ M$_\odot$, which we can not do with our data set. In order to study this low mass end of the stellar mass -- size relation, dwarf galaxies need to be considered. Many previous works have studied the connection between galaxies and dwarf galaxies. For instance, \cite{Forbes2008MNRAS} have shown that elliptical galaxies and the dwarf elliptical galaxies from \cite{Binggeli1998A&A} follow a continuous trend on the size--luminosity plane which flattens to a constant size of $\sim$1 kpc. More recently, \cite{Eigenthaler2018ApJ} and \cite{Yasna2018ApJ} have shown that there is a similar flattening for dwarf and elliptical galaxies in the Fornax cluster in the stellar mass -- size and size -- luminosity relations, respectively.  
\cite{Eigenthaler2018ApJ} investigate the properties of the low surface-brightness dwarf galaxy population in the core region ($\lesssim$ r$_{\mathrm{vir}}$/4) of the Fornax galaxy cluster, using data obtained as part of the Next Generation Fornax Survey (NGFS). In Figure \ref{fig:dwarfs}, we overplot their results onto the stellar mass -- size relation derived in this work. 
We note here that the effective radius is measured in the rest-frame g-band wavelength as opposed to the 5000\AA \ size. 
While this does not make a significant difference, as discussed in \S \ref{size_est}, we derive the stellar mass -- size relation on the rest-frame g-band for consistency with \cite{Eigenthaler2018ApJ}. Unfortunately, not enough information is available for the NGFS dwarf galaxies to reliably separate them into quiescent and star-forming galaxies. As this sample of dwarf galaxies is from the inner regions of the Fornax cluster, the dwarf galaxy morphology-density relation \citep[][]{Cote2009AJ} would suggest that most are dwarf ellipticals (DEs), which are usually quiescent. Although the majority of the  NGFS dwarf galaxies from \cite{Eigenthaler2018ApJ} are likely quiescent \citep[e.g.,][]{Venhola2019A&A, Carlsten2021arXiv}, we compare them to both of our star-forming and quiescent mass — size relations as we cannot properly separate them using the same criteria used to identify quiescent galaxies from HFF and CANDELS.

Comparing the dwarf galaxies from \cite{Eigenthaler2018ApJ}, it is interesting to see that they seem to align with \textit{both} of our sequences. This could be coincidental, or contain interesting physics. On one hand, the dwarf galaxies seem to turn off at intermediate masses, following the same $\sim$1 kpc size that we find for the flattening of the quiescent galaxies, which might indicate a causal connection between these dwarfs and the quiescent galaxies we find. On the other hand, at lower masses, they follow closely (but not perfectly), our relation of the star-forming galaxies, indicating that they could be just a continuation of the mass-size relation of star-forming galaxies that we already find over four orders of magnitude in mass, to lower masses. As the dwarf galaxy sample is most likely a mixture of quiescent and star-forming objects, both might be true. But without pushing our analysis to lower masses and larger samples on better data, we can not distinguish the different processes just yet.

\section{Summary} \label{sec:conclusion} 
By using a multi-wavelength modelling approach as well as the depth of the HFF survey and the large sample size of CANDELS, we have extended the evolution of the stellar mass -- size relation over $0.2\leq z \leq2$ to low stellar mass galaxies and high redshift. We measure the stellar mass and morphological parameters, including size and S{\'e}rsic index, in a consistent way across both HFF and CANDELS samples.
From these data, we find that the star-forming and quiescent galaxies follow entirely separate sequences on the stellar mass -- size plane regardless of whether star-forming and quiescent galaxies are separated based on star formation activity, colour, or S{\'e}rsic index. 
This result is best illustrated in Figure \ref{fig:MASS-SIZE}, where the quiescent sequence can be seen to flatten at lower stellar masses across all three selections.
We observationally confirm that this flattening at around 1 kpc is present out to at least $z \sim 1.0$, with some signatures out to $z \sim 2$.
These results confirm the findings of previous studies, but represent an important advance, since we have extended the stellar mass range over which we are measuring the stellar mass -- size relation. Our main findings are as follows: 

\begin{itemize}
    \item The stellar mass -- size relation of star-forming galaxies is well described by a single power law to at least $z=2$ even when low stellar mass galaxies (M$_* \geq 10^7$ M$_\odot$) are included. Star-forming galaxies have a slope of $\beta \sim$ 0.2 on the mass -- size plane, across all redshifts studied in this work. While the slope does not show a strong evolution with redshift, the zero-point of the models increases with cosmic time, indicating that galaxies of the same mass were more compact at high redshift than their present-day counterparts. We note however, that low mass galaxies with $\sim 10^{7} - 10^8$ M$_\odot$ occupy the same region of the stellar mass -- size relation regardless of redshift, as shown in Figure \ref{fig:SF_evolution}. This suggests that these low mass star-forming galaxies are not significantly growing in size with cosmic time. 
    
    \item The stellar mass -- size relation of quiescent galaxies, on the other hand, is well described by a double power law at stellar masses $> 10^7$M$_\odot$ due a flattening in the stellar mass -- size relation that occurs at sizes of $\sim$1 kpc.  Dwarf galaxy studies \citep[e.g.,][]{Eigenthaler2018ApJ, Yasna2018ApJ}, however,  suggest that this functional form might not hold at lower masses. High mass quiescent galaxies can be represented by a single power law, with a slope of $\sim 0.7$ over the redshift covered in this work. Our results at the high mass end are consistent with previous works as shown in Figure \ref{fig:Q_highM_M_Re}.

    \item We note that although the star-forming and quiescent galaxies models at the low mass end for the star-forming and quiescent sequences are different, the scatter around these relations is large at stellar masses of $10^7 - 10^{8.5}$ M$_{\odot}$.
%    the width of the relations are more significant in this regime. 
    Although this increased scatter could be, and probably largely is, measurement errors on faint objects, this result hints at quiescent and star-forming galaxies evolving via the same mechanisms, or at least via mechanisms that influence galaxy size and mass growth in the same way.
 
    \item Finally, we show that, while the trends that we have recovered represent the data above 10$^{7}$ M$_{\odot}$ well, extending the stellar mass -- size relation to include dwarf galaxies can give a more comprehensive picture of galaxy formation and raises additional, interesting questions that deserve further investigations.

\end{itemize}

This study presents the stellar mass -- size relation and its evolution down to $10^{7}$ M$_\odot$, but plenty of scope remains for future work. Exploring the stellar mass -- relation down to even lower masses could help us better understand the connection between dwarf galaxies and giant galaxies. Additionally, while understanding how environment impacts galaxy evolution is outside the scope of this paper, the HFF presents a great data set to study this effect. %\bh{comment: Maltby has done this on STAGES data, but the effect is, if present, very small}. 
Finally, with current facilities, observing low mass galaxies, particularly quiescent ones, at high redshift is challenging. Future studies with the James Webb Space Telescope, the Nancy Grace Roman Space Telescope, and Euclid present exciting possibilities for better understanding the stellar mass -- size relation at high redshift.

%%%%%%%%%%%%%%%%%%%%%%%%%%%%%%%%%%%%%%%%%%%%%%%%%%
\section{Acknowledgements} 

We thank the anonymous referee whose comments and suggestions have significantly improved the quality of this paper. KVN gratefully acknowledges support from the Kathryn A. McCarthy Graduate Fellowship in Physics at Tufts University. KVN also thanks ESO, Chile for a seven-month scientific visit, which was instrumental in making this work possible. KVN and DM acknowledge the very generous support of the National Science Foundation under Grant Number 1513473 and by HST-AR-14302 and HST-AR-14553, provided by NASA through a grant from the Space Telescope Science Institute, which is operated by the Association of Universities for Research in Astronomy, Incorporated, under NASA contract NAS5-26555. KEW wishes to acknowledge funding from the Alfred P. Sloan Foundation. The Cosmic Dawn Center is funded by the Danish National Research Foundation under grant No. 140.
%%%%%%%%%%%%%%%%%%%%%%%%%%%%%%%%%%%%%%%%%%%%%%%%%%
\section*{Data Availability}
The 3D-HST, HFF-DeepSpace, and NGFS data that support the findings of this study are openly available. Data directly related to this publication and its figures are available on request from the corresponding author.

%%%%%%%%%%%%%%%%%%%% REFERENCES %%%%%%%%%%%%%%%%%%
% The best way to enter references is to use BibTeX:

\bibliographystyle{mnras}
\bibliography{my} % if your bibtex file is called example.bib

%%%%%%%%%%%%%%%%% APPENDICES %%%%%%%%%%%%%%%%%%%%%

\appendix
\section{HFF Structural Catalogues}\label{HFF_catalog}
Along with this work, we publish catalogues of structural properties for HFF galaxies that have been measured with \textsc{Galapagos-2} and \textsc{GalfitM} \citep{MegaMorph}. We provide a total of 12 catalogues -- one for each cluster and parallel field. Although we use a combination of HFF and CANDELS galaxies for this study, we make available only HFF catalogues, as a catalogue of structural properties for CANDELS galaxies will be released in a forthcoming publication (H\"au\ss ler et al., in prep.).

\begingroup
\setlength{\tabcolsep}{4pt} % Default value: 6pt
\renewcommand{\arraystretch}{1} % Default value: 1

\begin{table*}
\centering
\caption{Columns from the released HFF structural catalogues. id$_\mathrm{HFF}$ corresponds to the id numbers from the HFF-DeepSpace catalogue \citep{Shipley2018}, while the remainder of the parameters are derived with the \textsc{GalfitM} and \textsc{Galapagos} codes. All columns for which the description begins with `\textbf{[7]}' are seven-element arrays. We provide all Re and Re error estimates in pixels as we apply size quality cuts based on the measurements in pixels. Hence, a pixel scale of 0.06$''/$pixel is needed to convert the measured Re into units of arcseconds. Position angles are defined such that 0 is `up', increasing anticlockwise. Most non-integer parameter values are truncated for the example galaxy and shown here only for guidance regarding the format.}
\begin{tabular}{ l l l}
\hline
Column & Description & Example Galaxy \\
\hline

id$_{\mathrm{HFF}}$ &  HFF-DeepSpace catalogue id & 657 \\
RA & \textsc{SExtractor} ALPHA\_J2000 & 342.3186149 \\
DEC & \textsc{SExtractor} DELTA\_J2000 & $-$44.5636569 \\
CXX\_IMAGE & \textsc{SExtractor} CXX\_IMAGE & 0.030213457 \\
CYY\_IMAGE & \textsc{SExtractor} CYY\_IMAGE & 0.11420833 \\
CXY\_IMAGE & \textsc{SExtractor} CXY\_IMAGE & 0.037952058 \\
THETA\_IMAGE & \textsc{SExtractor} THETA\_IMAGE  & $-$12.16 \\
ELLIPTICITY & \textsc{SExtractor} ELLIPTICITY & 0.53 \\
KRON\_RADIUS & \textsc{SExtractor} KRON\_RADIUS & 3.5 \\
BACKGROUND & \textsc{SExtractor} BACKGROUND & 9.684838E-5 \\
FLUX\_BEST & \textsc{SExtractor} FLUX\_BEST & 3.850794 \\
FLUXERR\_BEST & \textsc{SExtractor} FLUXERR\_BEST & 1.000767 \\
MAG\_BEST & \textsc{SExtractor} MAG\_BEST & 24.4761 \\
MAGERR\_BEST & \textsc{SExtractor} MAGERR\_BEST & 0.2822 \\
FWHM\_IMAGE & \textsc{SExtractor} FWHM\_IMAGE & 10.05 \\
FLAGS & \textsc{SExtractor} FLAGS & 3 \\
CLASS\_STAR & \textsc{SExtractor} CLASS\_STAR  & 0.029 \\
SKY\_GALA\_BAND &  \textbf{[7]} \textsc{Galapagos-2} sky values at each band & \{$-$1.77E-4, $-$7.49E-5, $-$2.19E-4, $-$1.47E-4, $-$1.19E-4, \\
 & &  $-$2.31E-4, $-$1.86E-4\}\\
SKY\_SIG\_BAND &\textbf{[7]} \textsc{Galapagos-2} sky values uncertainty. The sky value is fixed during & \{2.07E-5, 1.73E-5, 2.86E-5, 7.12E-6, 1.56E-5,  \\ 
& the fit, but these values can be potentially used to detect `difficult’ fits & 2.33E-5, 1.36E-5\}\\
SKY\_FLAG\_BAND &\textbf{[7]} \textsc{Galapagos-2} sky flag & \{0.0, 0.0, 0.0, 0.0, 0.0, 0.0, 0.0\} \\
FLAG\_GALFIT & Single-Sérsic fit flag: $-1=$ not enough bands with data; not attempted, & 2\\
 &  $0=$ not attempted, $1=$ fits started, but crashed, $2=$ fits completed  & \\
NEIGH\_GALFIT & Number of neighbouring profiles fit (or fixed) during the modelling & 7 \\
CHISQ\_GALFIT & \textsc{GalfitM} $\chi^2$ value & 66011.9 \\
CHISQ\_GALFIT\_PRIME & \textsc{GalfitM} $\chi^2$ value within the primary ellipse& 30272.7 \\
NDOF\_GALFIT & Degrees of freedom (DoF) allowed during the fit, includes  & 58203\\
& number of pixels. Used to derive $\chi^2$ and $\chi^2/\nu$ & \\
NDOF\_GALFIT\_PRIME  & same as NDOF\_GALFIT but for the fit within the primary ellipse & 18527 \\
CHI2NU\_GALFIT  &  \textsc{GalfitM} reduced $\chi^2$: $\chi^2/\nu$ & 1.13417 \\
CHI2NU\_GALFIT\_PRIME & same as CHI2NU\_GALFIT but for the fit within the primary ellipse & 1.63398 \\
use\_flag & modelling parameter flag: $0=$ at least one parameter has run into a  & 1 \\
& fitting constraint; $1=$ no parameters have run into constraints& \\
X\_GALFIT\_DEG & DoF of x-position; not allowed to vary with wavelength & 1 \\
X\_GALFIT\_BAND & \textbf{[7]} x-position for F125W, F435W, F606W, F814W, F105W, F140W,  & \{57.064, 57.064, 57.064, 57.064, 57.064, 57.064, 57.064\} \\
& and F160W, in this order & \\
... & & \\
MAG\_GALFIT\_DEG& DoF of apparent magnitude; full freedom has been allowed & 7 \\
MAG\_GALFIT\_BAND & \textbf{[7]} apparent magnitude measured at each band & \{24.35, 26.10, 25.22, 24.68, 24.46, 24.29, 24.26\}\\
MAGERR\_GALFIT\_BAND & \textbf{[7]} apparent magnitude uncertainties at each band & \{0.007, 0.017, 0.008, 0.005, 0.005, 0.006, 0.007\} \\
MAG\_GALFIT\_CHEB & \textbf{[7]} magnitude Chebyshev polynomial coefficients -- as we  & \{24.799, $-$0.784, 0.338, $-$0.134, 0.0450, $-$0.002, $-$0.003\}\\
& allow full freedom, all values are nonzero &  \\
MAGERR\_GALFIT\_CHEB & \textbf{[7]} uncertainties on the magnitude Chebyshev polynomial coefficients& \{0.002, 0.008, 0.005, 0.003, 0.003, 0.003, 0.005\} \\
RE\_GALFIT\_DEG & DoF of effective radius (Re) & 3 \\
RE\_GALFIT\_BAND & \textbf{[7]} Re [in pixels] at each band & \{5.07, 7.01, 6.44, 5.82, 5.31, 4.98, 4.97\} \\
REERR\_GALFIT\_BAND & \textbf{[7]} Re uncertainty [in pixels] at each band & \{0.08, 0.24, 0.14, 0.08, 0.08, 0.08, 0.12\} \\
RE\_GALFIT\_CHEB & \textbf{[7]} Re Chebyshev polynomial coefficients -- as we use a 2nd  & \{5.71, -1.02, 0.28, 0.0, 0.0, 0.0, 0.0\}\\
& order Chebyshev polynomial, the first 3 values are nonzero &  \\
REERR\_GALFIT\_CHEB & \textbf{[7]} uncertainties on the Re Chebyshev polynomial coefficients & \{0.03, 0.13, 0.09, 0.0, 0.0, 0.0, 0.0\}\\
N\_GALFIT\_DEG & DoF of Sérsic index ($n$)  & 3\\
N\_GALFIT\_BAND & \textbf{[7]} $n$  at each band & \{1.37, 0.86, 1.02, 1.18, 1.31, 1.39, 1.38\}\\
NERR\_GALFIT\_BAND &  \textbf{[7]} $n$  uncertainty at each band & \{0.05, 0.10, 0.05, 0.04, 0.05, 0.06, 0.09\} \\
N\_GALFIT\_CHEB & \textbf{[7]} $n$  Chebyshev polynomial coefficients & \{1.20, 0.26, -0.08, 0.0, 0.0, 0.0, 0.0\}\\
NERR\_GALFIT\_CHEB &  \textbf{[7]} uncertainties on the $n$ Chebyshev polynomial coefficients & \{0.02, 0.06, 0.05, 0.0, 0.0, 0.0, 0.0\}\\
Q\_GALFIT\_DEG & DoF of axis ratio; not allowed to vary with wavelength  & 1\\
Q\_GALFIT\_BAND & \textbf{[7]} axis ratio at each band & \{0.144, 0.144, 0.144, 0.144, 0.144, 0.144, 0.144\}\\
... & & \\
% QERR\_GALFIT\_BAND & \textbf{[7]} axis ratio uncertainty in each band & \{0.001, 0.001, 0.001, 0.001, 0.001, 0.001, 0.001\}\\
% Q\_GALFIT\_CHEB & \textbf{[7]} axis ratio Chebyshev polynomial coefficients & \{0.1438, 0.0, 0.0, 0.0, 0.0, 0.0, 0.0\}\\
% QERR\_GALFIT\_CHEB & \textbf{[7]} uncertainties on the axis ratio Chebyshev polynomial coefficients & \{0.0007, 0.0, 0.0, 0.0, 0.0, 0.0, 0.0\}\\
PA\_GALFIT\_DEG & DoF of position angle; not allowed to vary with wavelength  & 1\\
PA\_GALFIT\_BAND & \textbf{[7]} position angle at each band & \{76.21, 76.21, 76.21, 76.21, 76.21, 76.21, 76.21\}\\
... & & \\
% PAERR\_GALFIT\_BAND & \textbf{[7]} position angle uncertainty at each band & \{0.11, 0.11, 0.11, 0.11, 0.11, 0.11, 0.11\} \\
% PA\_GALFIT\_CHEB & \textbf{[7]} position angle Chebyshev polynomial coefficients &\{76.21, 0.0, 0.0, 0.0, 0.0, 0.0, 0.0\} \\
% PAERR\_GALFIT\_CHEB & \textbf{[7]} uncertainties on the position angle Chebyshev polynomial coefficients& \{0.05, 0.0, 0.0, 0.0, 0.0, 0.0, 0.0\}\\

\hline
\end{tabular}

\label{tab:HFFcat}
\end{table*}
\endgroup

In this appendix, we describe the properties of the HFF structural catalogues. In Table \ref{tab:HFFcat}, we list the columns with a description, as well as an example -- namely, galaxy id$_\mathrm{HFF}=657$ from the Abell1063 parallel field that is shown in Figures \ref{fig:Gala_params} and \ref{fig:galfitm}. It is important to note that this structural catalogue is 3D catalogue in the sense that many of the columns (e.g., MAG\_GALFIT\_BAND, MAGERR\_GALFIT\_BAND, etc.) consist of seven values. The parameters that are reported for each band (i.e., those ending in `\_BAND') are reported for the F125W, F435W, F606W, F814W, F105W, F140W, and F160W bands, in this order. These catalogues are available through the VizieR database: \url{http://cdsarc.u-strasbg.fr/viz-bin/qcat?J/MNRAS}, and all HFF data products are publicly available via the HFF-DeepSpace web page: \url{http://cosmos.phy.tufts.edu/~danilo/HFF/Download.html}.

\section{Comparison to Dimauro et al. (2019)}\label{Paola_comp}

 As briefly discussed in \S \ref{SF_HighMass}, we find results for the high mass star-forming galaxy population that do not agree as well as expected with the stellar mass -- size relations reported by \cite{Dimauro2019}. In particular, the slopes derived for the star-forming galaxies are at odds especially at $z>1$. This is particularly puzzling as the same data and largely identical software has been used to measure galaxy sizes. Here, we investigate a number of differences between our sample and the one used by \cite{Dimauro2019} to show how each impacts the derived stellar mass -- size relation. 
 
\begin{figure*}
    \centering
    \includegraphics[width=1\textwidth]{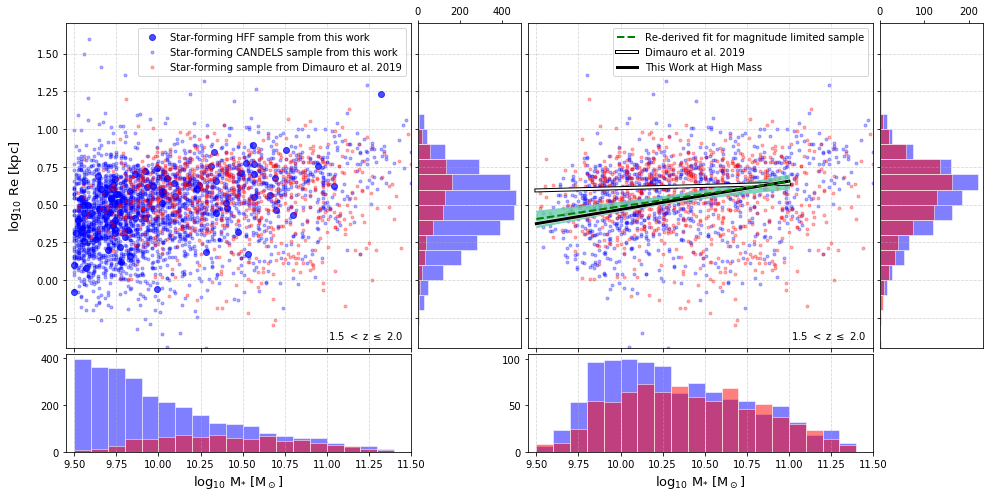}
    \caption{The stellar mass -- size relation of high mass star-forming galaxies from our sample shown in blue, where the size of the points indicates whether they come from the CANDELS fields or the HFF, and the \protect\cite{Dimauro2019} star-forming sample shown in red. In the left panels, we show the sample comparison using our original magnitude cut at 1 magnitude brighter than the 90\% detection completeness limit. In the right panels, we show the sample comparison using the same magnitude cut for both CANDELS samples i.e., m$_\mathrm{F160W}$ $\leq$ 23 following \protect\cite{Dimauro2018MNRAS}. The size distributions of our sample (blue) and the \protect\cite{Dimauro2019} sample (red) are shown as vertical histograms, while the stellar mass distributions are shown as horizontal histograms. The purple regions indicate where the two histograms overlap. In the right panel, we show the best-fit stellar mass -- size relations from this work and \protect\cite{Dimauro2019}, as indicated in the legend. Additionally, we re-derive the relation for our sample with a magnitude cut of m$_\mathrm{F160W}$ $\leq$ 23 and the result is shown in green, where the shaded green region indicates the $1\sigma$ confidence level. }
    \label{fig:dist_comp}
\end{figure*}

We begin by testing the size and mass distributions of the star-forming galaxies used in both works. These comparison are shown in Figure \ref{fig:dist_comp}, where we show only the highest redshift bin as this is where the largest difference in the two samples can be seen; however, the conclusions drawn for this bin also hold across our entire redshift range. From Figure \ref{fig:dist_comp}, it can be seen that at the high mass end (i.e., above stellar masses of $10^{9.5}$ M$_\odot$), our sample predominantly consists of galaxies from the larger CANDELS fields. This is expected, as we use the HFF galaxies mostly to probe the low stellar mass range, where the number of CANDELS galaxies becomes more limited.

From the right panels of Figure \ref{fig:dist_comp}, we note that while the number of galaxies is significantly different between the original samples (left panels), this difference is primarily caused by the stricter magnitude cut that \cite{Dimauro2019} use. We re-derive the best-fit lines for our sample when a magnitude cut of 23 in the F160W band is applied. This fit is shown in green and is consistent with the high mass stellar mass -- size relation that we derive for our sample, presented in Figure \ref{fig:SF_highM_M_Re}, and shown in Figure \ref{fig:dist_comp} as a black line. We therefore argue that the stellar mass -- size relation that we derive at the high mass end does not significantly change when we limit the sample to brighter galaxies. This suggests that the different results between this work and \cite{Dimauro2019} are not a direct consequence of a different magnitude cut. Additionally, from Figure \ref{fig:dist_comp}, it can be seen that when similar sample selections are applied to both samples, there are more galaxies with small sizes in our sample (vertical histogram on the right plot). The mass distributions of our sample (horizontal blue histogram on the right plot) are also skewed toward lower stellar masses compared to \cite{Dimauro2019}, such that below 10$^{10.3}$ M$_\odot$, we have a significantly larger number of galaxies. Above this stellar mass, the mass distributions of our samples are comparable. Since we have more compact, low mass objects than \cite{Dimauro2019}, this could possibly explain the different stellar mass -- size relations.

\begin{figure*}
    \centering
    \includegraphics[width=1\textwidth]{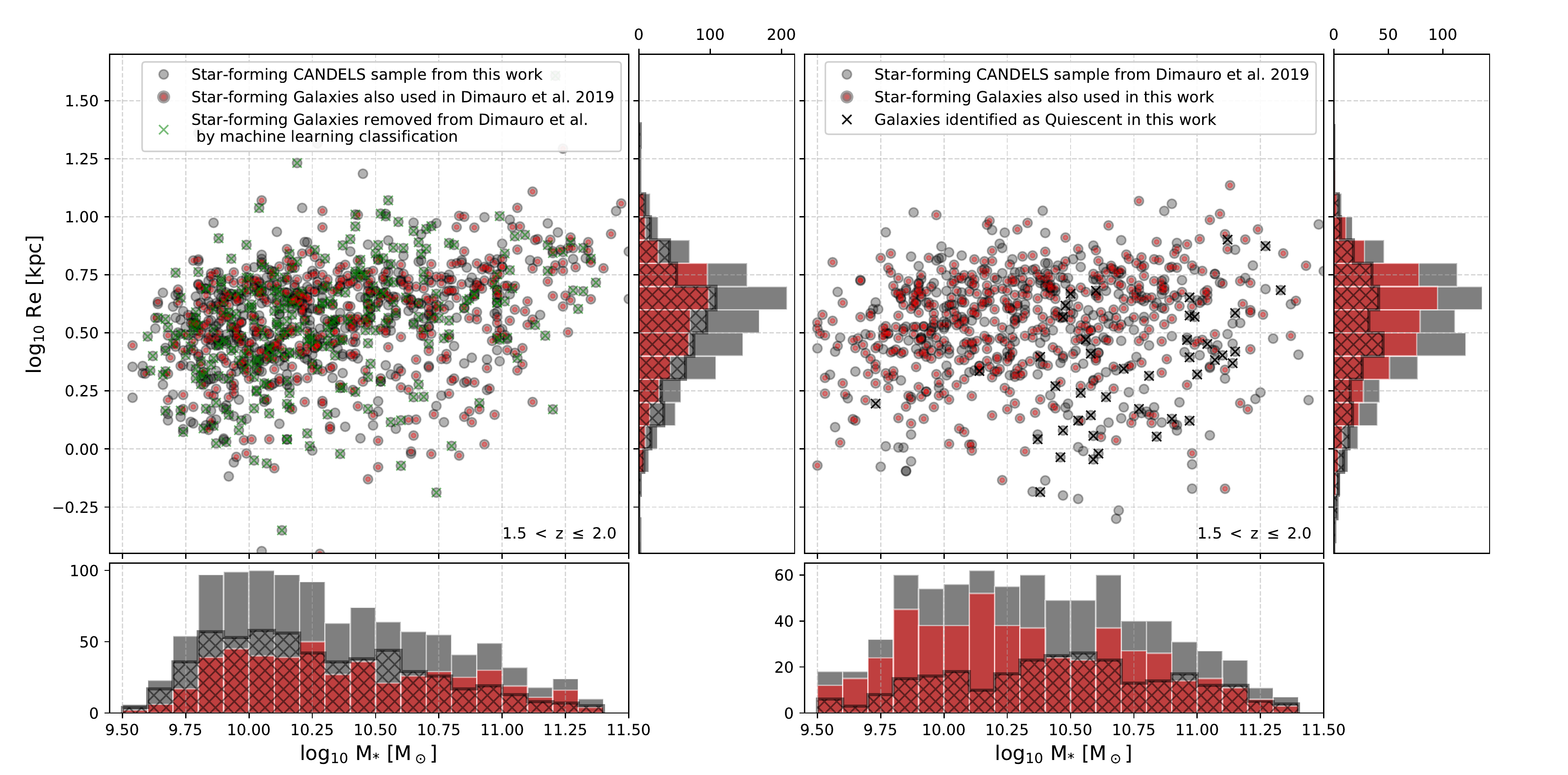}
    \caption{The stellar mass – size relation of high mass star-forming galaxies in the highest redshift bin from our sample (left panels) and the \protect\cite{Dimauro2019} (right panels). Galaxies included in both samples are shown in red. In the left panel, a large fraction of the sample that we use is not used in \protect\cite{Dimauro2019} as a result of the machine learning classification that they use (see \protect\cite{Dimauro2018MNRAS} for details). These objects are indicated in green.}
    \label{fig:matched_dist_comp}
\end{figure*}

To investigate this difference further, we match our sample to the \cite{Dimauro2019} sample in order to test the similarities between the galaxies that are used in both works versus those that are only used in one of the samples. The results are shown in Figure \ref{fig:matched_dist_comp}, where the left panels show our full sample in grey, where galaxies that are also used in \cite{Dimauro2019} are indicated with a red point. In the right panels, we show the same but for the \cite{Dimauro2019} sample (i.e., the \citealt{Dimauro2019} sample is shown in grey, where galaxies that are also used in this work are indicated in red). In the histograms, we show the size and stellar mass distributions of each work, where the grey histogram corresponds to the sample from each study (and in the left panel is identical to the blue histogram in Figure \ref{fig:dist_comp}). The red histograms are for galaxies that are used in both this work and \cite{Dimauro2019}, and the hashed histogram is for the objects that are not matched, i.e., only used in one of the works. 

For the star-forming galaxy population, we find that the stellar mass -- size relation has a slope of $\sim 0.2$ at all redshifts, while \cite{Dimauro2019} derive a shallower slope, especially at high redshift. Indeed, visually from Figure \ref{fig:matched_dist_comp}, the \cite{Dimauro2019} sample on the right appears flatter than ours on the left. Upon modelling the \cite{Dimauro2019} sample, we also obtain a shallow slope of $B = 0.03$ at $1.5 < z \leq 2$. Although the  fitting method used by both \cite{vdWel2014} and \cite{Dimauro2019} is different from ours, we find results that are consistent with \cite{vdWel2014} and we are able to reproduce the shallow results that \cite{Dimauro2019} find using their data. Likewise, if we fit our data with the same method as \cite{vdWel2014} and \cite{Dimauro2019}, by assuming the same functional form (i.e., Eq. \ref{eq:singlepl}), and computing the total likelihood for a set of six model parameters, we find very consistent parameters with those that we derive. These values are presented in Table \ref{tab:Paola_fit}, where our best-fit parameters for the high mass quiescent and star-forming galaxies from Tables \ref{tab:from_eq1} and \ref{tab:from_eq2} are shown in brackets for comparison. 

In the right panel of Figure \ref{fig:matched_dist_comp}, we also show the galaxies that are selected as quiescent by our UVJ diagram, but identified as star-forming by \cite{Dimauro2019}. These galaxies (shown as black crosses) are generally smaller and more massive compared to the overall star-forming sample, shown in grey, and therefore, including these objects in the fitting, would result in a flatter slope.

\begin{table*}
\caption{The estimated best-fit parameters for quiescent and star-forming galaxies following the same methods as \protect\cite{vdWel2014} and \protect\cite{Dimauro2019}  where $\sigma$ $\log_{10}$(Re) is the scatter in size. These results are closely consistent with the results from our own fitting method, which we report in brackets, and are shown in Tables \ref{tab:from_eq2} and \ref{tab:from_eq1} for the quiescent and star-forming samples, respectively.}
\centering
\begin{tabular}{ c c c c c c c}
\hline
\multicolumn{1}{l}{ \ } & \multicolumn{3}{c}{ Quiescent Galaxies} & \multicolumn{3}{c}{ Star-forming Galaxies} \\ 

$z$ & $\log_{10}$($A$) & $B$ & $\sigma \log_{10}$(Re)& $\log_{10}$($A$) & $B$ & $\sigma \log_{10}$(Re)\\
\hline 

0.2 $\leq z \leq$ 0.5 & {0.69} [0.61$\pm{0.01}$]
& {0.72} [0.68$\pm{0.04}$] & {0.19} & {0.78} [0.78$\pm{0.03}$] & 0.21 [0.22$\pm{0.03}$] & 0.22 \\
0.5 $<$ $z$ $\leq$ 1.0  & 
0.48 [0.45$\pm{0.01}$]& 0.72 [0.64$\pm{0.03}$] & 0.18 & 0.74 [0.74$\pm{0.01}$] & 0.21 [0.21$\pm{0.02}$]& 0.22 \\
1.0 $<$ $z$ $\leq$ 1.5  & 
0.31 [0.28$\pm{0.01}$] & 0.63 [0.63$\pm{0.04}$] & 0.20 & 0.66 [0.66$\pm{0.01}$]& 0.20 [0.21$\pm{0.01}$]& 0.22 \\
1.5 $<$ $z$ $\leq$ 2.0  &
0.24 [0.18$\pm{0.01}$] & 0.54 [0.61$\pm{0.05}$]& 0.25 & 0.59 [0.59$\pm{0.01}$] & 0.18 [0.20$\pm{0.02}$] & 0.23\\
\hline
\end{tabular}
\label{tab:Paola_fit}
\end{table*}

Finally, the way the uncertainties are obtained can significantly influence a model. We derive the size and mass uncertainties from \textsc{GalfitM} \citep{MegaMorph} and FAST \citep{Kriek2009}, respectively, while \cite{Dimauro2019} estimate the size uncertainties from their simulations 
%\bh{what are they simulating?}
and the stellar mass uncertainties are assumed to be proportional to the size uncertainties. We test this idea by fitting the matched objects from our work (i.e., the red points from the left panel in Figure \ref{fig:matched_dist_comp}) by using the uncertainties from \cite{Dimauro2019}. We find no significant change in the best-fit model when either set of uncertainties is used, therefore, the errors on the measurements are not driving the  discrepancy between our work and \cite{Dimauro2019}.

From all of the various tests carried out to investigate this discrepancy between our results, we are unable to say specifically which of these effects dominate the difference. However, we can rule out the fitting methods themselves as well as the weighting scheme used. From the visual impression in Figure \ref{fig:dist_comp}, it can be seen that the sample selection of quiescent and star-forming galaxies is different. For instance, the majority of the objects in the \cite{Dimauro2019} sample which are \textit{not} in our star-forming sample are massive and small, possibly indicating that they are elliptical quiescent galaxies. This leads us to the conclusion that the main difference is most likely a result of the sample selection.

%%%%%%%%%%%%%%%%%%%%%%%%%%%%%%%%%%%%%%%%%%%%%%%%%%
% Don't change these lines
\bsp	% typesetting comment
\label{lastpage}
\end{document}